\documentclass[onecolumn,12pt]{IEEEtran}
 \usepackage{graphicx}
\usepackage{amsmath}

\usepackage{graphics,graphicx,psfrag,color,float, hyperref}
\usepackage{epsfig,bbold,bbm}
\usepackage{bm}
\usepackage{tikz}
\usepackage{mathtools}
\usepackage{amsmath}
\usepackage{amssymb}
\usepackage{amsfonts}
\usepackage{amsthm, hyperref}
\usepackage{amsfonts}
\usepackage {times}
\usepackage{bbm}
\usepackage{soul}
\usepackage{titling}
\usepackage{amsmath}
\usepackage{layout}
\usepackage{amssymb}
\usepackage{amsfonts}
\usepackage{cleveref}
\usepackage{amsthm, hyperref}
\usepackage {times}
\usepackage{setspace}
\usepackage{amsfonts}
\usepackage{enumitem}
\newlist{steps}{enumerate}{1}
\setlist[steps, 1]{label = Step \arabic*}
\newlist{Cases}{enumerate}{1}
\setlist[Cases, 1]{label = Case \arabic*}

\newcommand{\abs}[1]{\left\vert#1\right\vert}

\newtheorem{fact}{Fact}

\newcommand{\beq}{\begin{equation}}
\newcommand{\enq}{\end{equation}}
\newcommand{\bel}{\begin{lemma}}
\newcommand{\enl}{\end{lemma}}
\newcommand{\bet}{\begin{theorem}}
\newcommand{\ent}{\end{theorem}}

\newcommand{\tr}{\mathrm{Tr}}

\newcommand{\ketbra}[1]{|#1\rangle\langle#1|}

\newcommand{\ceil}[1]{\left\lceil #1 \right\rceil}

\newcommand{\eps}{\varepsilon}
\newcommand*{\cC}{\mathcal{C}_{n}}
\newcommand*{\cA}{\mathcal{A}}

\newcommand*{\cH}{\mathcal{H}}
\newcommand*{\cM}{\mathcal{M}}

\newcommand*{\cB}{\mathcal{B}}
\newcommand*{\cD}{\mathcal{D}}

\newcommand*{\cN}{\mathcal{N}}
\newcommand*{\cS}{\mathcal{S}}

\newcommand*{\cT}{\mathcal{T}}
\newcommand*{\cX}{\mathcal{X}}

\newcommand*{\cE}{\mathcal{E}}

\newcommand*{\G}{\mbox{Good}}

\newcommand{\bra}[1]{\langle #1|}
\newcommand{\ket}[1]{|#1 \rangle}

\mathchardef\mhyphen="2D

\newcommand{\norm}[2]{{\left\lVert #1 \right\rVert}_{#2}}
\newcommand{\customfootnotetext}[2]{{% Group to localize change to footnote
  \renewcommand{\thefootnote}{#1}% Update footnote counter representation
  \footnotetext[0]{#2}}}
\makeatletter
\newcommand*{\rom}[1]{\expandafter\@slowromancap\romannumeral #1@}
\makeatother

\mathchardef\mhyphen="2D

\newtheorem{remark}{Remark}
\newtheorem{definition}{Definition}

\newtheorem{theorem}{Theorem}
\newtheorem{lemma}{Lemma}
\newtheorem{corollary}{Corollary}

\begin {document}
\title{Intersection and union of subspaces with applications to communication over authenticated classical-quantum channels and composite hypothesis testing}
\author{
Naqueeb Ahmad Warsi\textsuperscript{$*$} 
 and Ayanava Dasgupta\textsuperscript{$*$}
}
\customfootnotetext{$*$}{
Indian Statistical Institute,
Kolkata 700108, India.
Email: 
{\sf 
naqueebwarsi@isical.ac.in, ayanavadasgupta\_r@isical.ac.in
}
}
\date{}

\maketitle
\begin{abstract}
In information theory, we often use intersection and union of the typical sets to analyze various communication problems. However, in the quantum setting it is not very clear how to construct a measurement which behaves analogously to intersection and union of the typical sets. In this work, we construct a projection operator which behaves very similarly to intersection and union of the typical sets. Our construction relies on the Jordan's lemma. Using this construction we study the problem of communication over authenticated classical-quantum channels and derive its capacity. As another application of our construction, we also study the problem of quantum asymmetric composite hypothesis testing.
\end{abstract}
\section{Introduction}
\label{sec:introduction}
A fundamental bottleneck that often arises in quantum information theory is about calculating the probability of the intersection of two events.  Understanding this bottleneck will help us to understand many quantum network information theory problems in the one-shot setting, e.g., entanglement-assisted multiple access channel \cite{Hsieh_2008}, entanglement-assisted broadcast channel with common message in the one-shot setting \cite{Anshu_2019,Sen2021UnionsIA,Anshu_2022,Berta_2011}. Recently, the broadcast channel with a common message was studied for the case of a classical-quantum channel in \cite{Sen2021InnerB}. For details on one-shot setting see for example,  \cite{wang-renner-prl}, \cite{radhakrishnan-sen-warsi-archive-v1}, \cite{J.M.Renes}, and references therein. Sen also studied this problem from the point of view of communication over classical quantum multiple access channel \cite{Sen2021UnionsIA}, where he was able to resolve the intersection issue with respect to classical quantum states and their respective marginals.
% In this work, we make some progress towards understanding the intersection bottleneck for general quantum states and as an application, we give a proof for the converse of Quantum Chernoff Stein's lemma. Quantum Chernoff Stein's lemma is a well-studied problem, see for example, \cite{Ogawa_2005, https://doi.org/10.48550/arxiv.quant-ph/0307170}, \cite{Brandao:2010wh}. 

In the classical setting, when we calculate the probability of events under a particular probability distribution, it translates to the measurement of quantum states in a quantum setting. These measurements are in general POVM.  However, because of the Gelfand-Naimark's lemma \cite{MR0009426} it is enough to consider only projective measurements. Therefore, in this manuscript, we will only consider projective measurements. 

To understand the intersection issue in the quantum case, consider the following statement about classical probability distributions. Let $P,$ $Q_1$ and $Q_2$ be probability distributions over a set $\cX$ such that $\mbox{supp}(P) \subseteq \mbox{supp}(Q_1)$ and $\mbox{supp}(P) \subseteq \mbox{supp}(Q_2).$ 
Further, for ${k_1}, k_2 >0,$ let $\cA_1:=\{x: Q_1(x) \leq 2^{-{k_1}}P (x)\}$ and $\cA_2:=\{x: Q_2(x) \leq 2^{-{k_2}}P (x)\}$ with the property that $\Pr_P\{\cA_1\} \geq 1-\eps$ and $\Pr_P \{\cA_2\} \geq 1- \eps.$ Then, it is easy to see that for $\cA ^\star:= \cA_1 \cap \cA_2$, we have $\Pr_P\{\cA^\star\} \geq 1-2\eps,$ $\Pr_{Q_1}\{\cA^\star\} \leq 2^{-{k_1}}$  and $\Pr_{Q_2}\{\cA^\star\} \leq 2^{-{k_2}},$ where $\Pr_P\{\cA^\star\} \geq 1-2\eps,$ follows from the union bound for probability and the other two inequalities follows from the fact that $\cA ^\star \subseteq\cA_1$ and $\cA ^\star \subseteq \cA_2.$ We aim to prove a similar statement in the quantum setting.  To define the problem formally let us consider $\rho, \sigma_1$ and $\sigma_2 \in \mathcal{D}(\cH)$ such that $\mbox{supp}(\rho) \subseteq \mbox{supp}(\sigma_1)$ and $\mbox{supp}(\rho) \subseteq \mbox{supp}(\sigma_2)$. 
Further, for ${k_1}, k_2 >0,$ let $\Pi_1, \Pi_2$ be such that $\Pi_1:=\{\sigma_1 \preceq 2^{-{k_1}}\rho\}$ and  $\Pi_2:=\{\sigma_2 \preceq 2^{-{k_2}}\rho\}$ (see Section \ref{sec:bg} for the definition of $\{\cdot \preceq \cdot \}$) with the property that $\tr [\Pi_1 \rho] \geq 1-\eps$ and $\tr[\Pi_2\rho] \geq 1- \eps.$ Then, we want to find a projector $\Pi^\star$ such that $\tr[\Pi^{\star}\rho] \geq 1- f(\eps),$ $\tr[\Pi^{\star}\sigma_1] \leq 2^{-{k_1}}$ and $\tr[\Pi^{\star}\sigma_2] \leq 2^{-{k_2}},$ where $f(\eps)$ is some polynomial in $\eps$ (in the classical case it is $(1-2\eps)$).

We can also have a complementary version of the problem discussed above. Let $P,$ $Q_1$ and $Q_2$ be probability distributions over a set $\cX$ such that $\mbox{supp}(P) \subseteq \mbox{supp}(Q_1)$ and $\mbox{supp}(P) \subseteq \mbox{supp}(Q_2).$  Further, for $k_1,k_2 >0,$ let $\cA_1:=\{x: Q_1(x) \geq 2^{-{k_1}}P (x)\}$ and $\cA_2:=\{x: Q_2(x) \geq 2^{-{k_2}}P (x)\}$ with the property that $\Pr_P\{\cA_1\} \geq 1-\eps$ and $\Pr_P \{\cA_2\} \geq 1- \eps.$ Then, it is easy to see that for $\cA ^\star:= \cA_1 \cap \cA_2$, we have $\Pr_P\{\cA^\star\} \geq 1-2\eps,$ $\Pr_{Q_1}\{\cA^\star\} \geq 2^{-{k_1}} (1-2\eps)$  and $\Pr_{Q_2}\{\cA^\star\} \geq 2^{-{k_2}} (1-2\eps),$ where $\Pr_P\{\cA^\star\} \geq 1-2\eps,$ follows from the union bound for probability and the other two inequalities follows from the fact that $\cA ^\star \subseteq\cA_1$ and $\cA ^\star \subseteq \cA_2.$ We aim to prove a similar statement in the quantum setting. To define the problem formally let us consider $\rho, \sigma_1$ and $\sigma_2 \in \mathcal{D}(\cH)$ such that $\mbox{supp}(\rho) \subseteq \mbox{supp}(\sigma_1)$ and $\mbox{supp}(\rho) \subseteq \mbox{supp}(\sigma_2)$. Further, for $k_1, k_2 >0,$ let $\Pi_1, \Pi_2$ be such that $\Pi_1:=\{\sigma_1 \succeq 2^{-{k_1}}\rho\}$ and  $\Pi_2:=\{\sigma_2 \succeq 2^{-{k_2}}\rho\}$ with the property that $\tr [\Pi_1 \rho] \geq 1-\eps$ and $\tr[\Pi_2\rho] \geq 1- \eps.$ Then, we want to find a projector $\Pi^\star$ such that $\tr[\Pi^{\star}\rho] \geq 1- f(\eps),$ $\tr[\Pi^{\star}\sigma_1] \geq 2^{-{k_1}}f(\eps)$ and $\tr[\Pi^{\star}\sigma_2] \geq 2^{-{k_2}}f(\eps),$ where $f(\eps)$ is some polynomial in $\eps$ (in the classical case it is$(1-2\eps)$). 

We can have a more general version of the problem discussed above. Consider two finite collections of distributions $\cS_1:=\{P_i\}_{i=1}^{|\cS_1|}$ and $\cS_2:=\{Q_j\}_{j=1}^{|\cS_2|},$ defined over a set $\cX.$ Further, $\forall i \in \cS_1$ and $\forall j \in \cS_2,$ we are given a set $\cA_{ij} \subset \cX$ such that $\Pr_{P_i}\{A_{ij}\} \geq 1- \eps$ and $\Pr_{Q_j}\{A_{ij}\} \leq 2^{-k_{ij}},$ for some $k_{ij}>0.$ We want to construct an $\cA^\star$ such that $\forall i \in \cS_1, j \in \cS_2,$ we have $\Pr_{P_i}\{\cA^\star\} \geq 1- |\cS_2|\eps$ and $\Pr_{Q_j}\{\cA^\star\} \leq |\cS_1|.2^{-k},$ where $k:= \min_{i,j}k_{i,j.}$ It is easy to see that $\cA^\star:= \cap_{j=1}^{|\cS_2|}\cup_{i=1}^{|\cS_1|}\cA_{i,j}.$ We aim to prove a similar statement in the quantum setting. To define the problem formally let us consider two collections of quantum states $\Theta_{\cS_1} := \left\{\rho_i\right\}_{i=1}^{|\cS_1|}$ and $\Theta_{\cS_2} := \left\{\sigma_j\right\}_{j=1}^{|\cS_2|}$, defined over a Hilbert space $\cH$. Further for $\forall i \in \cS_1, j \in \cS_2$, we are given a projector $\Pi_{ij}$ such that $\tr[\Pi_{ij}\rho_i] \geq 1 - \eps$ and $\tr[\Pi_{ij}\sigma_j] \leq 2^{-k_{ij}}$ for some $k_{ij} > 0$. We want to find a projector $\Pi^{\star}$ which is similar in spirit to $\cA^\star:= \cap_{j=1}^{|\cS_2|}\cup_{i=1}^{|\cS_1|}\cA_{i,j}.$ 
%such that , we have $\forall i \in \cS_1, j \in \cS_2$ and a $\delta \in (0,1)$, we have $\tr[\Pi^{\star}\rho_i] \geq 1 - f(1/\delta)^{\log|\cS_2|}.(\eps + \log|
%\cS_2|.g(\delta^{1/2}))$ and $\tr[\Pi^{\star}\sigma_j] \leq f(1/\delta)^{\log|\cS_2|}.2^{-k},$ where $k:= \min_{i,j}k_{i,j.}, 
 %f(1/\delta)$ is a function of $1/\delta$ and $g(\delta^{1/2})$ is %a function of $\delta^{1/2}$. 
Such a $\Pi^{\star}$ will help us to study the problem of asymmetric composite hypothesis testing discussed in Lemma \ref{genmany}. 

In the classical case for two sets $\cA_1,\cA_2 \subseteq \cX$, it is well understood what does $\cA_{1 \cup 2} := \cA_1 \cup \cA_2$ and  $\cA_{1 \cap 2} := \cA_1 \cap \cA_2$ mean. Likewise, in the quantum case consider $\Pi_1,\Pi_2 \preceq \mathbb{I}_{\cH}$ which are projectors over some subspaces of Hilbert space $\cH$. Unlike the classical case, we do not have a very clear understanding of $\Pi_{1 \cup 2}$ and $\Pi_{1 \cap 2}$. One approach to construct $\Pi_{1 \cup 2}$, is by defining a projector on the span of the eigenvectors of the $\Pi_1$ and $\Pi_2$. But this approach has a drawback, because this span may be the whole Hilbert space $\cH$. Similarly, one may define $\Pi_{1 \cap 2}$, by first defining a projector on the span of the eigenvectors of $\Pi_1^c$ and $\Pi_2^c$ and then take its orthogonal compliment. This approach also has a drawback, because this span may be the whole Hilbert space $\cH$ and therefore $\Pi_{1 \cap 2}$ may not exist. Another approach for constructing $\Pi_{1 \cap 2}$ can be $\Pi_1\Pi_2\Pi_1.$ However, this construction also has an issue because $\Pi_1$ and $\Pi_2$ do not commute in general and sandwiching one projector with the other will depend on their ordering.
 
% A potential candidate for $\Pi^\star$ can be $\Pi_1\Pi_2\Pi_1.$ However, this construction has an issue because of the fact that $\Pi_1$ and $\Pi_2$ do not commute in general and sandwiching one projector by the other will depend on their ordering.

In this paper, we will address the issues discussed above. We will show that it is indeed possible to construct a $\Pi^\star$ such that it satisfies the properties that are very analogous to the intersection and union of subsets as discussed above. Our construction for $\Pi^\star$ heavily relies on Jordan's lemma \cite{BSMF_1875_3_103_2}.  This lemma accomplishes simultaneous block diagonalization of two projection operators. This remarkable observation from Jordan has found many applications in quantum information and computation, for example, \cite{MW-GAMES}, \cite{AJW-Compound} and also in post-quantum cryptography \cite{cryptoeprint:2021/334}.

We now mention the following lemma which we prove in this work. Almost all the results obtained in this manuscript are based on this lemma. For the notations used in the statement of the lemma below, see Section \ref{sec:bg}.
\begin{lemma} ({\bf{Upper-bound}})
\label{lbound}
{Let $\rho, \sigma_1$ and $\sigma_2 \in \mathcal{D}(\cH)$ such that $\mbox{supp}(\rho) \subseteq \mbox{supp}(\sigma_1)$ and $\mbox{supp}(\rho) \subseteq \mbox{supp}(\sigma_2)$. Further, for ${k_1}, k_2 >0,$ let $\Pi_1, \Pi_2$ be such that $\Pi_1:=\{\sigma_1 \preceq 2^{-{k_1}}\rho\}$ and  $\Pi_2:=\{\sigma_2 \preceq 2^{-{k_2}}\rho\}$ with the property that $\tr [\Pi_1 \rho] \geq 1-\eps$ and $\tr[\Pi_2\rho] \geq 1- \eps$,  where $\eps \in (0,1)$. Then, there exists a projector $\Pi^\star$ such that,} 
\begin{align}
&\tr[\Pi^{\star}\rho] \geq 1-2\eps^{\frac{1}{2}}, \label{r1}\\
     &\tr[\Pi^{\star}\sigma_1] \leq 2^{-{k_1}},\label{r2}\\
     &\tr[\Pi^{\star}\sigma_2] \leq 2^{-{k_2}}+4\sqrt{2}\eps^{\frac{1}{4}}.\label{r3}
\end{align} 
\end{lemma}
We also generalize the above lemma for the case when we have a collection $\{\sigma_i\}_{i=1}^{|\cS|},$ where $|\cS|< \infty.$
For details see Lemma \ref{generalubound} and its proof in subsection \ref{Generalised} in Appendix.

We use Lemma \ref{lbound} to study a problem of communication over a noisy classical-quantum channel in an adversarial setting.  In this setting, like the compound channel case \cite{WarsiCompound2019} we have a finite collection of channels $\{\cN^{(i)}_{X \to B}\}_{i=1}^{|\cS|}.$ However, unlike the compound channel scenario the goal here is that we aim to communicate successfully only if the channel $\cN^{(s_0)}_{X \to B}$ is being used. However, for any $s \neq s_0,$ we only want to detect that the channel $\cN^{(s_0)}_{X \to B}$ is not being used.

We can think of this scenario wherein a third party first authenticates the channel being used by inputting the symbol $s_0$ into the channel. However, if this third party cheats by inputting some other symbol $s \neq s_0$ in the channel. Then, we should either detect this or decode the message correctly. We define the capacity ($C_{\mbox{Auth}}$) for this model of communication and show that 
\begin{equation*}
C_{\mbox{Auth}}= \max_{p_X}I[X;B_{(s_0)}], 
\end{equation*}

where, $I[X;B_{(s_0)}]$ is calculated with respect to the channel $\cN^{(s_0)}_{X \to B}$ and defined in Lemma \ref{capacity_authentication}. It can be observed that this communication model is also about channel discrimination. In our achievability proof, we use Lemma \ref{lbound} to accomplish channel discrimination without disturbing the state a lot. Conditioned on the fact that we have detected the correct channel ($\cN^{(s_0)}_{X \to B}$) we then proceed to decode the message sent by the sender. We also give a converse for this problem.

A more general version of the above communication model was introduced and studied by Kosut and Kliewer in \cite{Kosut2018}. In this general version, parties are interested in communicating successfully only if the channel $\cN^{(s_0)}_{X \to B}$ is being used across all the $n$ channel uses. However, in other cases when the channel $\cN^{(s_0)}_{X \to B}$ is not being used across all the $n$ channel uses, i.e. if the channel statistic is changing across all the $n$ channel uses. Then, the receiver should either declare that  
the channel is behaving arbitrarily or decoding the message correctly.

With the help of the techniques used to prove Lemma $1$, we also give a proof for a complementary version of the problem discussed in Lemma \ref{lbound}, mentioned as Lemma \ref{mr} below.
\begin{lemma} ({\bf{Lower-bound, complementary version of Lemma \ref{lbound}}})
\label{mr}
{Let $\rho, \sigma_1$ and $\sigma_2 \in \mathcal{D}(\cH)$ such that $\mbox{supp}(\rho) \subseteq \mbox{supp}(\sigma_1)$ and $\mbox{supp}(\rho) \subseteq \mbox{supp}(\sigma_2)$. Further, for ${k_1}, k_2 >0,$ let $\Pi_1, \Pi_2$ be such that $\Pi_1:=\{\sigma_1 \succeq 2^{-{k_1}}\rho\}$ and  $\Pi_2:=\{\sigma_2 \succeq 2^{-{k_2}}\rho\}$ with the property that $\tr [\Pi_1 \rho] \geq 1-\eps$ and $\tr[\Pi_2\rho] \geq 1- \eps$,  where $\eps \in (0,1)$ and is arbitrarily close to zero. Then, there exists a projector $\Pi^\star$ such that,} 
\begin{align}
%\label{nr1}
%\Pi^\star &\preceq \Pi_1,\\
%\label{nr2}
%\Pi^\star &\preceq \Pi_2,\\
\label{rlb1}
\tr[\Pi^{\star}\rho] &\geq 1-2\eps^{\frac{1}{2}}, \\
\label{rlb2}
\tr[\Pi^{\star}\sigma_1] &\geq 2^{-{k_1}}( 1-2\eps^{\frac{1}{2}}),\\
\label{rlb3}
\tr[\Pi^{\star}\sigma_2] &\geq 2^{-{k_2}}( 1-10\eps^{\frac{1}{2}}) - 2\sqrt{2}\eps^{\frac{1}{4}}.
\end{align} 
\end{lemma}

The Corollary below follows from Lemma \ref{mr} and it can be viewed as a quantum generalization of Fact \ref{tcl}, mentioned in section \ref{sec:bg}. 
\begin{corollary}
\label{main}
\textit{ Let $\rho^{\otimes n}$ and $\sigma^{\otimes n}$ be two quantum states over the Hilbert space $\cH^{\otimes n}$ such that $D(\rho \|\sigma) < \infty$. Further, let $\Pi^{(2)}_n$ be a projector such  that $\tr[\Pi^{(2)}_n \rho^{\otimes n}] \geq 1-\eps.$ Then, 
\begin{align*}
\tr[\Pi^{(2)}_n\sigma^{\otimes n}]  \geq 2^{-n(D(\rho\| \sigma)+\delta)} f(\eps) - 2\sqrt{2}\eps^{\frac{1}{4}},
\end{align*}
where, $f(\eps) = (1-10\eps^{\frac{1}{2}})$ and $\delta \in (0,1)$ is arbitrary.}
\end{corollary}
However, unlike the classical result mentioned in Fact \ref{tcl}, we have a subtractive term in the quantum setting as mentioned in the above corollary.

\subsection{Comparison with previous works} 
 Sen in \cite{Sen2021UnionsIA}, resolved the intersection issue for the case of classical-quantum states and their marginals respectively, using ``tilting" and ``augmenting" the Hilbert space to a significantly larger Hilbert space. Furthermore, he mentions the following in  \cite{Sen2021UnionsIA},
 \begin{quote}
     ``The statement of our quantum joint typicality
lemma is not as strong as the classical statement. It can only
handle negative hypothesis states that are a tensor product of
marginals and at most one arbitrary quantum state. Proving
a quantum lemma that can handle arbitrary negative hypothesis states, as in the classical setting, is an important open
problem."
 \end{quote}
 
Unlike Sen \cite{Sen2021UnionsIA}, we get a small additive term (See \eqref{r3} in Lemma \ref{lbound}) because of the non-commutativity of the projection operators involved.

The problem of asymmetric composite hypothesis testing was also studied by Berta et al. in \cite{Berta2017OnCQ}. We make a comparison of our results (Lemma \ref{ssrho} and Lemma \ref{genmany}) with the results obtained in \cite{Berta2017OnCQ} in subsection \ref{bertasub}.

The rest of the paper is organized as follows. In Section \ref{sec:bg}, we mention the notations and facts used in this manuscript. Section \ref{proof_lbound} discusses a detailed proof for Lemma \ref{lbound} and also discusses two generalized versions of Lemma \ref{lbound}, which are mentioned as Lemma \ref{generalubound} and \ref{generalubound_seq}. In Section \ref{authenticated_section}, we study the problem of communication over authenticated classical-quantum channels and establish its capacity. In Section \ref{sec : Construc}, we give a proof for Lemma \ref{mr} and in Section \ref{sec:hypothesis_testing}, we study the problems of quantum asymmetric composite hypothesis testing under various settings.

\section{Notations and Facts}\label{sec:bg}
We use $\cH$ to denote a finite-dimensional Hilbert space and $\mathcal{D}(\cH)$ to represent the set of all quantum states acting on $\cH$. For any quantum state $\rho$, we define $\mbox{supp}(\rho):=\mbox{span}\{\ket{i}: \lambda_{i}> 0\},$ where $\left\{\lambda_i\right\}$ represent non-zero eigenvalues of $\rho$. Similarly, for any distribution $P$ defined over a set $\mathcal{X}$, $\mbox{supp}(P):=\{x\in \mathcal{X}: P(x)>0\}.$
Given two distribution $P \text{  and } Q$ over the set $\mathcal{X}$ such that $\mbox{supp}(P) \subseteq \mbox{supp}(Q)$, the relative entropy between $P$ and $Q$ is defined as,
\begin{equation*}
    D(P\|Q) := \sum_{x \in \mathcal{X}}P(x)\log\left(\frac{P(x)}{Q(x)}\right).
\end{equation*}

Given two quantum states $\rho, \sigma \in \mathcal{D}(\cH)$ such that $\mbox{supp}(\rho) \subseteq \mbox{supp}(\sigma)$, the quantum relative entropy (also known to be \emph{quantum KL-divergence}) between $\rho$ and $\sigma$ is defined as,
\begin{equation*}
    D(\rho\| \sigma):= \tr[\rho(\log(\rho)-\log(\sigma)].
\end{equation*}

For two normal operators $A$ and $B$ we will use the notation $\Pi :=\{A \succeq B\} \triangleq \{B \preceq A\}$ to denote a projection operator on the positive eigenspace of the operator $(A-B).$  We use $\rho^{\otimes n}$ over ${\cH}^{\otimes n}$, to represent $n$ independent copies of $\rho$.
\begin{fact}[{\cite[Lemma 11.8.1]{covertom}}]\label{tcl}
\textit{Let $B_n \subseteq \mathcal{X}^n$ be any set of sequences $x_1,x_2,\cdots,x_n$ such that $P_1(B_n) > 1-\eps.$ Let $P_2$ be any other distribution such that $D(P_1\|P_2) < \infty$. Then, $P_2(B_n) > (1- 2\eps)2^{-n(D(P_1\|P_2)+\delta)}$, where $\delta,\eps \in (0,1)$.}
\end{fact}
\begin{fact}[Jordan's lemma {\cite{BSMF_1875_3_103_2}\cite[Fact 6]{AJW-Compound}}]
\label{jordan}
For any two projectors $\Pi_{1}, \Pi_{2}$ there exists a set of orthogonal projectors $\{\mathbf{P}_{\alpha}\}_{\alpha=1}^{k}$ (each of dimension either one or two), for some natural number $k$, such that
\begin{enumerate}
    \item $\sum_{\alpha =1}^{k}\mathbf{P}_{\alpha} = \mathbb{I}$.
    \item $\mathbf{P}_{\alpha}\Pi_{i} = \Pi_{i}\mathbf{P}_{\alpha}$, for all $i \in \{1,2\}, \alpha \in [1:k]$.
    \item $\mathbf{P}_{\alpha}\Pi_{i}\mathbf{P}_{\alpha}$ is a one dimensional projector for $i \in \{1,2\}, \alpha \in [1:k]$.
\end{enumerate}
\end{fact}
\begin{fact}[Markov's inequality \cite{paul1971introduction}]
\label{markov}
    Let $X$ be any non-negative random variable. Then, for any $\lambda > 0$,
    \begin{align*}
        Pr\{X > \lambda \} \leq \frac{\mathbb{E}[X]}{\lambda}.
    \end{align*}
\end{fact}
\begin{fact}[{\cite[Theorem 2]{Nagaoka07}}]\label{F4}
Let $\rho^{\otimes n}, \sigma^{\otimes n} \in \cD(\cH^n)$.
\begin{itemize}
    \item[]  Consider 
    \begin{equation}
    \Pi_n:=\{\sigma^{\otimes n} \succeq 2^{-n(D(\rho\| \sigma)+\delta)}\rho^{\otimes n}\},\label{F41}
    \end{equation}
    
    where, $\delta \in (0,1)$. Then, 
    \begin{equation}
        \lim_{n \to \infty} \tr[\Pi_n\rho^{\otimes n}] = 1.\label{DM}
    \end{equation}
    \item[]  Consider 
    \begin{equation}
    \Pi_n:=\{\sigma^{\otimes n} \preceq 2^{-n(D(\rho\| \sigma)-\delta)}\rho^{\otimes n}\},\label{F42}
    \end{equation}
    
    where, $\delta \in (0,1)$. Then, 
    \begin{equation}
        \lim_{n \to \infty} \tr[\Pi_n\rho^{\otimes n}] = 1.\label{DM2}
    \end{equation}
\end{itemize}
\end{fact}
% \begin{proof}
%     Equations \eqref{DM}, \eqref{DM2} can be proved directly from \cite[Theorem 2]{Nagaoka07} which is stated as follows:
%     \begin{equation*}
%         \lim_{n \to \infty} \tr \left[\left(\rho^{\otimes n} - 2^{na} \sigma^{\otimes n}\underset{(=)}{\succ} 0\right )\rho^{\otimes n} \right] = \begin{cases}
%             1  & \text{if } a < D(\rho \| \sigma)\\
%             0  & \text{if
%             } a > D(\rho \| \sigma).
%         \end{cases}
%     \end{equation*}
%     If we consider $a := D(\rho \| \sigma) - \delta$ in the above equation, it directly follows that \eqref{DM2} holds. 
%     Now we can rewrite the above equation as follows:
%     \begin{equation*}
%         \lim_{n \to \infty} \tr \left[\left(\rho^{\otimes n} - 2^{nb} \sigma^{\otimes n}\underset{(=)}{\prec} 0\right )\rho^{\otimes n} \right] = \begin{cases}
%             0  & \text{if } b < D(\rho \| \sigma)\\
%             1  & \text{if
%             } b > D(\rho \| \sigma).
%         \end{cases}
%     \end{equation*}
%     If we consider $b := D(\rho \| \sigma) + \delta$ in the above equation, it directly follows that \eqref{DM} holds.
% \end{proof}
\begin{fact}[\cite{Datta-2008, marcobook}]\label{tp}
    Consider a classical quantum state $\rho_{XB} := \sum_{x}P_{X}(x)\ketbra{x}\otimes\rho_{x}^{B}.$ For every $x^n \in \cX^{n}$, $\Pi_{x^n,\delta} := \left\{\rho_{x^n}^{B^n} \succeq 2^{n\left(I[X;B] - \delta \right)}\rho^{B^{\otimes n}}\right\}$ (where $\delta \in (0,1)$), $\rho_{x^n}^{B^n}= \otimes_{i=1}^{n}\rho_{x_i}^{B}$ and $\rho^{B} := \mathbb{E}_{X} \left[\rho_{X}^{B}\right]$. Then, for $n$ large enough, 
    \begin{equation*}
        \mathbb{E}_{X^{n}} \tr\left[\Pi_{X^{n},\delta}\rho_{X^n}^{B^n}\right] \geq 1 - \eps,
    \end{equation*}
    
    where, $\eps \in (0,1).$
\end{fact}
% Let $\rho^{\otimes n}, \sigma^{\otimes n} \in \cD(\cH^n).$ Consider $\Pi_n:=\{\sigma^{\otimes n} \succeq 2^{-n(D(\rho\| \sigma)+\delta)}\rho^{\otimes n}\},$ where $\delta \in (0,1)$ Then, $\lim_{n \to \infty} \tr[\Pi_n\rho^{\otimes n}] \to 1.$ 

\begin{fact}[Gao's union bound \cite{Gao_2015,RyanRamgopal2021}
]\label{Gao}
Let $\Pi_1,\Pi_2, \cdots, \Pi_n$ be projectors over $\cH$ and $\rho \in \cD(\cH)$. Then,  
\begin{equation*}
\tr(\Pi_n \cdots \Pi_2\Pi_1 \rho \Pi_1\Pi_2 \cdots \Pi_n) \geq 1-4 \sum_{i=1}^n \tr[\Pi^c_i\rho],
\end{equation*}

where, $\Pi^c_i= \mathbb{I} - \Pi_i$.
\end{fact}
\begin{fact}[\cite{wilde-book}]
\label{trace_norm}
    Given a projective measurement $\Pi$ and two states $\rho_1,\rho_2 \in \mathcal{D}(\mathcal{H})$, we have,
    \begin{equation*}
        \tr[\Pi(\rho_1 - \rho_2)] \leq \norm{\rho_1 - \rho_2}{1},
    \end{equation*}
    
    where, $\norm{\rho_1 - \rho_2}{1} := \tr[\sqrt{(\rho_1 -\rho_2)^{\dagger}(\rho_1 -\rho_2)}]$.
\end{fact}
\begin{fact}[Hayashi Nagaoka's inequality \cite{HayashiNagaoka2003}]\label{hayashi_nagaoka}
    Consider $U$  and $V$ be two operators with $0 \preceq U \preceq \mathbb{I}$ and $ V \succeq 0$. Then,
    \begin{equation*}
        \mathbb{I} - {\left(U + V\right)}^{-\frac{1}{2}}U{\left(U + V\right)}^{-\frac{1}{2}} \preceq 2\left(\mathbb{I} - U\right) + 4V. 
    \end{equation*}
\end{fact}
\begin{fact}[Gentle measurement lemma under ensemble of states \cite{WinterG}]\label{gent_measurement}
    Let $\{p(x),\rho_x\}$ be an ensemble and let $\bar{\rho}:= \sum_{x}p(x)\rho_x$. If an operator $E$, where $0\preceq E \preceq I$, has high overlap with the expected state $\bar{\rho}$ i.e. $\tr[E\bar{\rho}] \geq 1 - \eps$, where $\eps \in (0,1)$. Then,
    \begin{equation*}
        \mathbb{E}_{X}\left[\norm{\sqrt{E}\rho_{X}\sqrt{E} - \rho_{X}}{1}\right]\leq 2\eps^{\frac{1}{2}}.
    \end{equation*}
\end{fact}
% \begin{claim} ({\bf{Main Result for two projectors}})
% \label{claim:mru} 
% \textit{Let $\rho, \sigma_1$ and $\sigma_2 \in \mathcal{D}(\cH)$ such that $\mbox{supp}(\rho) \subseteq \mbox{supp}(\sigma_1)$ and $\mbox{supp}(\rho) \subseteq \mbox{supp}(\sigma_2)$. Further, for ${k_1}$ and $ k_2 >0,$ let $\Pi_1$ and $\Pi_2$ be such that $\Pi_1:=\{\sigma_1 \preceq 2^{-{k_1}}\rho\}$ and  $\Pi_2:=\{\sigma_2 \preceq 2^{-{k_2}}\rho\}$ with the property that for some $\eps > 0$, $\tr [\Pi_1 \rho] \geq 1-\eps$ and $\tr[\Pi_2\rho] \geq 1- \eps.$ Then, there exists a projector $\Pi^{\star}$ such that for some $1/2 > 0$} 
% \begin{align}
% %\label{nr1}
% %\Pi^{\star} &\preceq \Pi_1,\\
% %\label{nr2}
% %\Pi^{\star} &\preceq \Pi_2,\\
% &\tr[\Pi^{\star}\rho] \geq 1-3\eps^{\frac{1}{2}} \label{r1}\\
%      &\tr[\Pi^{\star}\sigma_1] \leq 2^{-{k_1}}\label{r2}\\
%      &\tr[\Pi^{\star}\sigma_2] \leq 2^{-    {k_2}}+4\sqrt{2}\eps^{\frac{1}{4}}\label{r3}
% \end{align} 
% % \begin{align}
% % %\label{nr1}
% % %\Pi^\star &\preceq \Pi_1,\\
% % %\label{nr2}
% % %\Pi^\star &\preceq \Pi_2,\\
% % \label{r1}
% % \tr[\Pi^{\star}\rho] &\geq (1-4\eps^{\frac{1}{2}}), \\
% % \label{r3}
% % \tr[\Pi^{\star}\sigma_2] &\leq 2^{-{k_2}}+ 10\eps^{\frac{1}{4}},\\
% % \label{r2}
% % \tr[\Pi^{\star}\sigma_1] &\leq 2^{-{k_1}}
% % \end{align} 
% \end{claim}
\section{Proof of Lemma \ref{lbound}}\label{proof_lbound}
\subsection{Construction of the Intersection Projector $\Pi^{\star}$}\label{2_proj_inter}
Let $\Pi_1$ and $\Pi_2$ be as mentioned in Lemma \ref{lbound}. We will apply Fact \ref{jordan} on the pair $\{\Pi_1, \Pi_2\}.$ Consider $\left\{\mathbf{P}_{\alpha}\right\}_{{\alpha}=1}^{{\bar{k}}}$ (where $\bar{k}$ is some natural number depending on $\Pi_1,\Pi_2$) as the set of orthogonal projectors obtained by Fact \ref{jordan} applied with respect to the pair $\{\Pi_1, \Pi_2\}.$ Thus, $\sum_{{\alpha} =1}^{{\bar{k}}}\mathbf{P}_{\alpha} = \mathbb{I}.$ Furthermore, for every ${\alpha} \in [1:{\bar{k}}]$ we define  $\Pi_{1,{\alpha}} := \mathbf{P}_{\alpha}\Pi_1\mathbf{P}_{\alpha}$ as the following one dimensional projector: 
%%\vspace{5pt}
\begin{align}
\Pi_{1,{\alpha}} := \ket{v_{\alpha}}\bra{v_{\alpha}}\nonumber,
\end{align}
for some $\ket{v_{\alpha}}$ in the range of $\mathbf{P}_{\alpha}.$ Similarly, for every ${\alpha} \in [1:{\bar{k}}]$, we have
\begin{align}
\Pi_{2,{\alpha}} := \ket{w_{\alpha}}\bra{w_{\alpha}}\nonumber,
\end{align}
%%\vspace{5pt}
for some $\ket{w_{\alpha}}$ in the range of $\mathbf{P}_{\alpha}.$
Thus, it now follows from the property of $\{\ket{v_{\alpha}}\}_{{\alpha}=1}^{\bar{k}}$ and $\{\ket{w_{\alpha}}\}_{{\alpha}=1}^{\bar{k}}$ that 
%%\vspace{5pt}
\begin{align*}
\Pi_1 &:= \sum_{{\alpha} =1}^{{\bar{k}}} \ket{v_{\alpha}}\bra{v_{\alpha}},
\end{align*}
\begin{align*}
\Pi_2&:= \sum_{{\alpha} =1}^{{\bar{k}}} \ket{w_{\alpha}}\bra{w_{\alpha}}.
\end{align*}
%%\vspace{5pt}

Further, for every ${\alpha} \in [1:{\bar{k}}],$ let 
%%\vspace{5pt}
\begin{align}
\rho_{\alpha} &:= \mathbf{P}_{\alpha} \rho \mathbf{P}_{\alpha},\label{rhoalphal1}\\
\sigma_{1,{\alpha}} &:= \mathbf{P}_{\alpha} \sigma_1 \mathbf{P}_{\alpha},\\
\sigma_{2,{\alpha}} &:= \mathbf{P}_{\alpha} \sigma_2 \mathbf{P}_{\alpha}.
\end{align}
%%\vspace{5pt}
% Thus, from the definitions of $\Pi_1$ and $\Pi_2$, we have the following pair of inequalities for every ${\alpha} \in [1:{\bar{k}}],$
% %%\vspace{5pt}
% \begin{align}
% \label{property1}
% \bra{v_{\alpha}} \sigma_{1,{\alpha}}\ket{v_{\alpha}} &\leq 2^{-{k_1}} \bra{v_{\alpha}}\rho_{\alpha}\ket{v_{\alpha}}, \\
% \label{property2}
% \bra{w_{\alpha}} \sigma_{2,{\alpha}}\ket{w_{\alpha}} &\leq 2^{-{k_2}} \bra{w_{\alpha}}\rho_{\alpha}\ket{w_{\alpha}}, 
% \end{align}
% %%\vspace{5pt}

For our future discussions, it will be useful to make the following observations, for every ${\alpha} \in [1:{\bar{k}}],$ $\rho_{\alpha}\succeq 0$ and 
$\sum_{{\alpha}=1}^{{\bar{k}}}\tr[\rho_{\alpha}]= 1.$ Thus, $\{\tr[\rho_{\alpha}]\}_{{\alpha}=1}^{{\bar{k}}}$ forms a valid probability distribution over ${\alpha} \in [1:{{\bar{k}}}]$. 
To construct an intersection projector $\Pi^\star$ that satisfies \eqref{r1}, \eqref{r2}, \eqref{r3} we first make few observations. From Jordan's lemma (Fact \ref{jordan}) we have that for every ${\alpha} \in [1:\bar{k}]$ the vectors $\ket{w_{\alpha}}$ and $\ket{v_{\alpha}}$ lie in either a two-dimensional subspace or one-dimensional subspace. 
For the blocks where both $\ket{w_{\alpha}}$ and $\ket{v_{\alpha}}$ lie in a two-dimensional subspace we have the following,  
\begin{equation}
\label{valphatilda}
\ket{v_{\alpha}} = \cos(\theta_{\alpha})\ket{w_{\alpha}} + \sin(\theta_{\alpha})\ket{w^{\perp}_{\alpha}}.
\end{equation}

For the blocks, where both 
$\ket{w_{\alpha}}$ and $\ket{v_{\alpha}}$ lie in one dimensional subspace then in that case $\cos(\theta_{\alpha})=1,$ i.e., $\ket{w_{\alpha}} = \ket{v_{\alpha}}$ for these cases. Now we define the following set
%%\vspace{5pt}
\begin{align}
\G&:= \left\{{\alpha}: \cos^2(\theta_{\alpha}) \geq 1- 8\eps^{\frac{1}{2}}\right\},\label{good1}
\end{align}
where, $\cos^2(\theta_{\alpha})= {|\langle v_{\alpha}| w_{\alpha}\rangle|}^2$. The intuition for defining the set $\textnormal{\G}$ is that we only want to consider those blocks where $\ket{v_{\alpha}}$ and $\ket{w_{\alpha}}$ are almost the same, i.e., the angle between them is very small. 
From Lemma \ref{c23_gub} mentioned below, it follows that under the probability distribution $\{\tr[\rho_{\alpha}]\}_{\alpha=1}^k$,

\begin{equation}
    \Pr\{\textnormal{\G}\} \geq 1- \eps^{\frac{1}{2}}.\label{c23}
\end{equation}

\begin{lemma}
\label{c23_gub}
Consider two projectors $\Pi_1,\Pi_2$ over Hilbert space $\cH$ and a state $\rho \in \cD(\cH)$ such that $\tr[\Pi_1 \rho] \geq 1 - \eps_1$ and $\tr[\Pi_2 \rho] \geq 1 - \eps_2,$ where, $\eps_1,\eps_2 \in (0,1)$. Let $\{\rho_\alpha\}_{\alpha = 1}^{k}$ be the diagonal blocks of $\rho$ obtained from the Jordan decomposition of $\{\Pi_1,\Pi_2\}$ (see \eqref{rhoalphal1}). Let 
    $\textnormal{\G} := \left\{\alpha : \cos^2(\theta_{\alpha}) \geq 1- \delta\right\},$
where $\delta \in (0,1)$ and $\theta_\alpha$ is defined similarly as in \eqref{valphatilda}. Then,
$$\Pr\{\textnormal{\G}\} \geq 1- \frac{4(\eps_1 + \eps_2)}{\delta},$$
where the above probability is calculated with respect to the probability distribution $\{\tr[\rho_{\alpha}]\}_{\alpha=1}^k$.
\end{lemma}
\begin{proof}
    See Appendix  \ref{proofc23_gub}.
\end{proof}
%%\vspace{5pt}

To show the existence of $\Pi^{\star}$ which satisfies \eqref{r1}, \eqref{r2} and \eqref{r3} we construct $\Pi^{\star}$ as follows, 
%%\vspace{5pt}
\begin{equation}
\label{tilda}
\Pi^{\star}:= \sum_{{\alpha} \in \mbox{\G}} \ket{v_{\alpha}}\bra{v_{\alpha}}.
\end{equation}
%%\vspace{5pt} \\

We will now show that $\Pi^\star$ satisfies \eqref{r1}, \eqref{r2} and \eqref{r3} in the following subsection.
%%\vspace{5pt}
\subsection{Proof for the  properties of $\Pi^{\star}$}\label{pi_star_prop}
% We now show that $\Pi^{\star}$ has properties similar to that of $\cA^\star = \cA_1 \cap \cA_2$ as discussed in the introduction. In particular we show that $\Pi^{\star}$ satisfies \eqref{r1}, \eqref{r2}, \eqref{r3}. 
    \subsubsection*{Proof of \eqref{r1}}
%%\vspace{5pt}
Consider the following set of inequalities,

\begin{align*}
\tr[\Pi^{\star}\rho] &\overset{a}=\sum_{{\alpha}\in \mbox{\G}}\tr[\ket{v_{\alpha}}\bra{v_{\alpha}} \rho_{\alpha}]\\ 
&= \sum_{{\alpha}=1}^{{\bar{k}}}\tr[\ket{v_{\alpha}}\bra{v_{\alpha}} \rho_{\alpha}] - \sum_{{\alpha}\notin \mbox{\G}}
\tr[\ket{v_{\alpha}}\bra{v_{\alpha}} \rho_{\alpha}] \\
&\overset{b} \geq 1- \eps - \sum_{{\alpha}\notin\mbox{\G}}\tr[\rho_{\alpha}]\\
&\overset{c} \geq 1-\eps -\eps^{\frac{1}{2}}\\
&\geq 1 -2\eps^{\frac{1}{2}},
\end{align*}
where, $a$ follows from the definition of \eqref{tilda}, $b$ follows from the fact that $\sum_{{\alpha}=1}^{{\bar{k}}}\tr[\ket{v_{\alpha}}\bra{v_{\alpha}} \rho_{\alpha}] = \tr[\Pi_1 \rho] \geq 1 - \eps$ and $c$ follows from \eqref{c23}. 
%%\vspace{5pt}
\subsubsection*{Proof of \eqref{r2}}Consider the following set of inequalities,
\begin{align*}
\tr[\Pi^{\star}\sigma_1] &\overset{a}=
 \sum_{{\alpha}\in \mbox{\G}} \tr[\ket{v_{\alpha}}\bra{v_{\alpha}} \sigma_{1,{\alpha}} ] \\
&\overset{b} \leq 2^{-{k_1}}\sum_{{\alpha} \in \G}\tr[\ket{v_{\alpha}}\bra{v_{\alpha}} \rho_{\alpha}]\\
&\leq  2^{-{k_1}}\sum_{{\alpha} \in \G}\tr[ \rho_{\alpha}]\\
&\leq  2^{-{k_1}},
%&\geq  2^{-{k_2}}(1-\eps^{\frac{1}{2}})\sum_{{\alpha} \in \G}\cos(2\theta_{\alpha})\tr[\rho_{\alpha}]\\
% &\overset{d}\lgeq 2^{-{k_1}}
\end{align*}
where, $a$ follows from the definition of $\Pi^{\star}$ in $\eqref{tilda},$ $b$ follows from the property of the projector $\Pi_1$ mentioned in the statement of Lemma \ref{lbound}.
%\vspace{15pt}
\subsubsection*{Proof of \eqref{r3}}

 Consider the following set of inequalities,
\begin{align}
    \tr[\Pi^{\star}\sigma_2] &\overset{a}= \sum_{{\alpha}\in \mbox{\G}} \tr[\ket{v_{\alpha}}\bra{v_{\alpha}}\sigma_{2,{\alpha}}] \nonumber\\
&\overset{b}=\sum_{{\alpha} \in \G}\bigg(\cos^2(\theta_{\alpha})\tr[\ket{w_{\alpha}}\bra{w_{\alpha}}\sigma_{2,{\alpha}}] + \sin^2(\theta_{\alpha})\tr[\ket{w^{\perp}_{\alpha}}\bra{w^{\perp}_{\alpha}} \sigma_{2,{\alpha}}] \nonumber\\
&\hspace{60pt}+ \cos(\theta_{\alpha})\sin(\theta_{\alpha})\left(\tr[\ket{w_{\alpha}}\bra{w^{\perp}_{\alpha}} \sigma_{2,{\alpha}}] + \tr[\ket{w^{\perp}_{\alpha}}\bra{w_{\alpha}} \sigma_{2,{\alpha}}]\right) \bigg) \nonumber\hspace{80pt}
\end{align}
\begin{align}
&\overset{c} \leq \sum_{{\alpha} \in \G}\bigg(\tr[\ket{w_{\alpha}}\bra{w_{\alpha}} \sigma_{2,{\alpha}}] +  \sin^2(\theta_{\alpha})\tr[\ket{w^{\perp}_{\alpha}}\bra{w^{\perp}_{\alpha}} \sigma_{2,{\alpha}}] \nonumber\\
&\hspace{60pt}+ \cos(\theta_{\alpha})\sin(\theta_{\alpha})\left(\tr[\ket{w_{\alpha}}\bra{w^{\perp}_{\alpha}} \sigma_{2,{\alpha}}] + \tr[\ket{w^{\perp}_{\alpha}}\bra{w_{\alpha}} \sigma_{2,{\alpha}}]\right) \bigg) \nonumber\\
&\overset{d}{\leq}\tr[\Pi_2\sigma_2] + \sum_{{\alpha} \in \G}\bigg(8\eps^{\frac{1}{2}}\tr[\ket{w^{\perp}_{\alpha}}\bra{w^{\perp}_{\alpha}} \sigma_{2,{\alpha}}] \nonumber\\
&\hspace{60pt}+ \cos(\theta_{\alpha})\sin(\theta_{\alpha})\left(\tr[\ket{w_{\alpha}}\bra{w^{\perp}_{\alpha}} \sigma_{2,{\alpha}}] + \tr[\ket{w^{\perp}_{\alpha}}\bra{w_{\alpha}} \sigma_{2,{\alpha}}]\right) \bigg) \nonumber\\
% &\overset{e}{\leq}\tr[\Pi_2\sigma_2] + \sum_{{\alpha} \in \G}\bigg(2\sqrt{2}\eps^{\frac{1}{4}}\tr[\ket{w^{\perp}_{\alpha}}\bra{w^{\perp}_{\alpha}} \sigma_{2,{\alpha}}] \nonumber\\
% &\hspace{60pt}+ 2\sqrt{2}\eps^{\frac{1}{4}}\left(\tr[\ket{w_{\alpha}}\bra{w^{\perp}_{\alpha}} \sigma_{2,{\alpha}}] + \tr[\ket{w^{\perp}_{\alpha}}\bra{w_{\alpha}} \sigma_{2,{\alpha}}]\right) \bigg) \nonumber\\
&\overset{e}{\leq}\tr[\Pi_2\sigma_2] + \sum_{{\alpha} \in \G}\bigg(8\eps^{\frac{1}{2}}\tr[\ket{w^{\perp}_{\alpha}}\bra{w^{\perp}_{\alpha}} \sigma_{2,{\alpha}}] \nonumber + 2\sqrt{2}\eps^{\frac{1}{4}}\left(\tr[\sigma_{2,{\alpha}}] \right) \bigg)\hspace{65pt} \nonumber\\
&\overset{f}{\leq} 2^{-{k_2}}\tr[\Pi_2\rho] + \sum_{{\alpha} \in \G}\bigg(8\eps^{\frac{1}{2}}\tr[\sigma_{2,{\alpha}}] \nonumber + 2\sqrt{2}\eps^{\frac{1}{4}}\tr[\sigma_{2,{\alpha}}] \bigg) \nonumber\\
&\leq 2^{-{k_2}} + 2\sqrt{2}\eps^{\frac{1}{4}} + 8\eps^{\frac{1}{2}},\nonumber\\
&\overset{g}{\leq} 2^{-{k_2}} + 4\sqrt{2}\eps^{\frac{1}{4}},
\end{align}
% &\overset{h}\geq2^{-{k_2}}(1-\eps^{\frac{1}{2}})(1-16\eps^{\frac{1}{2}}) \sum_{{\alpha} \in \G}\left(\tr[\rho_{\alpha}]\right)\nonumber\\
% \label{mrf}
% %&\geq  2^{-{k_2}}(1-\eps^{\frac{1}{2}})\sum_{{\alpha} \in \G}\cos(2\theta_{\alpha})\tr[\rho_{\alpha}]\\
% &\overset{i}\geq 2^{-{k_2}}(1-\eps^{\frac{1}{2}})(1-16\eps^{\frac{1}{2}})(1-3\eps^{\frac{1}{2}})\nonumber\\
% &\geq 2^{-{k_2}}(1-20\eps^{\frac{1}{2}}),
% \\
where, $a$ follows from the definition of $\Pi^{\star}$ (see \eqref{tilda}), $b$ follows from \eqref{valphatilda}, $c$ follows because $\cos^2({\theta_{\alpha}}) \leq 1$, $d$ follows from fact that from \eqref{good1}, for all $\alpha \in \G$, $\sin^{2}(\theta_{\alpha}) \leq 8\eps^{\frac{1}{2}}$ , $e$ follows from the following series of inequalities:
\begin{align}
   &\cos(\theta_{\alpha})\sin(\theta_{\alpha})\left(\tr[\ket{w_{\alpha}}\bra{w^{\perp}_{\alpha}} \sigma_{2,{\alpha}}] + \tr[\ket{w^{\perp}_{\alpha}}\bra{w_{\alpha}} \sigma_{2,{\alpha}}]\right)\nonumber\\
   &\leq\abs{\cos(\theta_{\alpha})\sin(\theta_{\alpha})\left(\tr[\ket{w_{\alpha}}\bra{w^{\perp}_{\alpha}} \sigma_{2,{\alpha}}] + \tr[\ket{w^{\perp}_{\alpha}}\bra{w_{\alpha}} \sigma_{2,{\alpha}}]\right)}\nonumber\\
   &= \abs{\cos(\theta_{\alpha})\sin(\theta_{\alpha})}\abs{\left(\tr[\ket{w_{\alpha}}\bra{w^{\perp}_{\alpha}} \sigma_{2,{\alpha}}] + \tr[\ket{w^{\perp}_{\alpha}}\bra{w_{\alpha}} \sigma_{2,{\alpha}}]\right)}\nonumber\\
   &= \abs{\cos(\theta_{\alpha})}\abs{\sin(\theta_{\alpha})}\abs{\left(\tr[\ket{w_{\alpha}}\bra{w^{\perp}_{\alpha}} \sigma_{2,{\alpha}}] + \tr[\ket{w^{\perp}_{\alpha}}\bra{w_{\alpha}} \sigma_{2,{\alpha}}]\right)}\nonumber\\
   &\leq 2\sqrt{2}\eps^{\frac{1}{4}} \tr[\sigma_{2,{\alpha}}],\label{extra_term_bound}
\end{align}
where, the last inequality follows from the facts that for all $\alpha \in \G$, $|\sin(\theta_{\alpha})| \leq 2\sqrt{2}\eps^{\frac{1}{4}}, \abs{\cos(\theta_{\alpha})} \leq 1$ and from \eqref{positivity_res_1} and \eqref{positivity_res_2} of Lemma \ref{positivity} since $\sigma_{2,{\alpha}} \succeq 0$. Ineqality $f$ follows from the definition of $\Pi_2$ (see the statement of Lemma \ref{lbound}) and $g$ follows since $8\eps^{\frac{1}{2}} \in (0,1]$ is arbitrarily close to zero, $8\eps^{\frac{1}{2}} \leq \sqrt{8\eps^{\frac{1}{2}}} = 2\sqrt{2}\eps^{\frac{1}{4}}$.

 %{\footnote{For the case where, $\sin(\theta_{\alpha}) \geq  \cos(\theta_{\alpha})$ we will bound $d$ by lower bounding $\sin(\theta_{\alpha})$ by  $\cos(\theta_{\alpha}).$}}%

% where, $a$ follows from the definition of $\Pi^{\star}$ (see \eqref{tilda}), $b$ follows from \eqref{valphatilda}, $c$ follows because $\cos^2({\theta_{\alpha}}) \leq 1$ and for the sake of simplicity $\cos({\theta_{\alpha}}) \leq 1$, $d$ follows from \eqref{property2} and $e$ follows since in \eqref{good1}, $\eps$ is arbitrarily small, therefore $\theta_{\alpha} \in [0,\pi/4]$ and   $\sin^2(\theta_{\alpha}) \leq 8\eps^{\frac{1}{2}}$ and hence, $\sin(\theta_{\alpha}) \leq 2\sqrt{2}\eps^{\frac{1}{4}}$ %{\footnote{For the case where, $\sin(\theta_{\alpha}) \geq  \cos(\theta_{\alpha})$ we will bound $d$ by lower bounding $\sin(\theta_{\alpha})$ by  $\cos(\theta_{\alpha}).$}}%
% , $f$ follows because $\sigma_{2,{\alpha}}\succeq 0$ and from \eqref{positivity_res_1} of lemma \eqref{positivity}. 
\begin{lemma}
\label{positivity}
Let $A$ be a $2\times 2$ positive semi-definite matrix, i.e.,
$$A = \begin{bmatrix}
a & b \\
\bar{b} & d
\end{bmatrix},$$
where, $a, d \geq 0,$ $\bar{b}$ is the conjugate of $b$ and let $\mbox{Re}(b)$ represent the real part of $b.$ Then, we have the following two inequalities :
\begin{align}
    &\tr[A]  {\geq} 2 \mbox{Re}(b),\label{positivity_res_1}\\
    &\tr[A] + 2 \mbox{Re}(b){\geq} 0.\label{positivity_res_2}
\end{align}

Thus, $\abs{b + \bar{b}} \leq \tr[A].$
\end{lemma}
\textit{Proof of \eqref{positivity_res_1}:}
From the positivity of the matrix $A$ it follows that for any non-zero
-vector $|x\rangle$, $\langle x|A|x\rangle \geq 0$. Now if we multiply $A$ by  the row vector $[1,-1]$ on the left and  by the column vector $[1,-1]^T$ on the right, we will get $\tr[A] - 2 \mbox{Re}(b)\geq 0$, since $A$ is positive.

\textit{Proof of \eqref{positivity_res_2}:} Similarly if we multiply $A$ by  the row vector $[1,1]$ on the left and  by the column vector $[1,1]^T$ on the right, we will get $\tr[A] + 2 \mbox{Re}(b)\geq 0$, since $A$ is positive.
\begin{remark} 
    In the above proof of \eqref{r3}, if for all $\alpha \in \G$, the term $\cos(\theta_{\alpha})\sin(\theta_{\alpha})\big(\tr[\ket{w_{\alpha}}\bra{w^{\perp}_{\alpha}} \sigma_{2,{\alpha}}] + \tr[\ket{w^{\perp}_{\alpha}}\bra{w_{\alpha}} \sigma_{2,{\alpha}}]\big)$ is negative, we can ignore this term since we are finding an upper-bound of $\tr[\Pi^{\star}\sigma_2]$.
Hence, we get the following upper-bound for this particular case
\begin{equation*}
    \tr[\Pi^{\star}\sigma_2] \leq 2^{-{k_2}} + 8\eps^{\frac{1}{2}}. 
\end{equation*}
% In the above proof, if $\theta_{\alpha}$ remains with $[\pi/2,\pi]$, then $\sin(\theta_{\alpha})$ remains positive and $\cos(\theta_{\alpha})$ becomes negative, causing the term $\cos(\theta_{\alpha})\sin(\theta_{\alpha})$ becomes negative and we can ignore this term since we're will to find an upper-bound of $\tr[\Pi^{\star}\sigma_2]$.
% Hence, we get the following upper-bound for this particular case:
% \begin{equation*}
%     \tr[\Pi^{\star}\sigma_2] \leq 2^{-{k_2}} + 2\sqrt{2}\eps^{\frac{1}{4}} 
% \end{equation*}
% For the proof of generalized result in Appendix \ref{Generalised} we 
\end{remark}
The following lemma is a general version of Lemma \ref{lbound}, where there are more than two projectors.
\begin{lemma}[The case of more than two projectors]\label{generalubound}
    Let $\rho \in \mathcal{D}(\cH)$, and a finite set $\cS$, where $|\cS| = 2^t$, for some $t > 0$ and $\left\{\sigma_i \in \mathcal{D}(\cH)\right\}_{i \in \{1,2,\cdots,|\cS|\}}$ is a collection of $|\cS|$ states such that $\forall i \in \{1,2,\cdots,|\cS|\}$, $\mbox{supp}(\rho) \subseteq \mbox{supp}(\sigma_i)$. Now $\forall i \in \cS$, considering the pair $\left\{\rho, \sigma_i\right\}$ we have a projector $\Pi_i$ defined in the following way:\\\\
\begin{equation}
    \Pi_i:=\{\sigma_i \preceq 2^{-{k_i}}\rho\} \mbox{ \quad\quad for some } k_i>0,\label{gubeq}
\end{equation}
with the property that $\tr[\Pi_i\rho] \geq 1 - \eps$. Further, for this collection of $|\cS|$ projectors ${\{\Pi_i\}}_{i\in\{1,2,\cdots,|\cS|\}}$, there exists a projector $\Pi^{\star}$ such that,
\begin{align}
%\label{nr1}
%\Pi^{\star} &\preceq \Pi^{\star},\\
%\label{nr2}
%\Pi^{\star} &\preceq \Pi_2,\\
&\tr[\Pi^{\star}\rho] \geq 1-(t+1)\eps^{(1 - t\beta)}, \label{g1}\\
     % &\tr[\Pi^{\star}\sigma^{\otimes n}_1] \leq 2^{-nD(\rho || \sigma_1)}\label{g2}\\
     &\tr[\Pi^{\star}\sigma_i]  \leq 2^{-k_i} + c(t)\eps^{\frac{\beta}{2}}, \quad i \in \{1,2,\cdots,|\cS|\},\label{g3}
     % &\tr[\Tilde{\Pi}^n\sigma^{\otimes n}_i] \leq 2^{-n(D(\rho |\sigma_i)+\delta) +4\eps^{\frac{1}{4}}\label{g3}, \forall i \in \cS\setminus\{1\}
\end{align}
where, $\beta$ is such that $(t+1)\eps^{(1 - t\beta)} \ll 1$ and $c(t) \in [0,4\sqrt{2}\sum_{j=1}^{t}\sqrt{j}]$.
\end{lemma}
\begin{proof}
This proof is based on the proof of Lemma \ref{lbound}. Before giving the proof, note here that when we defined the set $\G$ in the proof of Lemma \ref{lbound},  we could have replaced $\eps^{\frac{1}{2}}$ by $\eps^{\beta}$ where $\beta \in (0,1)$. In the discussions below, for every pair of projectors, the $\G$ set will be defined in terms of $\eps^{\beta}$. 

\begin{figure}[h]
    \centering
    \resizebox{120mm}{85mm}{
    \begin{tikzpicture}
    \node at (0,0) {Step $0$}; \node at (1,0) {$\Pi_1$}; \node at (2,0) {$\Pi_2$}; \node at (3,0) {$\Pi_3$}; \node at (4,0) {$\Pi_4$};
    \draw[dashed] (4.3,0)--(5.5,0);
    \node at (6,0) {$\Pi_{2^t-3}$}; \node at (7,0) {$\Pi_{2^t-2}$}; \node at (8,0) {$\Pi_{2^t-1}$}; \node at (9,0) {$\Pi_{2^t}$};

  \draw (1,-0.2) -- (1.5,-0.86);\draw (2,-0.2) -- (1.5,-0.86);

  \draw (3,-0.2) -- (3.5,-0.86);\draw (4,-0.2) -- (3.5,-0.86);
  \draw (6,-0.2) -- (6.5,-0.86);\draw (7,-0.2) -- (6.5,-0.86);
  \draw (8,-0.2) -- (8.5,-0.86);\draw (9,-0.2) -- (8.5,-0.86);
    \node at (0,-1.2) {Step $1$}; 
    \node at (1.5,-1.2) {$\Pi_{\{1,2\}}$};
    \node at (3.5,-1.2) {$\Pi_{\{3,4\}}$};
    \draw[dashed] (4,-1.2)--(5.5,-1.2);
     
    \node at (6.5,-1.2) {$\Pi_{\{2^t-3,2^t-2\}}$};
    \node at (8.5,-1.2) {$\Pi_{\{2^t-1,2^t\}}$};
    \draw (1.5,-1.4) -- (2.5,-2.06);\draw (3.5,-1.4) -- (2.5,-2.06);
    \draw (6.5,-1.4) -- (7.5,-2.06);\draw (8.5,-1.4) -- (7.5,-2.06);
    \node at (0,-2.4)
    {Step $2$};
    \node at (2.5,-2.4) {$\Pi_{\{1,2,3,4\}}$};
    \node at (7.5,-2.4) {$\Pi_{\{2^t-3,2^t-2,2^t-1,2^t\}}$};
    \draw [dashed] (0,-2.8) -- (0,-4.8);
    \draw [dashed] (2.5,-2.8) -- (4,-4.8);
    \draw [dashed] (7.5,-2.8) -- (6,-4.8);
    \node at (0,-5.2)
    {Step $t$};\node at (0.7,-5.2)
    {$-1$};
    \node at (4,-5.2) {$\Pi_{\{1,\cdots,2^{t-1}\}}$};
    \node at (6,-5.2) {$\Pi_{\{2^{t-1}+1,\cdots,2^{t}\}}$};
    \draw (4,-5.4)--(5,-6.06);
    \draw (6,-5.4)--(5,-6.06);
    \node at (0,-6.4){Step $t$};
    \node at (5,-6.4) {$\Pi_{\{1,\cdots,2^{t}\}}$};
\end{tikzpicture}}
    \caption{A parallel construction of the intersection $\Pi^{\star}$ of the collection $\left\{\Pi_1,\Pi_2,\cdots,\Pi_{2^t}\right\}$ of projectors mentioned in the statement of Lemma $4$.}
    \label{fig:bottom_up_construction}
\end{figure}

We start with $2^{t}$ projectors and consider $2^{t-1}$ pairs of these projectors. We find the intersection of each pair of projectors in parallel, resulting in $2^{t-1}$ projectors. Next, we consider $2^{t-2}$ pairs of the $2^{t-1}$ projectors and find the intersection of each pair, ending up with $2^{t-2}$ intersection projectors.

We repeat this process, halving the number of projectors we consider at each step until we are left with only one projector. This bottom-up approach allows us to systematically work our way down from the initial $2^{t}$ projectors to the final single projector through a series of parallel intersection operations. However, at each step we change $\G$ which depends on the number of the step. See Figure \ref{fig:bottom_up_construction} for a better understanding of the above strategy.

 For further details of the proof see subsection \ref{Generalised} in the Appendix.
\end{proof}

We can derive different bounds as compared to the bounds obtained in Lemma \ref{generalubound} if instead of the parallel construction discussed in Figure \ref{fig:bottom_up_construction}, we use a sequential construction as discussed below.

We start with $2^{t}$ projectors and consider the first two projectors i.e. $\Pi_1$ and $\Pi_2$ from the collection. We find the intersection $\Pi_{\{1,2\}}$ of the pair $\{\Pi_{1},\Pi_{2}\}$. Next, we consider $\Pi_{\{1,2\}}$ and $\Pi_3$ and find the intersection $\Pi_{\{1,2,3\}}$.

We keep repeating this process until we are left with only one projector. This approach allows us to systematically work our way down from the initial $2^{t}$ projectors to a final single projector through a series of intersection operations sequentially. However, at each step, we change the set $\G$ which depends on the number of the step. See Figure \ref{fig:sequential_construction} below for a better understanding of the above strategy.
\begin{figure}[h]
     \centering
    \resizebox{105mm}{85mm}{\begin{tikzpicture}
    \node at (0,0) {Step $0$}; \node at (1,0) {$\Pi_1$}; \node at (2,0) {$\Pi_2$}; \node at (3,0) {$\Pi_3$}; 
    \draw[dashed] (3.3,0)--(6,0);
    \node at (6.5,0) {$\Pi_{2^t-1}$}; \node at (7.5,0) {$\Pi_{2^t}$};

  \draw (1,-0.2) -- (1.5,-0.86);\draw (2,-0.2) -- (1.5,-0.86);
  
    \node at (0,-1.2) {Step $1$}; 
    \node at (1.5,-1.2) {$\Pi_{\{1,2\}}$};
    \node at (3,-1.2) {$\Pi_{3}$};
    \draw (1.5,-1.4) -- (2.25,-2.06);\draw (3,-1.4) -- (2.25,-2.06);
    \draw[dashed] (3.3,-1.2)--(6,-1.2);
     
    \node at (6.5,-1.2) {$\Pi_{2^t-1}$}; \node at (7.5,-1.2) {$\Pi_{2^t}$};
    \draw [dashed] (6.5,-1.6) -- (6.5,-3.8);
    \draw [dashed] (7.5,-1.6) -- (7.5,-3.8);
    \node at (0,-2.4)
    {Step $2$};
    \node at (2.25,-2.4) {$\Pi_{\{1,2,3\}}$};
    \draw [dashed] (0,-2.8) -- (0,-3.8);
    \draw [dashed] (2.5,-2.8) -- (3.6,-3.8);
    \node at (0,-4.2)
    {Step $2^t$};\node at (0.9,-4.2)
    {$-3$};
    \node at (4,-4.2) {$\Pi_{\{1,\cdots,2^{t}-2\}}$};
    \node at (6.5,-4.2) {$\Pi_{2^t-1}$}; \node at (7.5,-4.2) {$\Pi_{2^t}$};
    \draw (4,-4.4)--(5.25,-5.06);
    \draw (6.5,-4.4)--(5.25,-5.06);
    \node at (0,-5.4)
    {Step $2^t$};\node at (0.9,-5.4)
    {$-2$};
    \node at (5.25,-5.4) {$\Pi_{\{1,\cdots,2^{t}-1\}}$};
    \node at (7.5,-5.4) {$\Pi_{2^t}$};
    \draw (5.25,-5.6)--(6.375,-6.26);
    \draw (7.5,-5.6)--(6.375,-6.26);
    \node at (0,-6.6)
    {Step $2^t$};\node at (0.9,-6.6)
    {$-1$};
    \node at (2,-6.6)
    {$= \abs{\cS}-1$};
    \node at (6.25,-6.6) {$\Pi_{\{1,\cdots,2^{t}\}}$};
    % \node at (0,-3) {\vdots};
    % \node at (0,-3.5) {\vdots};
    % \node at (0,-4) {\vdots};
    % \node at (0,-4.5) {\vdots};
\end{tikzpicture}}
\caption{A sequential construction of the intersection $\Pi^{\star}$ of the collection $\left\{\Pi_1,\Pi_2,\cdots,\Pi_{2^t}\right\}$ of projectors mentioned in the statement of Lemma $4$.}
    \label{fig:sequential_construction}
\end{figure}

The above sequential construction yields us the following lemma.

\begin{lemma}\label{generalubound_seq}
    Let $\rho \in \mathcal{D}(\cH)$, and a finite set $\cS$, where $|\cS| = 2^t$, for some $t > 0$ and $\left\{\sigma_i \in \mathcal{D}(\cH)\right\}_{i \in \{1,2,\cdots,|\cS|\}}$ is a collection of $|\cS|$ states such that $\forall i \in \{1,2,\cdots,|\cS|\}$, $\mbox{supp}(\rho) \subseteq \mbox{supp}(\sigma_i)$. Now $\forall i \in \cS$, considering the pair $\left\{\rho, \sigma_i\right\}$ we have a projector $\Pi_i$ defined in the following way:\\\\
\begin{equation}
    \Pi_i:=\{\sigma_i \preceq 2^{-{k_i}}\rho\} \mbox{ \quad\quad for some } k_i>0,\label{gubeq}
\end{equation}
with the property that $\tr[\Pi_i\rho] \geq 1 - \eps$. Further, for this collection of $|\cS|$ projectors ${\{\Pi_i\}}_{i\in\{1,2,\cdots,|\cS|\}}$, there exists a projector $\Pi^{\star}$ such that,
\begin{align}
%\label{nr1}
%\Pi^{\star} &\preceq \Pi^{\star},\\
%\label{nr2}
%\Pi^{\star} &\preceq \Pi_2,\\
&\tr[\Pi^{\star}\rho] \geq 1-2\eps^{(1 - (2^t-1)\beta)}, \label{g1seq}\\
     % &\tr[\Pi^{\star}\sigma^{\otimes n}_1] \leq 2^{-nD(\rho || \sigma_1)}\label{g2}\\
     &\tr[\Pi^{\star}\sigma_i]  \leq 2^{-k_i} + d(i,\abs{\cS})\eps^{\frac{\beta}{2}}, \quad i \in \{1,2,\cdots,|\cS|\},\label{g3seq}
     % &\tr[\Tilde{\Pi}^n\sigma^{\otimes n}_i] \leq 2^{-n(D(\rho |\sigma_i)+\delta) +4\eps^{\frac{1}{4}}\label{g3}, \forall i \in \cS\setminus\{1\}
\end{align}
where, $\beta$ is such that $2\eps^{(1 - (2^t-1)\beta)} \ll 1$ and $d(i,\abs{\cS}) = 4\sqrt{2}\sum_{j=i}^{\abs{\cS}-1}\sqrt{j}, \forall i \in  \{1,\cdots,\abs{\cS}-1\}$ and $d(\abs{\cS},\abs{\cS}) := 0$.
\end{lemma}
\begin{proof}
    See Appendix \ref{proof_generalubound_seq} for the proof.
\end{proof}
\begin{remark}
    It is important to note that the parallel construction mentioned in the proof of Lemma \ref{generalubound} is more efficient than the sequential construction mentioned in the proof of Lemma \ref{generalubound_seq}. This is because, in sequential construction, the RHS of equation \eqref{g1seq} in Lemma \ref{generalubound_seq} decreases exponentially with increasing $t$. However, in parallel construction, the RHS of equation $\eqref{g1}$ in Lemma \ref{generalubound} decreases polynomially with increasing $t$. Anshu et al. used a similar parallel construction to prove \cite[Lemma 3]{AJW-Compound}.
\end{remark}
\section{Communication Over Authenticated Classical-Quantum channel}\label{authenticated_section} 
Consider a finite collection of classical-quantum channels $\{\cN^{(s)}_{X \to B}\}_{s=1}^{|\cS|}.$  Now $\cN^{(s)}_{X \to B}$ corresponding to state $s$ can be defined as a map from a classical system $X$ with alphabet $\cX$ and orthonormal basis$\{ \ket{x} : x \in \cX\}$ to a quantum system $B$ with an output Hilbert space $\cH_{B}$ with a specified set of quantum states $\{\rho_{x}^{(s)} : x \in \cX\}$.
We can represent the above mapping using a classical quantum state $\rho^{(s)}_{XB}$ corresponding to a state $s \in \cS$, where $|\cS| < \infty$ as follows,
\[\rho^{(s)}_{XB} := \sum_{x}P_{X}(x)\ketbra{x}\otimes\rho_{x}^{(s)}.\]
% Alice sends a message$M$ from a message set $\{1,2,\cdots2^{nR}\}$ and Bob decodes it to be $\hat{M}$. Our goal is the following:\\\\
% \tab If $s = s_0$ then the adversary is ineffective or absent (defaultly setting the state $s_0$) and we should reliably decode the message.\\\\

In this setting, both the sender and receiver are not aware of which one amongst these $|\cS|$ channels will be used for communication. However, they are interested in communicating successfully (with high probability) only when the $s_0$ channel is used for communication. For all other cases, they should \emph{either detect that the $s_0$ channel is not being used for communication or they should be able to communicate successfully} (with high probability).

\begin{definition}
A $(2^{nR},n)$-code $\cC,$ for communication over authenticated classical-quantum channel\\ $\{\cN^{(s)}_{X \to B}\}_{s=1}^{|\cS|}$ consists of 
\begin{itemize}
\item A message set $\cM_n := \{1,\cdots,2^{nR}\}$.
\item An encoding operation $\cE^{(n)}: \cM_n \to \cX^n$ .
\item A decoding operation $\cD^{(n)}: \cS(\cH^{\otimes n}_B) \to \cM_n\cup \{\perp\}.$ Here $\{\perp\}$ represents the event that channel corresponding to $s_0$ is not being used for communication. 
\end{itemize}

Given a message $m$, after the encoding operation, we have an $n$-length encoded-word $X^n(m) := \cE^{(n)}(m)$ which we send over the $n$ uses of the channel. For $m \in \{1,\cdots,2^{nR}\}$ and $s \in \cS,$ let $\hat{M}$ be the decoded message. Then, the probability of error is defined as 
\[
    e (m, \cC, s )= 
\begin{cases}
  \Pr\{\hat{M} \neq m |\cC, s \},& \text{if } s= s_0, \\
    \Pr\{\hat{M} \neq m \cap \neg{\perp }|\cC,  s \},& \text{if } s\neq s_0.            \end{cases}
\]
\end{definition}
 We define the average error probability of a code $\cC$,for a given $s \in \cS$, $\Pr\{\cC,s\}$ as follows,
\beq
\Pr\{\cC,s\}: =\frac{1}{2^{nR}}\sum_{m=1}^{2^{nR}}e (m, \cC, s).\nonumber
\enq 

For some $R>0$, for every $\bar{\eps} \in (0,1)$,
$\delta>0$, and sufficiently large $n$ , let there exist a ($2^{nR},n$)-code $\cC$ such that
\beq
\frac{\log|\cM_n|}{n} > R -\delta, ~~ \max_{s \in \cS} \Pr\{\cC,s\} < \bar{\eps}. 
\enq

The supremum over all such $R$ is called the authentication capacity $C_{\mbox{Auth}}$
of the classical-quantum authenticated channel.

This communication model can be considered as a relaxed version of the compound channel wherein irrespective of the channel being used the parties should be able to communicate successfully with high probability. Before further discussion on the analysis of error, we first state the following lemma.
\begin{lemma}(Authentication Capacity)\label{capacity_authentication}
    The authentication capacity of the classical-quantum authentication channel is given as follows,
    \begin{equation*}
        C_{\mbox{\textnormal{Auth}}} = \max_{P_{X}}I[X;B_{(s_0)}],
    \end{equation*}
    where, $I[X;B_{(s_0)}] = S(\sum_{x}P_{X}(x)\rho_{x}^{(s_0)})-\sum_{x}P_{X}(x)S(\rho^{(s_0)}_x).$
\end{lemma}
\begin{proof}
{(\textbf{Achievability})}
% The event $M \neq \hat{M}$ can be written as following:
% \begin{equation*}
%    \Pr\{ M \neq \hat{M}\} = \frac{1}{2^{nR}}\sum_{m = 1}^{2^{nR}}\Pr(\hat{M} \neq m | M = m)
% \end{equation*}
In this section, we show that for every $R \leq C_{\mbox{Auth}}$ there exists a $(2^{nR},n)$-code $\cC$ such that $\lim_{n \to \infty}\max_{s \in \cS} \Pr\{\cC,s\}$ goes to zero.\\
\subsubsection*{Codebook Generation}
We fix a probability distribution, denoted as $P_{X}$ over $\cX$, that achieves the capacity $C_{\mbox{\textnormal{Auth}}}$. Subsequently, Alice randomly generates $2^{nR}$ sequences of length $n$, denoted as $X^n(m) \in \cX^n$, for each $m$ in the set $\{1, \ldots, 2^{nR}\}$ according to $P_{X^n}(X^n(m)) = \prod_{i=1}^{n}P_{X}(X_i(m))$, where $X^n(m) = \left(X_1(m),X_2(m),\cdots,X_n(m)\right)$. We then declare a codebook $\cC := \left\{X^n(m)\right\}_{m=1}^{2^{nR}}$, where 
\begin{equation*}
    \Pr\{\cC\} = \prod_{m=1}^{2^{nR}}\prod_{i=1}^{n}P_{X}(X_i(m)).
\end{equation*}

This codebook $\cC$ is shared between Alice and Bob.

\subsubsection*{Encoding Strategy}
If Alice has to send the message $m.$ Then, she encodes $m$ to $X^n(m)$ and transmits it over the channel.
\subsubsection*{Decoding Strategy}
Following is the decoding strategy by Bob upon receiving the state $\rho^{(s)}_{X^n(m)} \in \cH^{\otimes n}_B$, where $\rho^{(s)}_{X^n(m)} := \bigotimes_{i=1}^{n}\rho^{(s)}_{X_{i}(m)}$ is the state induced at the Bob's end by the channel $s$.
In the discussion below, the event $\{\neg{\perp}\}$ implies that the $s_0$ channel has been used for communication.
\begin{steps}
\item\label{s1}
Bob performs the measurements $\{\Pi_{(n)}^{\neg{\perp}}, \mathbb{I} - \Pi_{(n)}^{\neg{\perp}}\}$ on $\rho^{(s)}_{X^n(m)}$, where the properties of $\Pi_{(n)}^{\neg{\perp}}$ are discussed in Corollary \ref{decode_s0}. The outcome of this measurement detects whether the channel corresponding to $s_0$  has been used for the transmission or not. If the measurement outcome detects that a channel corresponding to some $s \neq s_0$ is being used. Then, Bob declares that the channel has been wrongly authenticated and he stops there without decoding the message.
However, if the measurement outcome corresponds to $s_0$. Then, Bob proceeds to \ref{s2} mentioned below. 
\item\label{s2}
Assuming that the POVM in \ref{s1} detects the channel $s_0$. Bob has the following state in his possession:
\begin{equation}
    \hat{\rho}^{(s)}_{X^n(m)} := \frac{\Pi_{(n)}^{\neg{\perp}}\rho^{(s)}_{X^n(m)} \Pi_{(n)}^{\neg{\perp}}}{\tr[\Pi_{(n)}^{\neg{\perp}}\rho^{(s)}_{X^n(m)} ]}.\label{ds}
\end{equation}

Bob, applies the following POVM on $\hat{\rho}^{(s)}_{X^n(m)}$ to decode the transmitted message:
\begin{equation}
    \Lambda^{(s_0)}_{(m)} := {\left(\sum_{m'=1}^{2^{nR}}\Pi^{(s_0)}_{X^n(m'),\delta'}\right)}^{-\frac{1}{2}}\Pi^{(s_0)}_{X^n(m),\delta'}{\left(\sum_{m'=1}^{2^{nR}}\Pi^{(s_0)}_{X^n(m'),\delta'}\right)}^{-\frac{1}{2}},\nonumber
\end{equation}
where, $\Pi^{(s_0)}_{X^n(m),\delta'} := \Bigg\{\rho^{(s_0)}_{X^n(m)} \succeq 2^{n\left(I[X;B_{(s_0)}] - \delta' \right)}\rho^{(s_0)}\Bigg\}$, where $\delta' \in (0,1)$.\\
\end{steps}

% If in the previous step, the channel corresponding to $s_0$ is detected then we further proceed to decode the message. Given a message $m$ sent over the channel ${\cN}_{X \to B}^{(s_0)\otimes n}$, we have the quantum state $\rho^{(s_0)}_{X^n(m)}$
% Now to decode the message properly, for a particular code $X^n(m)$ we define a typical projector $\Pi^{(s_0)}_{X^n(m),\delta'}$ in the following way:
% \begin{equation}
%     \Pi^{(s_0)}_{X^n(m),\delta'} := \Bigg\{\rho^{(s_0)}_{X^n(m)} \succeq 2^{n\left(I[X;B_{(s_0)}] - \delta' \right)}\rho^{(s_0)}\Bigg\}
% \end{equation}
% where, $\rho^{(s_0)} := \mathbb{E}_{X^n}\left[\rho^{(s_0)}_{X^n}\right]$. Using the law of large numbers, we can say that for an $\eps \in (0,1)$
% \begin{equation}
%     \tr\left[\Pi^{(s_0)}_{X^n(m),\delta'}\rho^{(s_0)}_{X^n(m)}\right] \geq 1 -\eps \label{typical_set_prob}
% \end{equation}
% We will decode the messages using 
% $\left\{\Lambda^{(s_0)}_{(m)}\right\}_{m=1}^{2^{nR}}$, where $\Lambda^{(s_0)}_{(m)}$ is defined as follows
% \begin{equation}
%     \Lambda^{(s_0)}_{(m)} := {\left(\sum_{m'=1}^{2^{nR}}\Pi^{(s_0)}_{X^n(m'),\delta'}\right)}^{-\frac{1}{2}}\Pi^{(s_0)}_{X^n(m),\delta'}{\left(\sum_{m'=1}^{2^{nR}}\Pi^{(s_0)}_{X^n(m'),\delta'}\right)}^{-\frac{1}{2}}.
% \end{equation}
% Now our decoding strategy is that upon receiving a quantum state $\rho^{(s_0)}_{X^n}$, we measure it with $\left\{\Lambda^{(s_0)}_{(m)}\right\}_{m=1}^{2^{nR}}$ and if the outcome is $\hat{m}$ i.e. the output state corresponds to the POVM $\Lambda^{(s_0)}_{(\hat{m})}$ then Bob declares $\hat{m}$ to be the decoded message.
\subsubsection*{Error Analysis}
We now calculate the following:
% Upon inputting $X^n(m)$ over the channel ${\cN}_{X \to B}^{(s_0)\otimes n}$ we get a corresponding quantum state $\rho^{(s_0)}_{X^n(m)} \in \cH^{\otimes n}_B$. After measuring the state with $\Pi_{(n)}^{\neg{\perp}}$, the post measurement state cane be given as following
% \begin{equation}
%     \hat{\rho}^{(s_0)}_{X^n(m)} := \frac{\Pi_{(n)}^{\neg{\perp}}\rho^{(s_0)}_{X^n(m)} \Pi_{(n)}^{\neg{\perp}}}{\tr[\Pi_{(n)}^{\neg{\perp}}\rho^{(s_0)}_{X^n(m)} ]}.
% \end{equation}

% \subsubsection{Decoding Strategy for a point-to-point channel}

% Our decoding strategy is that upon receiving a $\rho^{(s_0)}_{X^n}$, we use the following randomized decoding strategy.

\begin{align}
    \mathbb{E}_{\cC} \left[\sum_{s \in \cS}\Pr\{\cC,s\}\right] = \mathbb{E}_{\cC} [\Pr\{\cC,s_0\}] + \mathbb{E}_{\cC} \left[\sum_{\substack{s' \neq s_0}}\Pr\{\cC,s'\}\right].\label{sum_error}
\end{align}

We first upper-bound the first term in r.h.s. of \eqref{sum_error}. In the discussions below, $\cE_{m} := \{\hat{M} \neq m\}$,
\begin{align}
    \mathbb{E}_{\cC} [\Pr\{\cC,s_0\}]&= \frac{1}{2^{nR}}\sum_{m = 1}^{2^{nR}}\mathbb{E}_{\cC} \left[e(m, \cC, s_0)\right]\nonumber\\
    &= \frac{1}{2^{nR}}\sum_{m = 1}^{2^{nR}}\mathbb{E}_{\cC} \left[\Pr\{{\cE_{m}|m,\cC, s_0}\}\right]\hspace{250pt}\nonumber
        \end{align}
\begin{align}
    \hspace{40pt}&= \frac{1}{2^{nR}}\sum_{m = 1}^{2^{nR}}\mathbb{E}_{\cC} \left[\Pr\{{\cE_m \cap \perp|m,\cC, s_0}\} + \Pr\{{\cE_m \cap \neg{\perp}|m,\cC, s_0}\}\right]\nonumber\\
    &= \frac{1}{2^{nR}}\sum_{m = 1}^{2^{nR}}\mathbb{E}_{\cC} \left[\Pr\{{\cE_{m} |\perp,m,\cC,s_0}\}.\Pr\{\perp | m, \cC, s_0\}\right]\nonumber\\
    &+ \frac{1}{2^{nR}}\sum_{m = 1}^{2^{nR}}\mathbb{E}_{\cC} \left[\Pr\{{\cE_{m} |\neg{\perp},m,\cC,s_0}\}.\Pr\{\neg{\perp} |m, \cC, s_0\}\right]\hspace{100pt}\nonumber\\
    &{\leq} \frac{1}{2^{nR}}\sum_{m = 1}^{2^{nR}}\mathbb{E}_{\cC} \left[\Pr\{{\cE_{m}|\neg{\perp}, m,\cC, s_0}\}\right] + \frac{1}{2^{nR}}\sum_{m = 1}^{2^{nR}}\mathbb{E}_{\cC} \left[\Pr\{{\perp|m,\cC, s_0}\}\}\right]\nonumber\\
    &{=} \frac{1}{2^{nR}}\sum_{m = 1}^{2^{nR}}\mathbb{E}_{\cC} \left[\tr\left[\left(\mathbb{I} - \Lambda^{(s_0)}_{(m)}\right)\hat{\rho}^{(s_0)}_{X^n(m)}\right]\right] + \frac{1}{2^{nR}}\sum_{m = 1}^{2^{nR}}\mathbb{E}_{\cC} \left[\Pr\{{\perp|m,\cC, s_0}\}\right],\label{error_s0}
\end{align}
% Further,
% \begin{align*}
%     \Pr(\cE_{m}|\neg{\perp}, m,\cC, s_0) &= \tr\left[\left(\mathbb{I} - \Lambda^{(s_0)}_{(m)}\right)\hat{\rho}^{(s_0)}_{X^n(m)}\right] \\
%     % &\leq 2\tr\left[\left(\mathbb{I} - \Pi^{(s_0)}_{X^n(m),\delta'}\right)\rho^{(s_0)}_{X^n(m)}\right] + 4\sum_{m' \neq m}\tr\left[\left(\Pi^{(s_0)}_{X^n(m'),\delta'}\right)\rho^{(s_0)}_{X^n(m)}\right]
% \end{align*}
% where, $\cE_{m} := \{\hat{M} \neq m\}$, $a$ follows from the fact that $\Pr(\neg{\perp} | m, \cC, s_0) = \tr\left[\Pi_{(n)}^{\neg{\perp}} \rho_{X^n(m)}^{(s_0)}\right]$, which can further be bounded by $1$ and the term $\Pr({\cE_{m} |\perp,m,\cC,s_0})$ is the decoding error of the channel corresponding to the state $s_0$ but actually assuming that the channel corresponding to the $s_0$ is not being used for communication, which we also can upper-bound by 1 and $b$ follows from the fact that $\Pr({\cE_{m}|\neg{\perp}, m,\cC, s_0})\}$ is the decoding error of the channel corresponding to the state $s_0$, when the channel corresponding to $s_0$ is actually being used which can be given as $\tr\left[\left(\mathbb{I} - \Lambda^{(s_0)}_{(m)}\right)\hat{\rho}^{(s_0)}_{X^n(m)}\right]$, where $\hat{\rho}^{(s_0)}_{X^n(m)}$ is defined in \eqref{ds}.
where, \eqref{error_s0} follows from the definition of $\Pr({\cE_{m}|\neg{\perp}, m,\cC, s_0})\}$.
Now observe that,
\begin{align}
    \mathbb{E}_{\cC}\left[\tr\left[\left(\Lambda^{(s_0)}_{(m)}\right)\hat{\rho}^{(s_0)}_{X^n(m)}\right]\right] &= \mathbb{E}_{\cC}\left[\frac{\tr\left[\left(\Lambda^{(s_0)}_{(m)}\right)\Pi_{(n)}^{\neg{\perp}}\rho^{(s_0)}_{X^n(m)} \Pi_{(n)}^{\neg{\perp}}\right]}{\tr[\Pi_{(n)}^{\neg{\perp}}\rho^{(s_0)}_{X^n(m)} ]}\right]\nonumber\\
    & \overset{a}{\geq} \mathbb{E}_{\cC}\left[\tr\left[\left(\Lambda^{(s_0)}_{(m)}\right)\Pi_{(n)}^{\neg{\perp}}\rho^{(s_0)}_{X^n(m)} \Pi_{(n)}^{\neg{\perp}}\right]\right]\nonumber\\
    &\overset{b}{\geq} \mathbb{E}_{\cC}\left[\tr\left[\left(\Lambda^{(s_0)}_{(m)}\right)\rho^{(s_0)}_{X^n(m)}\right]\right] - \mathbb{E}_{\cC}\left[\norm{\rho^{(s_0)}_{X^n(m)} - \Pi_{(n)}^{\neg{\perp}}\rho^{(s_0)}_{X^n(m)} \Pi_{(n)}^{\neg{\perp}}}{1}\right]\nonumber\\
    &\overset{c}{\geq}\mathbb{E}_{\cC}\left[\tr\left[\left(\Lambda^{(s_0)}_{(m)}\right)\rho^{(s_0)}_{X^n(m)}\right]\right] - 2\sqrt{\left(\ceil{\log{(|\cS| -1)}}+1\right)\eps^{\left({1 - \ceil{\log(|\cS|-1)}}\beta\right)}},\label{post_state_lb}
\end{align}
where, $a$ follows from the fact $\tr[\Pi_{(n)}^{\neg{\perp}}\rho^{(s_0)}_{X^n(m)}] \leq 1$, $b$ follows from  Fact \ref{trace_norm} and $c$ follows by invoking Fact \ref{gent_measurement}  on \eqref{decode_s0_eq1} of Corollary \ref{decode_s0} for some $\beta \in (0,1)$ . 

The first term in \eqref{error_s0} is upper-bounded as follows,
\begin{align}
&\hspace{17pt}\mathbb{E}_{\cC}\left[\tr\left[\left(\mathbb{I} - \Lambda^{(s_0)}_{(m)}\right)\hat{\rho}^{(s_0)}_{X^n(m)}\right]\right]\nonumber\\
&\overset{a}{\leq} \mathbb{E}_{\cC}\left[\tr\left[\left( \mathbb{I} - \Lambda^{(s_0)}_{(m)}\right)\rho^{(s_0)}_{X^n(m)}\right]\right] + 2\sqrt{\left(\ceil{\log{(|\cS| -1)}}+1\right)\eps^{\left({1 - \ceil{\log(|\cS|-1)}}\beta\right)}}\hspace{130pt}\nonumber
\end{align}
\begin{align}
&\overset{b}{\leq} 2\mathbb{E}_{X^n(m)}\left[\tr\left[\left(\mathbb{I} - \Pi^{(s_0)}_{X^n(m),\delta'}\right)\rho^{(s_0)}_{X^n(m)}\right] \right] + 4\sum_{m' \neq m}\mathbb{E}_{X^n(m')}\mathbb{E}_{X^n(m)}\left[ \tr\left[\left(\Pi^{(s_0)}_{X^n(m'),\delta'}\right)\rho^{(s_0)}_{X^n(m)}\right]\right]\nonumber\\
&+ 2\sqrt{\left(\ceil{\log{(|\cS| -1)}}+1\right)\eps^{\left({1 - \ceil{\log(|\cS|-1)}}\beta\right)}}\nonumber\\
&\overset{c}{\leq} 2(1-1+\varepsilon) + 4\sum_{m' \neq m}\mathbb{E}_{X^n(m')}\left[ \tr\left[\left(\Pi^{(s_0)}_{X^n(m'),\delta'}\right)\rho^{(s_0)}\right]\right] + 2\sqrt{\left(\ceil{\log{(|\cS| -1)}}+1\right)\eps^{\left({1 - \ceil{\log(|\cS|-1)}}\beta\right)}}\nonumber\\
&\leq 2\varepsilon + 4. 2^{-n\left(I[X;B_{(s_0)}] - \delta' \right)} \sum_{m' \neq m}\mathbb{E}_{X^n(m')}\left[ \tr\left[\left(\Pi^{(s_0)}_{X^n(m'),\delta'}\right)\rho^{(s_0)}_{X^n(m')}\right]\right]\nonumber\\
&+ 2\sqrt{\left(\ceil{\log{(|\cS| -1)}}+1\right)\eps^{\left({1 - \ceil{\log(|\cS|-1)}}\beta\right)}}\nonumber\\
&\overset{d}{\leq} 2\varepsilon + 4. 2^{-n\left(I[X;B_{(s_0)}] - \delta' \right)}2^{nR} + 2\sqrt{\left(\ceil{\log{(|\cS| -1)}}+1\right)\eps^{\left({1 - \ceil{\log(|\cS|-1)}}\beta\right)}}\nonumber\hspace{150pt}\\
&= 2\varepsilon + 4. 2^{-n\left(I[X;B_{(s_0)}] - \delta' - R\right)} + 2\sqrt{\left(\ceil{\log{(|\cS| -1)}}+1\right)\eps^{\left({1 - \ceil{\log(|\cS|-1)}}\beta\right)}},\label{err_s0_1st_term}
\end{align}
where, $a$ follows from \eqref{post_state_lb}, $b$ follows from Fact \ref{hayashi_nagaoka}. $c$ follows from Fact \ref{tp} and $d$ follows from the fact that for any $m'$ , $\mathbb{E}_{X^n(m')}\left[ \tr\left[\left(\Pi^{(s_0)}_{X^n(m'),\delta'}\right)\rho^{(s_0)}_{X^n(m')}\right]\right] \leq 1$. Now for the other term in \eqref{error_s0}, from \eqref{decode_s0_eq1}, we can upper-bound it in the following way,
\begin{align}
    \mathbb{E}_{\cC}\left[\Pr({\perp|m,\cC, s_0})\right] &= \mathbb{E}_{X^n(m)}\left[\tr\left[\left( \mathbb{I} -\Pi_{(n)}^{\neg{\perp}}\right) \rho_{X^n(m)}^{(s_0)}\right]\right] \leq \left(\ceil{\log{(|\cS| -1)}}+1\right)\eps^{\left({1 - \ceil{\log(|\cS|-1)}}\beta\right)}\label{err_s0_2nd_term}.
\end{align}

Thus, combining equation \eqref{err_s0_1st_term} and \eqref{err_s0_2nd_term} we can write,
\begin{align}
    &\mathbb{E}_{\cC}\left[\Pr\{\cC, s_0\}\right]   = \frac{1}{2^{nR}}\sum_{m = 1}^{2^{nR}}\mathbb{E}_{\cC} \left[\Pr\{{\cE_{m}|\neg{\perp}, m,\cC, s_0}\}\right] + \frac{1}{2^{nR}}\sum_{m = 1}^{2^{nR}}\mathbb{E}_{\cC} \left[\Pr\{{\perp|m,\cC, s_0}\}\right]\nonumber\\
    &\leq \frac{1}{2^{nR}}\sum_{m = 1}^{2^{nR}}\bigg\{2\varepsilon + 4. 2^{-n\left(I[X;B_{(s_0)}] - \delta' - R\right)} + 2{\sqrt{\left(\ceil{\log{(|\cS| -1)}}+1\right)\eps^{\left({1 - \ceil{\log(|\cS|-1)}}\beta\right)}}}\nonumber\\
    &+ \left(\ceil{\log{(|\cS| -1)}}+1\right)\eps^{\left({1 - \ceil{\log(|\cS|-1)}}\beta\right)}\bigg\}\nonumber\\
    &\overset{a}{\leq} 2\varepsilon + 4. 2^{-n\left(I[X;B_{(s_0)}] - \delta' - R\right)} + 3{\sqrt{\left(\ceil{\log{(|\cS| -1)}}+1\right)\eps^{\left({1 - \ceil{\log(|\cS|-1)}}\beta\right)}}}\label{err_s0_term},
\end{align}
where, $a$ follows because the term $\left(\ceil{\log{(|\cS| -1)}}+1\right)\eps^{\left({1 - \ceil{\log(|\cS|-1)}}\beta\right)} \in (0,1)$ since $\eps$ can be made arbitrarily small because of \eqref{DM2} of Fact \ref{F4}. We now upper-bound $\mathbb{E}_{\cC} [\Pr\{\cC,s\}]$, where $s \neq s_0$ as follows,
\begin{align}
    \mathbb{E}_{\cC} [\Pr\{\cC,s\}]
    &= \frac{1}{2^{nR}}\sum_{m = 1}^{2^{nR}}\mathbb{E}_{\cC} \left[e(m, \cC, s)\right]\nonumber\\
    &= \frac{1}{2^{nR}}\sum_{m = 1}^{2^{nR}}\mathbb{E}_{\cC} \left[\Pr\{{\cE_{m} \cap \neg{\perp} |m, \cC, s}\}\right]\hspace{130pt}\nonumber\hspace{80pt}\\
    &= \frac{1}{2^{nR}}\sum_{m = 1}^{2^{nR}}\mathbb{E}_{\cC} \left[ \Pr\{{\cE_{m} |\neg{\perp},m,\cC,s}\}.\Pr\{\neg{\perp} | m,\cC,s\}\right]\nonumber\\
    &\leq \frac{1}{2^{nR}}\sum_{m = 1}^{2^{nR}}\mathbb{E}_{\cC} \left[ \Pr\{\neg{\perp} | m,\cC,s\}\right]\nonumber
        \end{align}
    \begin{align}
    &= \frac{1}{2^{nR}}\sum_{m = 1}^{2^{nR}}\mathbb{E}_{X^n(m)} \left[\tr\left[\Pi_{(n)}^{\neg{\perp}} \rho_{X^n(m)}^{(s)}\right]\right]\nonumber\\
    &\overset{a}{\leq} \frac{1}{2^{nR}}\sum_{m = 1}^{2^{nR}}\left(2^{-n\left(\underset{s \in \cS : s \neq s_0}{\min}D(\rho^{(s_0)} || \rho^{(s)}) - \delta\right)} + O\left(\left(\ceil{\log(|\cS|-1)}\right)^{\frac{3}{2}}\right)\eps^{\frac{\beta}{2}}\right)\nonumber\\
    &= 2^{-n\left(\underset{s \in \cS : s \neq s_0}{\min}D(\rho^{(s_0)} || \rho^{(s)}) - \delta\right)} + O\left(\left(\ceil{\log(|\cS|-1)}\right)^{\frac{3}{2}}\right)\eps^{\frac{\beta}{2}}\label{err_s_term},
\end{align}
where, $a$ follows from the properties of $\Pi_{(n)}^{\neg{\perp}}$ as discussed in the \ref{s1} which we import from \eqref{decode_s0_eq2} of Corollary \ref{decode_s0}, where $\forall s \in \cS, \rho^{(s)} := \mathbb{E}_{X}\left[\rho_{X}^{(s)}\right]$ , $\delta\in(0,1)$ and $\delta$ is chosen in such a way that $\underset{s \in \cS : s \neq s_0}{\min}D(\rho^{(s_0)} || \rho^{(s)}) - \delta > 0$.

From \eqref{err_s0_term} and \eqref{err_s_term}, it follows that as  $n \to \infty$, and if $R < I[X:B_{(s_0)}] - \delta'$, we have $\mathbb{E}_{\cC} \left[\sum_{s \in \cS}\Pr\{\cC,s\}\right] $ $\to 0$. 
Thus, there exists a deterministic code such that $\max_{s}\Pr\{M \neq \hat{M}| s\} \to 0$. This completes the proof of achievability.

{(\textbf{Converse})}
Since any code for the authentication channel model discussed above, can also  be used for communication over the channel
 $\cN^{(s_0)}_{X \to B}$. Thus, for any transmission rate $R>0$ which is achievable for the communication via the authentication channel model is also achievable for the communication via the point-to-point channel $\cN^{(s_0)}_{X \to B}$. Therefore, $C_{\mbox{Auth}} \leq \max_{P_{X}}I[X;B_{(s_0)}]$. 
\end{proof}
\section{Proof of Lemma \ref{mr}}
\label{sec : Construc}
Let $\Pi_1$ and $\Pi_2$ be as mentioned in Lemma \ref{mr}. We will apply Fact \ref{jordan} on the pair $\{\Pi_1, \Pi_2\}.$ Consider $\left\{\mathbf{P}_{\alpha}\right\}_{{\alpha}=1}^k$ (where $k$ is some natural number depending on $\Pi_1,\Pi_2$) as the set of orthogonal projectors obtained by Fact \ref{jordan} applied with respect to the pair $\{\Pi_1, \Pi_2\}.$ Thus, $\sum_{{\alpha} =1}^k \mathbf{P}_{\alpha} = \mathbb{I}.$ Furthermore, for every ${\alpha} \in [1:k]$ we define $\Pi_{1,{\alpha}} := \mathbf{P}_{\alpha}\Pi_1\mathbf{P}_{\alpha}$ and $\Pi_{2,{\alpha}} := \mathbf{P}_{\alpha}\Pi_2\mathbf{P}_{\alpha}$ as the following one dimensional projector: 
\begin{align*}
\Pi_{1,{\alpha}} := \ket{v_{\alpha}}\bra{v_{\alpha}},
\end{align*}
for some $\ket{v_{\alpha}}$ in the range of $\mathbf{P}_{\alpha}.$ Similarly, for every ${\alpha} \in [1:k]$, we have
\begin{align*}
\Pi_{2,{\alpha}} := \ket{w_{\alpha}}\bra{w_{\alpha}},
\end{align*}
for some $\ket{w_\alpha}$ in the range of $\mathbf{P}_{\alpha}.$
Thus, it is now follows from the property of $\{\ket{v_{\alpha}}\}_{{\alpha}=1}^k$ and $\{\ket{w_{\alpha}}\}_{{\alpha}=1}^k$ that 
\begin{align*}
\Pi_1 &:= \sum_{{\alpha} =1}^k \ket{v_{\alpha}}\bra{v_{\alpha}},\\
\Pi_2&:= \sum_{{\alpha} =1}^k \ket{w_{\alpha}}\bra{w_{\alpha}}.
\end{align*}

Further, for every ${\alpha} \in [1:k],$ let 
\begin{align}
\rho_{\alpha} &:= \mathbf{P}_{\alpha} \rho \mathbf{P}_{\alpha},\nonumber\\
\sigma_{1,{\alpha}} &:= \mathbf{P}_{\alpha} \sigma_1 \mathbf{P}_{\alpha},\nonumber\\
\sigma_{2,{\alpha}} &:= \mathbf{P}_{\alpha} \sigma_2 \mathbf{P}_{\alpha}.\label{sigma_2_decom}
\end{align}

Thus, from the definitions of $\Pi_1$ and $\Pi_2$, we have the following pair of inequalities for every ${\alpha} \in [1:k],$ 
\begin{align}
\label{property}
\bra{v_{\alpha}} \sigma_{1,{\alpha}}\ket{v_{\alpha}} &\geq 2^{-{k_1}} \bra{v_{\alpha}}\rho_{\alpha}\ket{v_{\alpha}}, \\
\label{property2}
\bra{w_{\alpha}} \sigma_{2,{\alpha}}\ket{w_{\alpha}} &\geq 2^{-{k_2}} \bra{w_{\alpha}}\rho_{\alpha}\ket{w_{\alpha}}. 
\end{align}

For our future discussions, it will be useful to make the following observations, for every ${\alpha} \in [1:k],$ $\rho_{\alpha}\succeq 0$ and 
$\sum_{{\alpha}=1}^k\tr[\rho_{\alpha}]= 1.$ Thus, $\{\tr[\rho_{\alpha}]\}_{{\alpha}=1}^k$ forms a valid probability distribution over ${\alpha} \in [1:k]$. 
To construct an intersection projector $\Pi^\star$ that satisfies \eqref{rlb1}, \eqref{rlb2}, \eqref{rlb3} we first make few observations. From Jordan's lemma (Fact \ref{jordan}) we have that for every ${\alpha} \in [1:k]$ the vectors $\ket{w_{\alpha}}$ and $\ket{v_{\alpha}}$ lie in either a two-dimensional subspace or one-dimensional subspace.
For the blocks where both $\ket{w_{\alpha}}$ and $\ket{v_{\alpha}}$ lie in a two-dimensional subspace we have the following,  
\begin{equation}
\label{valpha}
\ket{v_{\alpha}} = \cos(\theta_{\alpha})\ket{w_{\alpha}} + \sin(\theta_{\alpha})\ket{w^{\perp}_{\alpha}}.
\end{equation}

For the blocks, where both 
$\ket{w_{\alpha}}$ and $\ket{v_{\alpha}}$ lie in one dimensional subspace then in that case $\cos(\theta_{\alpha})=1,$ i.e., $\ket{w_{\alpha}} = \ket{v_{\alpha}}$ for these cases. 
Now we define the following set: 
\begin{align}
{\G}&:=\left\{{\alpha}: \cos^2(\theta_{\alpha}) \geq 1- 8\eps^{\frac{1}{2}}\right\},
\label{good_lb}
\end{align}
% The intuition for defining the set $\textnormal{\G}$ is that we only want to consider those blocks where, $\ket{v_{\alpha}}$ and $\ket{w_{\alpha}}$ are almost the same, i.e., the angle between them is very small.
% \begin{claim}
% \label{c1}
% The probability of the set $\cE_1$ under the distribution $\tr[\rho_{\alpha}]$ satisfies the following lower bound: $\Pr\{\cE_1\} \geq 1- \eps^{\frac{1}{2}}.$ 
% \end{claim}
% \begin{claim}
% \label{c2}
% The probability of the set $\cE_2$ under the distribution $\tr[\rho_{\alpha}]$ satisfies the following lower bound: $\Pr\{\cE_2\} \geq 1- \eps^{\frac{1}{2}}.$ 
% \end{claim}
% \begin{claim}
% \label{c4}
% The probability of the set $\cE_3$ under the distribution $\tr[\rho_{\alpha}]$ satisfies the following lower bound: $\Pr\{\cE_3\} \geq 1- \eps^{\frac{1}{2}}.$ 
% \end{claim}
where, $\cos^2(\theta_{\alpha}) = {|\langle v_{\alpha}| w_{\alpha}\rangle|}^2$. From \eqref{c23}, the probability of the set ${\G}$ under the distribution $\tr[\rho_{\alpha}]$ satisfies the following lower bound: 
\begin{equation}
    \Pr\{{\G}\} \geq 1- \eps^{\frac{1}{2}}. \label{pgl}
\end{equation}

% In the discussions below, since $\eps$ is arbitrarily close to zero (as given in the  statement of Lemma \ref{mr}) and therefore, we assume that
% \begin{equation}
%     \forall \alpha \in \G, \theta_{\alpha} \in [0,\pi/4]\cup[-\pi/4,0].\label{agr}
% \end{equation}

% \begin{claim}
% \label{clb2}
% The probability of the set ${\cE_2}$ under the distribution $\tr[\rho_{\alpha}]$ satisfies the following lower bound: $\Pr\{{\cE_2}\} \geq 1- \eps^{\frac{1}{2}}.$ 
% \end{claim}
To show the existence of $\Pi^\star$ which satisfies \eqref{rlb1}, \eqref{rlb2} and \eqref{rlb3} we construct $\Pi^\star$ as follows, 
\begin{equation}
\label{starlb}
\Pi^\star:= \sum_{{\alpha} \in {\G}} \ket{v_{\alpha}}\bra{v_{\alpha}}.
\end{equation}

Now we will show that $\Pi^\star$ satisfies \eqref{rlb1}, \eqref{rlb2} and \eqref{rlb3} in the following subsections.
\subsection{Proof for the  properties of $\Pi^\star$}
\subsubsection*{Proof of \eqref{rlb1}}\label{subsec_rlb1}
Consider the following set of inequalities,
\begin{align}
\tr[\Pi^\star\rho] &\overset{a}=\sum_{{\alpha}\in  {\G}}\tr[\ket{v_{\alpha}}\bra{v_{\alpha}} \rho_{\alpha}]\nonumber\\ 
&= \sum_{\alpha = 1}^{k}\tr[\ket{v_{\alpha}}\bra{v_{\alpha}} \rho_{\alpha}] - \sum_{{\alpha}\notin   {\G}}\tr[\ket{v_{\alpha}}\bra{v_{\alpha}} \rho_{\alpha}]\nonumber\\
&\overset{b}\geq 1 - \eps - \sum_{{\alpha}\notin {\G}}\tr[ \rho_{\alpha}]\nonumber\\
&\overset{c}\geq 1 - \eps - \eps^{\frac{1}{2}}\nonumber\\
&\geq 1-2\eps^{\frac{1}{2}},\label{prop_rlb1}
\end{align}
where, $a$ follows from the definition of \eqref{starlb}, $b$ follows from the fact that $\sum_{{\alpha}=1}^{{k}}\tr[\ket{v_{\alpha}}\bra{v_{\alpha}} \rho_{\alpha}] = \tr[\Pi_1 \rho] \geq 1 - \eps$ and $c$ follows from \eqref{pgl}.

\subsubsection*{Proof of \eqref{rlb2}}
Consider the following set of inequalities,
\begin{align*}
\tr[\Pi^\star\sigma_1] &\overset{a}= \sum_{{\alpha}\in  {\G}}\tr[\ket{v_{\alpha}}\bra{v_{\alpha}} \sigma_{1,{\alpha}}]\hspace{55pt} 
\end{align*}
\begin{align*}
&\overset{b} \geq 2^{-{k_1}}\sum_{{\alpha} \in  {\G}}\tr[\ket{v_{\alpha}}\bra{v_{\alpha}} \rho_{\alpha}]\\
&\overset{c}\geq  2^{-{k_1}}(1-2\eps^{\frac{1}{2}}),
%&\geq  2^{-{k_2}}(1-\eps^{\frac{1}{2}})\sum_{{\alpha} \in \G}\cos(2\theta_{\alpha})\tr[\rho_{\alpha}]\\
\end{align*}
where, $a$ follows from the definition of $\eqref{starlb},$ $b$ follows from \eqref{property}, $c$ follows from \eqref{prop_rlb1}.
\subsubsection*{Proof of \eqref{rlb3}}
\label{subsec : proof_rlb3}
Consider the following set of inequalities,
\begin{align}
&\tr[\Pi^\star\sigma_2] \overset{a}= \sum_{{\alpha}\in {\G}}\tr[\ket{v_{\alpha}}\bra{v_{\alpha}} \sigma_{2,{\alpha}}] \nonumber\\
&\overset{b}=\sum_{{\alpha} \in  {\G}}\bigg(\cos^2(\theta_{\alpha})\tr[\ket{w_{\alpha}}\bra{w_{\alpha}}\sigma_{2,{\alpha}}] + \sin^2(\theta_{\alpha})\tr[\ket{w^{\perp}_{\alpha}}\bra{w^{\perp}_{\alpha}} \sigma_{2,{\alpha}}] + \cos(\theta_{\alpha})\sin(\theta_{\alpha})\tr[\ket{w_{\alpha}}\bra{w^{\perp}_{\alpha}} \sigma_{2,{\alpha}}]\nonumber \\
&\hspace{60pt}+ \cos(\theta_{\alpha})\sin(\theta_{\alpha})\tr[\ket{w^{\perp}_{\alpha}}\bra{w_{\alpha}} \sigma_{2,{\alpha}}] \bigg) \nonumber\\
&\overset{c} \geq \sum_{{\alpha} \in  {\G}}\bigg((1 - 8\eps^{\frac{1}{2}})\tr[\ket{w_{\alpha}}\bra{w_{\alpha}} \sigma_{2,{\alpha}}] + \cos(\theta_{\alpha})\sin(\theta_{\alpha})\left(\tr[\ket{w_{\alpha}}\bra{w^{\perp}_{\alpha}} \sigma_{2,{\alpha}}] + \tr[\ket{w^{\perp}_{\alpha}}\bra{w_{\alpha}} \sigma_{2,{\alpha}}]\right)\bigg) \nonumber\\
&\overset{d}\geq \sum_{{\alpha} \in  {\G}}\bigg((1 - 8\eps^{\frac{1}{2}})\tr[\ket{w_{\alpha}}\bra{w_{\alpha}} \sigma_{2,{\alpha}}] -\abs{ \cos(\theta_{\alpha})\sin(\theta_{\alpha})\left(\tr[\ket{w_{\alpha}}\bra{w^{\perp}_{\alpha}} \sigma_{2,{\alpha}}] + \tr[\ket{w^{\perp}_{\alpha}}\bra{w_{\alpha}} \sigma_{2,{\alpha}}]\right)}\bigg) \nonumber\\
&\overset{e}\geq \sum_{{\alpha} \in  {\G}}\bigg((1 - 8\eps^{\frac{1}{2}})\tr[\ket{w_{\alpha}}\bra{w_{\alpha}} \sigma_{2,{\alpha}}] - 2\sqrt{2}\eps^{\frac{1}{4}}\tr[\sigma_{2,{\alpha}}]\bigg) \nonumber\\
% &= 2^{-{k_2}}\cos(2\theta_{\alpha})\sum_{{\alpha} \in  {\G}}\left(\tr[\ket{w_{\alpha}}\bra{w_{\alpha}} \rho_{\alpha}]\right)\\
&\overset{f}\geq  2^{-k_2}\bigg((1 - 8\eps^{\frac{1}{2}})\sum_{{\alpha} \in  {\G}}\tr[\ket{w_{\alpha}}\bra{w_{\alpha}} \rho_{{\alpha}}]\bigg) - 2\sqrt{2}\eps^{\frac{1}{4}}\nonumber\\
&\overset{g}{\geq} 2^{-k_2}(1 - 8\eps^{\frac{1}{2}})(1-2\eps^{\frac{1}{2}}) - 2\sqrt{2}\eps^{\frac{1}{4}}\nonumber\hspace{290pt}\\
&\geq 2^{-{k_2}}(1-10\eps^{\frac{1}{2}})  - 2\sqrt{2}\eps^{\frac{1}{4}},\nonumber
\label{mrf}
%&\geq  2^{-{k_2}}(1-\eps^{\frac{1}{2}})\sum_{{\alpha} \in  {\G}}\cos(2\theta_{\alpha})\tr[\rho_{\alpha}]\\
\end{align}
% \begin{align*}
%     \sum_{{\alpha} \in  {\G}}\tr[\sigma_{2,{\alpha}}] &= \sum_{\alpha = 1}^{k}\tr[\sigma_{2,{\alpha}}] - \sum_{{\alpha} \not\in  {\G}}\tr[\sigma_{2,{\alpha}}]\\
%     &= 1 - \sum_{{\alpha} \not\in  {\G}}\tr[\sigma_{2,{\alpha}}]\\
%     &\leq 1 - \sum_{{\alpha} \not\in  {\G}}\tr[\ket{v_{\alpha}}\bra{v_{\alpha}}\sigma_{2,{\alpha}}]\\
%     &= 1 - \tr[(\mathbb{I} - \Pi^\star)\sigma_2]
% \end{align*}
where, $a$ follows from the definition of $\Pi^\star$ (see \eqref{starlb}), $b$ follows from \eqref{valpha}, $c$ follows from $\cos^2(\theta_{\alpha}) \geq (1 - 8\eps^{\frac{1}{2}})$ for every $\alpha \in {\G}$ due to \eqref{good_lb} and the trivial bound $\sin^2(\theta_{\alpha})\tr[\ket{w^{\perp}_{\alpha}}\bra{w^{\perp}_{\alpha}} \sigma_{2,{\alpha}}] \geq 0$, $d$ follows from facts that for any  real number $x$, $x \geq -\abs{x}$ and $\cos(\theta_{\alpha})\sin(\theta_{\alpha})(\tr[\ket{w_{\alpha}}\bra{w^{\perp}_{\alpha}} $ $\sigma_{2,{\alpha}}]  + \tr[\ket{w^{\perp}_{\alpha}}\bra{w_{\alpha}} \sigma_{2,{\alpha}}])$ is a real number, $e$ follows from \eqref{extra_term_bound}, $f$ holds because of \eqref{property2} and the fact that $\sum_{\alpha \in \G}\tr[\sigma_{2,\alpha}] \leq \sum_{\alpha = 1}^{k}\tr[\sigma_{2,\alpha}] = 1,$ and $g$ follows from the fact that $\sum_{{\alpha} \in  {\G}}\tr[\ket{w_{\alpha}}\bra{w_{\alpha}} \rho_{{\alpha}}] = \tr[\Pi^{\star}\rho] \geq 1 - 2\eps^{\frac{1}{2}}$ due to \eqref{prop_rlb1}.
\section{Asymmetric Composite Hypothesis Testing}
\label{sec:hypothesis_testing}

\subsection{Asymmetric Composite Hypothesis Testing $\big($The case of $\{\rho_s\}_{s=1}^{|\cS|}$ and $\sigma \big)$}\label{acht}
% This problem is a generalization of the binary asymmetric hypothesis testing. Unlike Lemma \ref{lbound}, here we have a state 
Consider a collection $\{\rho_s\}_{s=1}^{|\cS|}$ and $\sigma$ where $s \in \cS,|\cS| < \infty.$ and $\forall s,  \rho_{s},\sigma \in \cD(\cH)$. The aim here is to accept states from the collection $\{\rho_s\}_{s=1}^{|\cS|}$ with high probability and the state $\sigma$ is wrongly identified as a state in the collection $\{\rho_s\}_{s=1}^{|\cS|}$ with very low probability. Formally, this problem can be defined  as follows,
\begin{definition}\label{chtdef}
    Consider $\sigma \in \cD\left(\cH\right)$ and let $\Theta_{\cS} := \left\{\rho_{s}\right\}_{s=1}^{|\cS|}$ be a collection of states, where $\forall s \in \cS, \rho_s \in \cD(\cH)$. For an $\eps \in (0,1)$ and a positive integer $n \in \mathbb{N}$, we define,
    \begin{equation}
        \beta_{(n,\eps)}\left(\Theta_{\cS},\sigma\right):= \min_{\substack{0\prec\Lambda\preceq\mathbb{I_n}:\\
        \forall s \in \cS, \tr[\Lambda\rho_s^{\otimes n}] \geq 1 -\eps}}\tr[\Lambda\sigma^{\otimes n}],\label{betacht}
    \end{equation}
    where, $\mathbb{I}_n$ is an identity operator on $\cH^{\otimes n}$. 
    \end{definition}
    
In this subsection, we will give an upper-bound (achievability) on $\frac{1}{n}\log\beta_{(n,\eps)}\left(\Theta_{\cS},\sigma\right)$. However, before giving this upper-bound, we first discuss the following classical problem. 
% The general case can be handled similarly. Before proving Lemma \ref{gen_cht}, we consider a classical version of the above problem.
Consider a finite collection of distributions $\{P_i\}_{i=1}^{|\cS|}$ and a distribution $Q$ defined over a set $\cX.$ Further, $\forall i \in \cS,$ we are given a set $\cA_{i} \subset \cX$ such that $\Pr_{P_i}\{A_{i}\} \geq 1- \eps$ and $\Pr_{Q}\{A_{i}\} \leq 2^{-k_{i}},$ for some $k_{i}>0.$ It is easy to show that $\forall i \in \cS_,$  we have $\Pr_{P_i}\{\cA^\star\} \geq 1- |\cS|\eps$ and $\Pr_{Q_j}\{\cA^\star\} \leq 2^{-k},$ where $k:= \min_{i}k_{i,}$ and $\cA^\star:= \cup_{i=1}^{|\cS|}\cA_{i}.$. Observe that we can re-write $\cA^\star:= \cX \setminus \left\{\cap_{i=1}^{|\cS|}\cA_{i}^{c}\right\}$. In Lemma \ref{gen_cht}, we construct a $\Pi_{n}^{\star}$ which behaves very similar to $\cA^\star:= \cX \setminus \left\{\cap_{i=1}^{|\cS|}\cA_{i}^{c}\right\}$.
This lemma will help us to give the achievability result. Before proving Lemma \ref{gen_cht}, we first prove the lemma below, which is for the case $|\cS| = 2$.
\begin{lemma}
\label{comp2}
Consider $\rho_1, \rho_2 \text{ and }\sigma \in \cD(\cH)$ such that $\mbox{supp}(\rho_i) \subseteq \mbox{supp}(\sigma),$ for every $i \in \{1,2\}$. Let $\Pi_n^{(1)}\text{ and }\Pi_n^{(2)}$ be two projectors obtained from \eqref{F42} of Fact \ref{F4} with respect to $\{\rho_1, \sigma\}$ and $\{\rho_2, \sigma\}$ respectively. Then, $\forall \eps,\delta,\delta' \in (0,1)$ and large enough $n$, we can construct a projector $\Pi_{n}^{\star}$ such that,
\begin{align}
    &\min_{i \in \{1,2\}}\tr[\Pi_{n}^{\star}\rho_i^{\otimes n}] \geq 1 - \eps- 2\delta^{\frac{1}{2}},\label{cht1}\\
    &\frac{1}{n}\log\left(\tr[\Pi_{n}^{\star}\sigma^{\otimes n}]\right) \leq -\min_{i \in \{1,2\}}D(\rho_i || \sigma)+ \delta' +  \frac{1}{n}\log\left(\frac{9}{\delta}\right).\label{cht2}
\end{align}
\end{lemma}
\begin{proof}
    The proof follows using techniques similar to the one used in the proof of Lemma \ref{lbound}. For the projectors $\bar{\Pi}_n^{(i)} := \mathbb{I} - \Pi_n^{(i)}, \forall i \in \{1,2\},$ we have

    \begin{align}
        &\tr[\bar{\Pi}_n^{(i)} \rho_i^{\otimes n}] \leq \eps,\nonumber\\
        &\tr[\bar{\Pi}_n^{(i)} \sigma^{\otimes n}] \geq 1 - 2^{-n \left(D(\rho_i || \sigma) - \delta'\right)}.\label{pi2c}
    \end{align}
    
    Let, $\{\ket{v_{n,\alpha}}\}_{{\alpha}=1}^k$ and $\{\ket{w_{n,\alpha}}\}_{{\alpha}=1}^k$ be the eigenvectors of $\bar{\Pi}_n^{(1)} \text{ and } \bar{\Pi}_n^{(2)}$ respectively obtained after the Jordan decomposition.
For $\delta \in (0,1)$, consider the following set:
\begin{align*}
{\G}&:=\left\{{\alpha}: \sin^2(\theta_{\alpha}) \leq \delta \right\},
% \label{good_lb_corollary}
\end{align*}
where, $\cos^2(\theta_{n,\alpha}) = {|\langle v_{n,\alpha}| w_{n,\alpha}\rangle|}^2$. Then, fixing $\eps_1 = 2^{-n 
    \left(D(\rho_1 || \sigma) - \delta'\right)}, \eps_2 = 2^{-n 
    \left(D(\rho_2 || \sigma) - \delta'\right)}$ in Lemma \ref{c23_gub} we have, 
\begin{equation}
    \sum_{\alpha \in \G}\tr[\sigma^{n}_{\alpha}] \geq 1- \frac{8}{\delta}.2^{-n 
    \left(\underset{i \in \{1,2\}}{\min}D(\rho_i || \sigma) - \delta'\right)}. \label{pgl1}
\end{equation}

We now construct $\Pi_{n}^{\star}$ as claimed in the statement of the lemma as follows, 
%%\vspace{5pt}
\begin{equation}
\label{starcht}
\Pi_{n}^{\star}:= \mathbb{I} - \sum_{{\alpha} \in \mbox{\G}} \ket{v_{n,\alpha}}\bra{v_{n,\alpha}}.
\end{equation}

We first prove \ref{cht2} using the following set of inequalities,
% \subsection{Proof of \eqref{cht1}}
% Consider the following set of inequalities,
\begin{align*}
    \tr[\Pi_{n}^{\star}\sigma^{\otimes n}] &=1 - \sum_{\alpha \in \G}\tr\left[\ket{v_{n,\alpha}}\bra{v_{n,\alpha}}\sigma^{n}_{\alpha}\right]\\
    &= 1 -\sum_{\alpha=1}^{k}\tr\left[\ket{v_{n,\alpha}}\bra{v_{n,\alpha}}\sigma^{n}_{\alpha}\right] + \sum_{\alpha \notin \G}\tr\left[\ket{v_{n,\alpha}}\bra{v_{n,\alpha}}\sigma^{n}_{\alpha}\right]\\
    &\leq 1 - \tr[\bar{\Pi}_n^{(1)}\sigma^{\otimes n}] + \sum_{\alpha \notin \G}\tr\left[\sigma^{n}_{\alpha}\right]\\
   &\overset{a}{\leq} 2^{-n \left(D(\rho_1 || \sigma)-\delta'\right)} + \frac{8}{\delta}.2^{-n 
    \left(\underset{i \in \{1,2\}}{\min}D(\rho_i || \sigma) - \delta'\right)}
    \end{align*}
\begin{align*}
    &\leq \frac{9}{\delta}.2^{-n 
    \left(\underset{i \in \{1,2\}}{\min}D(\rho_i || \sigma) - \delta'\right)},\hspace{100pt}
\end{align*}
where, $a$ follows from the \eqref{pi2c} and \eqref{pgl1}. So we have given a proof for \eqref{cht2}. 
Now we prove \eqref{cht1} in the two following cases. First, we show the following,
\begin{align*}
    \tr[\Pi_{n}^{\star}\rho_1^{\otimes n}] &=  1 - \sum_{\alpha \in \G}\tr\left[\ket{v_{n,\alpha}}\bra{v_{n,\alpha}}\rho^{\otimes n}_{1,\alpha}\right]\\
    &\geq 1 - \tr[\bar{\Pi}_n^{(1)} \rho_1^{\otimes n}]\\
    &\geq 1 -\eps.
\end{align*}

Then, similar to the proof of \eqref{r3}, we obtain the following,
\begin{equation*}
    \tr[\Pi_{n}^{\star}\rho_2^{\otimes n}] \geq 1 - \eps -  2\delta^{\frac{1}{2}}.
\end{equation*}
\end{proof}

We now prove a general version of Lemma \ref{comp2} below, wherein for the sake of simplicity, we assume that $|\cS|=2^t$ for some $t>0.$ The case when $|\cS|$, is not of the form $2^t$ can be handled similarly.
\begin{lemma}\label{gen_cht}
Let $\{\rho_s\}_{s=1}^{|\cS|}$ be a collection of states and $\sigma$ be another state such that, $\forall s \in \cS,$ $\mbox{supp}(\rho_s) \subseteq \mbox{supp}(\sigma).$ Then, $\forall \eps,\delta,\delta' \in (0,1)$ and large enough $n$, we can construct a projector $\Pi_{n}^{\star}$ such that,
\begin{align}
    &\min_{s \in \cS}\tr[\Pi_{n}^{\star}\rho_s^{\otimes n}] \geq 1 - \eps- 2\delta^{\frac{1}{2}}\log|\cS|,\label{many_rho_1_sigma_1}\\
    &\frac{1}{n}\log\left(\tr[\Pi_{n}^{\star}\sigma^{\otimes n}]\right) \leq -\min_{s \in \cS}D(\rho_s || \sigma) + \delta' + \frac{1}{n}\log\left(\left(\frac{9}{\delta}\right)^{\log |\cS|}\right).\label{many_rho_1_sigma_2}
\end{align}
\end{lemma}
\begin{proof} The proof follows using techniques similar to that used in the proofs of Lemma \ref{comp2} and Lemma \ref{generalubound} by replacing the set $\G^{(i)}$ in \eqref{goodi} with the set $\G^{(i)} :=\left\{{\alpha}: \sin^2(\theta_{\alpha}) \leq \delta \right\}$ for all $i \in [\log\abs{\cS}]$ i.e. keeping the set $\G^{(i)}$ fixed throughout the steps.
\end{proof}

For the case when $\mathbf{|\cS|}$ is not in the form of ${2^{t}}$, for $t > 0$, then the term $\log|\cS|$ in \eqref{many_rho_1_sigma_1} and \eqref{many_rho_1_sigma_2} becomes $\ceil{\log|\cS|}$. 

A similar result was also proved by Anshu et al. in \cite[Lemma 3]{AJW-Compound} using Jordan's lemma. Even though {we also use Jordan's lemma} in our construction, there is a difference both in the proof technique and the bounds obtained between these two results.

{The definition of the set $\mbox{Near}$ in \cite[Proof of Lemma $2$]{AJW-Compound} is not similar to that of the set $\mbox{Good}$ in our case } and the construction of {$\Pi^{\star}_{n}$ in Lemma \ref{gen_cht} is different than that of $\Pi^{\star}$ in \cite[Lemma 3]{AJW-Compound}}. This is because, in Lemma \ref{gen_cht}, for each pair of projectors,{ we find the complementary subspace of the intersection of the complementary subspaces of individual projectors}, whereas in  \cite[Lemma 3]{AJW-Compound}, for each pair of projectors, {the union of the projectors is computed directly}.

While there is a difference in strategy, the result of Lemma \ref{gen_cht} is slightly different than \cite[Lemma 3]{AJW-Compound} as well. The lower-bound mentioned in the RHS of \eqref{many_rho_1_sigma_1} of Lemma \ref{gen_cht}, decreases with a rate $ O(\delta^{\frac{1}{2}})$ with increasing $\delta$  whereas, in \cite[Lemma 3]{AJW-Compound} the lower-bound decreases with a rate $O(\delta)$ with increasing $\delta$ and is smaller than $O(\delta ^{\frac{1}{2}})$. However, the upper-bound mentioned in the RHS of \eqref{many_rho_1_sigma_2} performs better than that of \cite[Lemma 3]{AJW-Compound}, since in the RHS of \eqref{many_rho_1_sigma_2}, the upper-bound decreases with a rate $O(\frac{1}{\delta})$ which is smaller than $O(\frac{1}{\delta^2})$ in \cite[Lemma 3]{AJW-Compound}. Thus it is important to observe that, there is a trade-off between Lemma \ref{gen_cht}  and \cite[Lemma 3]{AJW-Compound}. 

    Using Lemma \ref{gen_cht} and Definition \ref{chtdef}, we now give the following upper-bound for the quantity $\beta_{(n,\eps)}\left(\Theta_{\cS},\sigma\right)$ defined in \eqref{betacht},
    \begin{theorem}(Upper bound)
        Fix a set $\Theta_{\cS} \subseteq \cD(\cH)$ and consider $\eps,\delta,\delta' \in (0,1)$. Then,
        \begin{equation*}
        \lim_{n \to \infty}\frac{1}{n}\log\beta_{\left(n,f\left(\eps,\delta, |\cS|\right)\right)}(\Theta_{\cS},\sigma) \leq -\min_{\rho \in \Theta_{\cS}}D(\rho || \sigma) + \delta',
    \end{equation*}
    where, $f\left(\eps,\delta, |\cS|\right) = \eps + 2\log|\cS|\delta^{\frac{1}{2}}.$
    \end{theorem}
    
\subsection{Asymmetric Composite Hypothesis Testing $\big($The case of $\rho$ and $\{\sigma_s\}_{s=1}^{|\cS|}\big)$}
Consider $\rho$ and a collection $\{\sigma_s\}_{s=1}^{|\cS|}$ where $s \in \cS,|\cS| < \infty.$ and $\forall s,  \rho,\sigma_{s} \in \cD(\cH)$. The aim here is to accept the state $\rho$  with high probability and the probability that any state from the collection $\{\sigma_s\}_{s=1}^{|\cS|}$ is wrongly identified as $\rho$ should be very small. Formally, this problem can be defined  as follows,
\begin{definition}\label{chtdef}
    Consider $\rho \in \cD\left(\cH\right)$ and let $\Theta_{\cS} := \left\{\sigma_{s}\right\}_{s=1}^{|\cS|}$ be a collection of states, where $\forall s \in \cS, \sigma_s \in \cD(\cH)$. For an $\eps \in (0,1)$ and a positive integer $n \in \mathbb{N}$, we define,
    \begin{equation}
        \min_{\substack{0\prec\Lambda\preceq\mathbb{I_n}:\\
        \tr[\Lambda\rho^{\otimes n}] \geq 1 -\eps}}\max_{s \in \cS}\tr[\Lambda\sigma_s^{\otimes n}],\label{betacht1}
    \end{equation}
    where, $\mathbb{I}_n$ is an identity operator on $\cH^{\otimes n}$. 
    \end{definition}
    
In this subsection, we will give an upper-bound (achievability) on $\tr[\Lambda\sigma_s^{\otimes n}]$. However, before giving this upper-bound, we first discuss the following classical problem. 
% The general case can be handled similarly. Before proving Lemma \ref{gen_cht}, we consider a classical version of the above problem.
Consider a distribution $P$ and a finite collection of distributions $\{Q_i\}_{i=1}^{|\cS|}$, defined over a set $\cX.$ Further, $\forall i \in \cS,$ we are given a set $\cA_{i} \subset \cX$ such that $\Pr_{P}\{\cA_{i}\} \geq 1- \eps$ and $\Pr_{Q_i}\{\cA_{i}\} \leq 2^{-k_{i}},$ for some $k_{i}>0.$ It is easy to show that $\forall i \in \cS_,$  we have $\Pr_{P_i}\{\cA^\star\}
 \geq 1- |\cS|\eps$ and $\Pr_{Q_j}\{\cA^\star\} \leq 2^{-k},$ where $k:= \min_{i}k_{i,}$ and $\cA^\star:= \cap_{i=1}^{|\cS|}\cA_{i}$. In Lemma \ref{ssrho}, we construct a $\Pi_{n}^{\star}$ which behaves very similar to $\cA^\star:= \cap_{i=1}^{|\cS|}\cA_{i}$.
This lemma will help us to prove the achievability result below. 
\begin{lemma}\label{ssrho}
    Consider a state $\rho \in \cD\left(\cH\right)$ and a collection of quantum states $\{\sigma_s\}_{s=1}^{|\cS|}$ such that for $\eps,\delta \in (0,1) ,\forall s \in \cS$ and for $n$ large enough,
%     \begin{align*}
% \tr\left[\Pi^{(s)}_{(n)}\rho_{X^n}^{(s_0)}\right]\right] &\geq 1 - \eps,\\
%         \mathbb{E}_{X^n}\left[\tr\left[\Pi^{(s)}_{(n)}\rho_{X^n}^{(s)}\right]\right] &\leq 2^{-n(D(\rho^{(s_0)} || \rho^{(s)}) + \delta)},
%     \end{align*}
    we can construct a projector $\Pi_{n}^{\star}$ which satisfies the following,
    \begin{align}
        % \mathbb{E}_{X^n}\left[P_{Y^n|X^n}^{(s_0)}(D^{(s_0)})\right] &\geq 1 - (\ceil{\log|\cS|} + 1)\eps^{1 - \ceil{\log|\cS|}.\alpha},\\
       \tr\left[\Pi_{n}^{\star}\rho^{\otimes n}\right] &\geq 1 -\left(\ceil{\log|\cS|}+1\right)\eps^{\left({1 - \ceil{\log|\cS|}\beta}\right)},\nonumber\\
        % \mathbb{E}_{X^n}\left[P_{Y^n|X^n}^{(s)}(D^{(s_0)})\right] &\leq 
        % 2^{-n(D(P_{Y^n}^{(s_0)}||P_{Y^n}^{(s)}) + \delta)}+O(\eps^{\frac{1}{4}}), \quad \forall s \in \cS
        \tr\left[\Pi_{n}^{\star}\sigma_s^{\otimes n}\right] &\leq 
        2^{-n(\min_{s \in \cS}D(\rho || \sigma_s) + \delta)} + c\left(\ceil{\log|S|}\right)\eps^{{\frac{\beta}{2}}}.\label{ssrhoeq}
    \end{align}
where, $\beta$ is such that $\left(\ceil{\log|\cS|}+1\right)\eps^{\left({1 - \ceil{\log|\cS|}\beta}\right)} \ll 1$ and $c(\cdot)$ is defined in the statement of Lemma \ref{generalubound}.
\end{lemma}
\begin{proof}
    See subsection \ref{ssrhop} in Appendix.
\end{proof}
\subsection{Asymmetric Composite Hypothesis Testing $\big($The case of $\{\rho_s\}_{s=1}^{|\cS_1|}$ and $\{\sigma_t\}_{t=1}^{|\cS_2|}\big)$}
We now deal with a more general case wherein we have two collections of states $\{\rho_s\}_{s=1}^{|\cS_1|} \text{ and } \{\sigma_t\}_{t=1}^{|\cS_2|},$ where $s \in \cS_1,t \in \cS_2$ and $|\cS_1|,|\cS_2| < \infty.$ The aim here is to accept any state from the collection $\{\rho_s\}_{s=1}^{|\cS_1|}$  with high probability and the probability that any state from the collection $\{\sigma_s\}_{s=1}^{|\cS|}$ is wrongly identified as a state in  $\{\rho_s\}_{s=1}^{|\cS_1|}$ should be very small. To be more precise consider the statement of Lemma \ref{genmany}. For the sake of simplicity, we will assume  ${|\cS_1|:=2^u,|\cS_2|:=2^v}$ in the lemma below for some ${u,v>0}$. The general case can be analyzed similarly. Before proving Lemma \ref{genmany}, we consider a classical version of the above problem. Consider two finite collections of distributions $\cS_1:=\{P_i\}_{i=1}^{|\cS_1|}$ and $\cS_2:=\{Q_j\}_{j=1}^{|\cS_2|},$ defined over a set $\cX.$ Further, $\forall i \in \cS_1$ and $\forall j \in \cS_2,$ we are given a set $\cA_{ij} \subset \cX$ such that $\Pr_{P_i}\{A_{ij}\} \geq 1- \eps$ and $\Pr_{Q_j}\{A_{ij}\} \leq 2^{-k_{ij}},$ for some $k_{ij}>0.$ We want to construct an $\cA^\star$ such that $\forall i \in \cS_1, j \in \cS_2,$ we have $\Pr_{P_i}\{\cA^\star\} \geq 1- |\cS_2|\eps$ and $\Pr_{Q_j}\{\cA^\star\} \leq 2^{-k},$ where $k:= \min_{i,j}k_{i,j.}$ It is easy to see that $\cA^\star:= \cap_{j=1}^{|\cS_2|}\cup_{i=1}^{|\cS_1|}\cA_{i,j}.$ In Lemma \ref{genmany}, we construct a $\Pi_{n}^{\star}$ which behaves very similar to $\cA^\star:= \cap_{j=1}^{|\cS_2|}\cup_{i=1}^{|\cS_1|}\cA_{i,j}.$
This lemma will help us to prove the achievability result below. 
\begin{lemma}\label{genmany}
Let $\{\rho_s\}_{s=1}^{|\cS_1|}$ and $\{\sigma_t\}_{t=1}^{|\cS_2|}$ be two collections of states such that, $\forall s \in \cS_1,t \in \cS_2$, $\mbox{supp}(\rho_s) \subseteq \mbox{supp}(\sigma_t).$ Then, $\forall \eps,\delta,\delta' \in (0,1)$ and large enough $n$, we can construct a projector $\Pi_{n}^{\star}$ such that,
\begin{align}
&\tr[\Pi_{n}^{\star}\rho_s^{\otimes n}] \geq 1 - {\left(\frac{9}{\delta}\right)}^{\log|\cS_2|}\left(\eps+ 2\delta^{\frac{1}{2}}\log|\cS_1|\right), \quad \forall s \in \cS_1,\label{genmanyeq2}\\
&\tr[\Pi_{n}^{\star}\sigma_t^{\otimes n}] \leq \left(\frac{9}{\delta}\right)^{\log|\cS_1|} 2^{-n\left(\underset{\substack{s \in \cS_1, \\ t \in \cS_2}}{\min}D(\rho_s || \sigma_t) - \delta'\right)} + 2\delta^{1/2}\log|\cS_2|, \quad \forall t \in \cS_2.\label{genmanyeq}
\end{align}
\end{lemma}
\begin{proof}
The proof follows using techniques similar to that used in the proofs Lemma \ref{gen_cht} and Lemma \ref{ssrho} while defining the set $\G^{(i)} :=\left\{{\alpha}: \cos^2(\theta_{\alpha}) \geq 1 - \delta \right\}$ in the proof of Lemma \ref{ssrho}, for all $i \in [\log\abs{\cS_2}]$.
\end{proof}

\subsection{Comparison with \cite[Theorem $1.1$]{Berta2017OnCQ}}\label{bertasub}
In \cite{Berta2017OnCQ}, Berta et al. studied the following problem. They considered two sets of quantum states $\cS$ and $\cT.$  Further, they defined the sets $\cS_n:= \mbox{conv}\{\rho^{\otimes n}; \rho \in \cS\}$ and $\cT_n:= \mbox{conv}\{\sigma^{\otimes n}; \sigma \in \cT\},$ The aim here is to design a positive operator $\Lambda_n \preceq \mathbb{I}$, such that for all $\rho^{\otimes n} \in \cS_n$,  we have, $\tr (\Lambda_n \rho^{\otimes n})\geq 1-\eps,$ where $\eps \in (0,1)$ and for every  $\sigma^{\otimes n} \in \cT_n$, $\tr (\Lambda_n \sigma^{\otimes n})$ is as small as possible. For this problem in the case when $\cS$ is a singleton set. \cite[Theorem 1.1]{Berta2017OnCQ} implies existence of a POVM such that $\tr (\Lambda_n \rho^{\otimes n})\geq 1-\eps,$ with the property that for every for every  $\sigma^{\otimes n} \in \cT_n$ 
\beq
\label{berta}
\tr (\Lambda_n \sigma^{\otimes n}) \leq 2^{- \underset{\mu \in \cT}{\inf} D(\rho^{\otimes n} \| \int \sigma^{\otimes n}d\mu(\sigma))}. 
\enq

 As observed in \cite{Berta2017OnCQ}, the exponent in \eqref{berta} is not in single letter form, which is due to the following reason:
 \begin{equation}
     \lim_{  n \to \infty}\frac{1}{n} \underset{\mu \in \cT}{\inf} D(\rho^{\otimes n} \| \int \sigma^{\otimes n}d\mu(\sigma)) \neq \underset{\sigma \in \cT}{\inf} D(\rho \| \sigma).\label{non_convergence}
 \end{equation}
 
 As compared to \eqref{berta}, in \eqref{ssrhoeq} of Lemma \ref{ssrho}, the exponent of the leading term is in single letter form. However, because of the non-commutativity of the projection operators involved, we get an additive term, which can be made arbitrarily small. 
 
 The case, when $\cS$ is not a singleton set, \cite[Theorem 1.1]{Berta2017OnCQ} implies existence of a POVM such that $\underset{\nu \in \cT}{\sup}\tr (\Lambda_n  \int \rho^{\otimes n}d\nu(\rho))\geq 1-\eps,$ with the property that for every for every  $\sigma^{\otimes n} \in \cT_n$, 
\beq
\label{berta2}
\tr (\Lambda_n \sigma^{\otimes n}) \leq 2^{\underset{\substack{\hspace{2pt}\rho \in \cS, \mu \in \cT}}{-\inf} D(\rho^{\otimes n} \| \int \sigma^{\otimes n}d\mu(\sigma))}. 
\enq

The exponent in \ref{berta2} is not in single letter form, whereas, in \eqref{genmanyeq} of Lemma \ref{genmany}, the exponent of the leading term is in single letter form due to \eqref{non_convergence}. However, we get an additive term in terms of $\delta$. This is because proof of Lemma \ref{genmany} follows from the result of Lemma \ref{ssrho}, which faces the same issue discussed in the above paragraph. Therefore, we can not make the probability that any state from the collection $\{\sigma_s\}_{s=1}^{|\cS|}$ is wrongly identified as a state in  $\{\rho_s\}_{s=1}^{|\cS_1|}$ arbitrarily small, since there is a trade-off between \eqref{genmanyeq2} and \eqref{genmanyeq} in terms of $\delta$. 

However, note that Anshu et al. in \cite[Theorem $5$]{AJW-Compound}, using the results of \cite[Lemma $3$]{AJW-Compound}, provide a proof of \eqref{berta2}, when $\cT$ is a finite set. As observed earlier, apart from the constants, the result mentioned in Lemma \ref{gen_cht} is structurally similar to that of \cite[Lemma $3$]{AJW-Compound}. Therefore, one can recover a proof of ($51$) in the case when $\cT$ is a finite set using Lemma $7$.
% \begin{lemma}\label{asym_hypo_test_achievL}
% For the asymmetric composite hypothesis testing problem there exists a binary projective measurement $\{\Pi^\star, \mathbb{I}-\Pi^\star\}$ such that for $n$ large enough, we have
% \begin{align*}
% \tr[\Pi^\star \rho^{\otimes n}] \geq 1- \left(\ceil{\log\left(|\cS|\right)}+1\right){\eps^{\left({1 -\ceil{\log\left(|\cS|\right)}.\beta}\right)}_n},
% \end{align*}
% where, $\lim_{n \to \infty}\eps_n \to 0.$ Further, for every $i \in \{1, \cdots, |\mathcal{S}|\}$, 
% \begin{align*}
% \tr[\Pi^\star \sigma^{\otimes n}_i] \leq 2^{-n (\cD(\rho\| \sigma_i) -\delta)} +O({\eps^{\ceil{\log\left(|\cS|\right)}.\frac{\beta}{2}}_n)
% \end{align*}
% where, $\delta \in (0,1)$ is arbitrary.
% \end{lemma}
% \begin{proof}
%     Consider a collection of projectors $\left\{\Pi_i\right\}_{i \in \{1,2,\cdots,|\cS|\}}$, where for every $i \in \{1, \cdots, |\cS|\}$, for some $\delta \in (0,1), \Pi_i$ is defined in the following way:
%     \begin{equation}
%         \Pi_i := \left\{\sigma^{\otimes n}_i \preceq 2^{-n\left(\cD(\rho||\sigma_i)-\delta\right)}\rho^{\otimes n}\right\}
%     \end{equation}
%     Now applying the method used to proof the result in Appendix \ref{Generalised} on this collection $\left\{\Pi_i\right\}_{i \in \{1,2,\cdots,|\cS|\}}$, directly gives us a proof of Lemma \ref{asym_hypo_test_achievL}
% \end{proof}

\section{Conclusion and Acknowledgements}
In this work, we have provided a notion of intersection and union of a collection of projection operators. Our method is based on Jordan's lemma and is arguably intuitive. An observation that we make while constructing a projection operator which behaves analogous to intersection of subsets is to consider only those sub-spaces that have very small angles between them. Using our technique we study a communication model, wherein, a third party authenticates the communication channel. We derive the capacity of this communication model. This communication model is similar to the compound channel case. However, unlike the compound channel case, the parties are only interested to communicate successfully only if the particular $s_o$ channel in the collection is used. In all other cases, the receiver should either declare that some channel other than $s_0$ channel is being used or it should decode the message correctly. We also study the problem of asymmetric composite hypothesis testing using the techniques developed in this manuscript.

As a future work, one can study a more general version of the communication model studied in this paper. In a more general version, the parties are interested to communicate successfully if the $s_0$ channel is being used across all the $n$ channel uses. In other cases when the $s_o$ channel is not being used across all the $n$ channel uses, i.e., if the channel statistic is changing across all the $n$ channel uses. Then, the receiver should either declare that the channel is behaving arbitrarily or decode the message correctly. Further, it will be interesting to find more applications of the intersection and union projectors developed in this work. 

We thank the anonymous referees and the editor for their helpful suggestions which have helped us to greatly improve the presentation and the technical content of this manuscript. We are grateful to one of the referees for the changes suggested in the proofs of Lemma \ref{lbound}, \ref{mr} and \ref{generalubound}.

The work of N. A. Warsi was supported in part by MTR/2022/000814, DST/INT/RUS/RSF/P-41/2021 from the Department of Science \& Technology, Govt. of India and DCSW grant provided by the Indian Statistical Institute.

\bibliographystyle{ieeetr}
\bibliography{master}

% As a future work, 
% it will be an interesting problem to prove a result similar to our main result for the reverse operator inequality. To elaborate more on this, consider $\rho, \sigma_1$ and $\sigma_2 \in \mathcal{D}(\cH)$ such that $\mbox{supp}(\rho) \subseteq \mbox{supp}(\sigma_1)$ and $\mbox{supp}(\rho) \subseteq \mbox{supp}(\sigma_2)$. Suppose for some ${k_1}, k_2 >0,$ let $\Pi_1, \Pi_2$ be such that $\Pi_1:=\{\sigma_1 \preceq 2^{-{k_1}}\rho\}$ and  $\Pi_2:=\{\sigma_2 \preceq 2^{-{k_2}}\rho\}$ with the property that $\tr [\Pi_1 \rho] \geq 1-\eps$ and $\tr[\Pi_2\rho] \geq 1- \eps.$ Then, finding a projector $\Pi^\star$ such that $\tr[\Pi^{\star}\rho] \geq 1- f(\eps),$ $\tr[\Pi^{\star}\sigma_1] \leq 2^{-{k_1}}f(\eps)$ and $\tr[\Pi^{\star}\sigma_2] \leq 2^{-{k_2}}f(\eps),$ where $f(\eps)$ is some function of $\eps$. 
%  \\\\\\
\section{Appendix}
\subsection{Proof of Lemma \ref{c23_gub}:}\label{proofc23_gub}
Using the basis obtained from Jordan's lemma (fact \ref{jordan}) we have the following representation of  $\Pi_1\Pi_2\Pi_1,$
\begin{equation}
\label{intersection_gub}
\Pi_1\Pi_2\Pi_1 := \sum_{{\alpha} =1}^{\bar{k}} \cos^2(\theta_{\alpha})\ket{v_{\alpha}}\bra{v_{\alpha}},
\end{equation}
where, $\cos^2(\theta_{\alpha}):= {|\langle v_{\alpha}| w_{\alpha}\rangle|}^2$.
We now have the following set of inequalities,  
\begin{align}
1 - 4(\eps_1+\eps_2) & \overset{a}\leq \tr[\Pi_1\Pi_2\Pi_1 \rho] \nonumber\\
 &\overset{b}= \sum_{{\alpha} =1}^{\bar{k}} \cos^2(\theta_{\alpha})\tr[\ket{v_{\alpha}}\bra{v_{\alpha}} \rho_{\alpha}] \nonumber\\
 \label{expec_gub}
 & \overset{c}\leq \sum_{{\alpha} =1}^{\bar{k}}\cos^2(\theta_{\alpha})\tr[\rho_{\alpha}],
\end{align}
where, $a$ follows from the Gao's union bound (Fact \ref{Gao}) and the fact that $\tr[\Pi_1 \rho] \geq 1 - \eps_1$ and $\tr[\Pi_2 \rho] \geq 1 - \eps_2$, $b$ follows from \eqref{intersection_gub} and $c$ follows from the property of the trace and the fact that $\ket{v_{\alpha}}\bra{v_{\alpha}}$ is a projector. Thus, it follows from \eqref{expec_gub} that 
\begin{align}
\label{expectation_gub}
\mathbb{E}[\cos^2(\theta_{\alpha})] \geq 1-4(\eps_1+\eps_2),
\end{align}
where, in the above the expectation is calculated with respect to the distribution $\tr[\rho_{\alpha}].$ Thus, 
\begin{align}
\Pr\{\sin^2(\theta_{\alpha}) > \delta\} & \overset{a} \leq \frac{\mathbb{E}[\sin^2(\theta_{\alpha})] }{\delta}\nonumber\\
&\overset{b} \leq \frac{4(\eps_1 + \eps_2)}{\delta},\label{c23_gub_equation}
% 1 - \Pr\{\G\} &\leq \eps^{\frac{1}{2}}\nonumber\\
%  \Pr\{\G\}&\geq 1 - \eps^{\frac{1}{2}}\nonumber
 \end{align}
 where, $a$ follows from the Fact \ref{markov} and $b$ follows from \eqref{expectation_gub}. This completes the proof of Lemma \ref{c23_gub}.\hfill\qed

\subsection{Proof of Lemma \ref{generalubound}}\label{Generalised}

We will prove Lemma \ref{generalubound} using `proof by induction'. We start the collection of $|\cS|= 2^{t}$ projectors ${\{\Pi_i\}}_{i\in\{1,2,\cdots,|\cS|\}}$. We then, consider the pairs of projectors $\left\{\Pi_{1},\Pi_{2}\right\}, \left\{\Pi_{3},\Pi_{4}\right\},\cdots\left\{\Pi_{2^t-1},\Pi_{2^t}\right\}$.  For each $i \in \{1,2,\cdots,2^{t-1}\}$, we have the following properties for $\left\{\Pi_{2i-1},\Pi_{2i}\right\}$,
\begin{align*}
    &\tr[\Pi_{2i-1}\rho] \geq 1- \eps,\\
     &\tr[\Pi_{2i-1}\sigma_{2i-1}] \leq 2^{-k_{2i-1}},\\
      &\tr[\Pi_{2i}\rho] \geq 1- \eps,\\
     &\tr[\Pi_{2i}\sigma_{2i}] \leq 2^{-k_{2i}}.
\end{align*}

Using a similar approach used to prove Lemma \ref{lbound}, we construct intersection projectors $\Pi_{\{1,2\}},\Pi_{\{3,4\}},$\quad$\cdots,\Pi_{\{2^{t}-1,2^{t}\}}$ corresponding to the pairs $\left\{\Pi_{1},\Pi_{2}\right\}, \left\{\Pi_{3},\Pi_{4}\right\},\cdots\left\{\Pi_{2^t-1},\Pi_{2^t}\right\}$ respectively using the following set,
\begin{equation}
    \G^{(1)}:= \left\{{\alpha}: \cos^2(\theta_{\alpha}^{(1)}) \geq 1- 8\eps^{\beta}\right\},\label{goodstep1}
\end{equation}
where, $\theta_\alpha^{(1)}$ depends on the choice of the pair of the projectors at step $1$. Thus, $\forall i \in \{1,2,\cdots,2^{t-1}\}$, $\Pi_{\{2i-1,2i\}}$ has the following properties,
\begin{align*}
    \tr[\Pi_{\{2i-1,2i\}}\rho] &\geq 1 - 2\eps^{(1 - \beta)}, \\
    % \tr[\Pi_{\{2i-1,2i\}}\sigma^{\otimes n}_] &\leq \tr[\Pi_1 \sigma^{\otimes n}_1]\\
    % &\leq 2^{-nD(\rho || \sigma_1)}\\
    \tr[\Pi_{\{2i-1,2i\}}\sigma_{j}] 
    &\leq 2^{-k_{j}} + c(1)\eps^{\frac{\beta}{2}}, \forall j \in \{2i-1,2i\},
\end{align*}
where, $c(1) \in [0,4\sqrt{2}]$. Thus, the above discussion proves the base case. Note that in this step for each pair of projectors, we define $\G^{(1)}$ by replacing $\eps^{\frac{1}{2}}$ by $\eps^\beta$ in \eqref{good1}. In the next step we again construct intersection projectors $\Pi_{\{1,2,3,4\}},\Pi_{\{5,6,7,8\}},\cdots,\Pi_{\{2^{t}-3,2^{t}-2, 2^{t}-1, 2^{t}\}}$ corresponding to the pairs $\left\{\Pi_{\{1,2\}},\Pi_{\{3,4\}}\right\},$ $\left\{\Pi_{\{5,6\}},\Pi_{\{7,8\}}\right\},\cdots,\left\{\Pi_{\{2^{t}-3,2^{t}-2\}}, \Pi_{\{2^{t}-1,2^{t}\}}\right\}$ respectively.

We keep doing this for $t-1$ times and at each $k$-th step we define $\G^{(k)}$ for the pair of projectors as follows,
\begin{align}
\G^{(k)}&:= \left\{{\alpha}: \cos^2(\theta_{\alpha}^{(k)}) \geq 1- 8k\eps^{\beta}\right\},\label{goodi}
\end{align}
where, $\theta_\alpha^{(k)}$ depends on the choice of the pair of the projectors at $k$-th step similar to that of \eqref{valphatilda}. Finally, we are left with the pair $\left\{\Pi_{\{1,\cdots,2^{t-1}\}},\Pi_{\{2^{t-1}+1,\cdots,2^{t}\}}\right\}$. We assume that for $t-1$-th step the statement of Lemma \ref{generalubound} holds i.e. $\Pi_{\{1,\cdots,2^{t-1}\}}$ and $\Pi_{\{2^{t-1} + 1,\cdots,2^{t}\}}$ satisfy the following:

\begin{align}
    \tr[\Pi_{\{1,\cdots,2^{t-1}\}}\rho] &\geq 1 - t\eps^{1-(t-1)\beta},\label{ind_gub_1}\\
    \tr[\Pi_{\{1,\cdots,2^{t-1}\}}\sigma_u] &\leq 2^{-k_u} + c({t-1})\eps^{\frac{\beta}{2}}, \forall u \in \{1,\cdots,2^{t-1}\},\label{ind_gub_2}\\
    \tr[\Pi_{\{2^{t-1} + 1,\cdots,2^{t}\}}\rho] &\geq 1 - t\eps^{1-(t-1)\beta},\label{ind_gub_3}\\
    \tr[\Pi_{\{2^{t-1} + 1,\cdots,2^{t}\}}\sigma_v] &\leq 2^{-k_v} + c({t-1})\eps^{\frac{\beta}{2}}, \forall v \in \{2^{t-1} + 1,\cdots,2^{t}\},\label{ind_gub_4}
\end{align}
where, $c({t-1}) \in [0,4\sqrt{2}\sum_{i=1}^{t-1}\sqrt{i}].$ For simplicity of notations, we denote $\Pi_{\{1,\cdots,2^{t-1}\}}$ and $\Pi_{\{2^{t-1} + 1,\cdots,2^{t}\}}$ as $\Pi_{A}$ and $\Pi_{B}$. We will apply Fact \ref{jordan} on the pair $\{\Pi_A, \Pi_B\}.$ Consider $\left\{\mathbf{P}_{\alpha}\right\}_{{\alpha}=1}^{{\bar{k}}}$ (where $\bar{k}$ is some natural number depending on $\Pi_A,\Pi_B$) as the set of orthogonal projectors obtained by Fact \ref{jordan} applied with respect to the pair $\{\Pi_A, \Pi_B\}.$ Thus, $\sum_{{\alpha} =1}^{{\bar{k}}}\mathbf{P}_{\alpha} = \mathbb{I}.$ Furthermore, for every ${\alpha} \in [1:{\bar{k}}]$ we define  $\Pi_{A,{\alpha}} := \mathbf{P}_{\alpha}\Pi_A\mathbf{P}_{\alpha}$ as the following one dimensional projector: 
%%\vspace{5pt}
\begin{align}
\Pi_{A,{\alpha}} := \ket{v_{\alpha}}\bra{v_{\alpha}}\nonumber,
\end{align}
for some $\ket{v_{\alpha}}$ in the range of $\mathbf{P}_{\alpha}.$ Similarly, for every ${\alpha} \in [1:{\bar{k}}]$, we have
\begin{align}
\Pi_{B,{\alpha}} := \ket{w_{\alpha}}\bra{w_{\alpha}}\nonumber,
\end{align}
%%\vspace{5pt}
for some $\ket{w_{\alpha}}$ in the range of $\mathbf{P}_{\alpha}.$
Thus, it now follows from the property of $\{\ket{v_{\alpha}}\}_{{\alpha}=1}^{\bar{k}}$ and $\{\ket{w_{\alpha}}\}_{{\alpha}=1}^{\bar{k}}$ that 
%%\vspace{5pt}
\begin{align*}
\Pi_A &:= \sum_{{\alpha} =1}^{{\bar{k}}} \ket{v_{\alpha}}\bra{v_{\alpha}},
\end{align*}
\begin{align*}
\Pi_B&:= \sum_{{\alpha} =1}^{{\bar{k}}} \ket{w_{\alpha}}\bra{w_{\alpha}}.
\end{align*}
%%\vspace{5pt}

Further, for every ${\alpha} \in [1:{\bar{k}}],$ let 
%%\vspace{5pt}
\begin{align}
\rho_{\alpha} &:= \mathbf{P}_{\alpha} \rho \mathbf{P}_{\alpha},\label{rhoalpha}\\
\sigma_{j,{\alpha}} &:= \mathbf{P}_{\alpha} \sigma_j \mathbf{P}_{\alpha}, \forall j \in {1,\cdots,\abs{\cS}}.
\end{align}
%%\vspace{5pt}
% Thus, from the definitions of $\Pi_1$ and $\Pi_2$, we have the following pair of inequalities for every ${\alpha} \in [1:{\bar{k}}],$
% %%\vspace{5pt}
% \begin{align}
% \label{property1}
% \bra{v_{\alpha}} \sigma_{1,{\alpha}}\ket{v_{\alpha}} &\leq 2^{-{k_1}} \bra{v_{\alpha}}\rho_{\alpha}\ket{v_{\alpha}}, \\
% \label{property2}
% \bra{w_{\alpha}} \sigma_{2,{\alpha}}\ket{w_{\alpha}} &\leq 2^{-{k_2}} \bra{w_{\alpha}}\rho_{\alpha}\ket{w_{\alpha}}, 
% \end{align}
% %%\vspace{5pt}

% For our future discussions it will be useful to make the following observations, for every ${\alpha} \in [1:{\bar{k}}],$ $\rho_{\alpha}\succeq 0$ and 
% $\sum_{{\alpha}=1}^{{\bar{k}}}\tr[\rho_{\alpha}]= 1.$ Thus, $\{\tr[\rho_{\alpha}]\}_{{\alpha}=1}^{{\bar{k}}}$ forms a valid probability distribution over ${\alpha} \in [1:{{\bar{k}}}]$. 
% To construct an intersection projector $\Pi^\star$ that satisfies \eqref{r1}, \eqref{r2}, \eqref{r3} we first make few observations. 
From Jordan's lemma (Fact \ref{jordan}) we have that for every ${\alpha} \in [1:\bar{k}]$ the vectors $\ket{w_{\alpha}}$ and $\ket{v_{\alpha}}$ lie in either a two-dimensional subspace or one-dimensional subspace. 
For the blocks where both $\ket{w_{\alpha}}$ and $\ket{v_{\alpha}}$ lie in a two-dimensional subspace we have the following,  
\begin{equation}
\label{valphatilda_gub}
\ket{v_{\alpha}} = \cos(\theta_{\alpha}^{(t)})\ket{w_{\alpha}} + \sin(\theta_{\alpha}^{(t)})\ket{w^{\perp}_{\alpha}}.
\end{equation}
For the blocks, where both 
$\ket{w_{\alpha}}$ and $\ket{v_{\alpha}}$ lie in one dimensional subspace then in that case $\cos(\theta_{\alpha}^{(t)})=1,$ i.e., $\ket{w_{\alpha}} = \ket{v_{\alpha}}$ for these cases. We now consider the following set
%%\vspace{5pt}
\begin{align}
\G^{(t)}&:= \left\{{\alpha}: \cos^2(\theta_{\alpha}^{(t)}) \geq 1- 8t\eps^{\beta}\right\},\label{good1_gub}
\end{align}
where, $\cos^2(\theta_{\alpha}^{(t)})= {|\langle v_{\alpha}| w_{\alpha}\rangle|}^2$.  
Then, from Lemma \ref{c23_gub}, fixing $\eps_1 =\eps_2 = t\eps^{1 -(t-1)\beta}$ and $\delta = 8t\eps^{\beta}$, the probability of the set $\G^{(t)}$ (defined in \eqref{good1_gub}) under the distribution $\{\tr[\rho_\alpha]\}$ has the following lower bound,
\begin{equation}
    \sum_{{\alpha}\in\mbox{\G}^{(t)}}\tr[\rho_{\alpha}] = \Pr\left\{\G^{(t)}\right\} \geq 1 - \eps^{1 - t\beta}.\label{Good_prob_gub}
\end{equation}
We now construct $\Pi^{\star}$ as follows, 
%%\vspace{5pt}
\begin{equation}
\label{tilda_gub}
\Pi^{\star}:= \sum_{{\alpha} \in \mbox{\G}^{(t)}} \ket{v_{\alpha}}\bra{v_{\alpha}}.
\end{equation}
%%\vspace{5pt} \\
In the following discussions, we will show that $\Pi^\star$ satisfies \eqref{g1} and \eqref{g3}.
%%\vspace{5pt}
% We now show that $\Pi^{\star}$ has properties similar to that of $\cA^\star = \cA_1 \cap \cA_2$ as discussed in the introduction. In particular we show that $\Pi^{\star}$ satisfies \eqref{r1}, \eqref{r2}, \eqref{r3}. 
    \subsubsection*{Proof of \eqref{g1}}
%%\vspace{5pt}
Consider the following set of inequalities,
\begin{align}
\tr[\Pi^{\star}\rho] &\overset{a}=\sum_{{\alpha}\in \mbox{\G}^{(t)}}\tr[\ket{v_{\alpha}}\bra{v_{\alpha}} \rho_{\alpha}]\nonumber\\ 
&= \sum_{{\alpha}=1}^{{\bar{k}}}\tr[\ket{v_{\alpha}}\bra{v_{\alpha}} \rho_{\alpha}] - \sum_{{\alpha}\notin \mbox{\G}^{(t)}}\tr[\ket{v_{\alpha}}\bra{v_{\alpha}} \rho_{\alpha}] \nonumber\\
&\overset{b} \geq 1- t\eps^{1-(t-1)\beta} - \sum_{{\alpha}\notin\mbox{\G}^{(t)}}\tr[\rho_{\alpha}]\nonumber\\
&\overset{c} \geq 1-t\eps^{1-(t-1)\beta} -\eps^{1-t\beta}\nonumber\\
&\geq 1 - (t+1)\eps^{1-t\beta},\hspace{125pt}\label{gub_lowb}
\end{align}
where, $a$ follows from the definition of \eqref{tilda_gub}, $b$ follows from the fact that $\sum_{{\alpha}=1}^{{\bar{k}}}\tr[\ket{v_{\alpha}}\bra{v_{\alpha}} \rho_{\alpha}] = \tr[\Pi_A \rho] \geq 1 - t\eps^{1-(t-1)\beta}$ (where the last inequality follows from \eqref{ind_gub_1}) and $c$ follows from  \eqref{Good_prob_gub}. 
%%\vspace{5pt}
\subsubsection*{Proof of \eqref{g3}}
We provide the proof of \eqref{g3} under two cases. We first consider a state $\sigma_j$, where $j \in \{1,\cdots,2^{t-1}\}$. Then the following set of inequalities holds,
\begin{align}
\tr[\Pi^{\star}\sigma_j] &\overset{a}=
 \sum_{{\alpha}\in \mbox{\G}^{(t)}} \tr[\ket{v_{\alpha}}\bra{v_{\alpha}} \sigma_{j,{\alpha}} ]\nonumber\\
&\overset{b}{\leq}  \sum_{\alpha=1}^{\bar{k}} \tr[\ket{v_{\alpha}}\bra{v_{\alpha}} \sigma_{j,{\alpha}} ]\nonumber\\
&= \tr[\Pi_{A}\sigma_j]\\
&\overset{c}{\leq} 2^{-k_j} + c({t-1})\eps^{\frac{\beta}{2}}\nonumber\\
&{\leq} 2^{-k_j} + c({t})\eps^{\frac{\beta}{2}},\label{gub_case1}
%&\geq  2^{-{k_2}}(1-\eps^{\frac{1}{2}})\sum_{{\alpha} \in \G}\cos(2\theta_{\alpha}^{(t)})\tr[\rho_{\alpha}]\\
% &\overset{d}\lgeq 2^{-{k_1}}
\end{align}
% \begin{align*}
% \tr[\Pi^{\star}\sigma_1] &\overset{a}= \sum_{{\alpha}\in \mbox{\G}}\tr[\ket{v_{\alpha}}\bra{v_{\alpha}} \sigma_{1,{\alpha}}] \\
% &\overset{b} \leq 2^{-{k_1}}\sum_{{\alpha} \in \G}\tr[\ket{v_{\alpha}}\bra{v_{\alpha}} \rho_{\alpha}]\\
% &\overset{c}\leq  2^{-{k_1}}\sum_{{\alpha} \in \G}\tr[\rho_{\alpha}]\\
% %(1-\eps^{\frac{1}{2}})(1-3\eps^{\frac{1}{2}})\nonumber\\
% &\overset{d}\leq 2^{-{k_1}}
% ,
% \end{align*}
where, $a$ follows from the definition of $\Pi^{\star}$ in $\eqref{tilda_gub},$ $b$ follows from the fact that $\G^{(t)} \subseteq \{1,\cdots,\bar{k}\}$, $c$ follows from \eqref{ind_gub_2} and in \eqref{gub_case1}, $c(t)$ is some constant within the range $[0,4\sqrt{2}\sum_{i=1}^{t}\sqrt{i}].$
%\vspace{15pt}
We now consider a state $\sigma_j$, where $j \in \{2^{t-1}+1,\cdots,2^t\}$. Then the following set of inequalities holds,
\begin{align}
    \tr[\Pi^{\star}\sigma_j] &\overset{a}= \sum_{{\alpha}\in \mbox{\G}^{(t)}} \tr[\ket{v_{\alpha}}\bra{v_{\alpha}}\sigma_{j,{\alpha}}] \nonumber\\
&\overset{b}=\sum_{{\alpha} \in \G^{(t)}}\bigg(\cos^2(\theta_{\alpha}^{(t)})\tr[\ket{w_{\alpha}}\bra{w_{\alpha}}\sigma_{j,{\alpha}}] + \sin^2(\theta_{\alpha}^{(t)})\tr[\ket{w^{\perp}_{\alpha}}\bra{w^{\perp}_{\alpha}} \sigma_{j,{\alpha}}] \nonumber\\
&\hspace{60pt}+ \cos(\theta_{\alpha}^{(t)})\sin(\theta_{\alpha}^{(t)})\left(\tr[\ket{w_{\alpha}}\bra{w^{\perp}_{\alpha}} \sigma_{j,{\alpha}}] + \tr[\ket{w^{\perp}_{\alpha}}\bra{w_{\alpha}} \sigma_{j,{\alpha}}]\right) \bigg)\hspace{150pt} \nonumber\\
&\overset{c} \leq \sum_{{\alpha} \in \G^{(t)}}\bigg(\tr[\ket{w_{\alpha}}\bra{w_{\alpha}} \sigma_{j,{\alpha}}] +  \sin^2(\theta_{\alpha}^{(t)})\tr[\ket{w^{\perp}_{\alpha}}\bra{w^{\perp}_{\alpha}} \sigma_{j,{\alpha}}] \nonumber\\
&\hspace{60pt}+ \cos(\theta_{\alpha}^{(t)})\sin(\theta_{\alpha}^{(t)})\left(\tr[\ket{w_{\alpha}}\bra{w^{\perp}_{\alpha}} \sigma_{j,{\alpha}}] + \tr[\ket{w^{\perp}_{\alpha}}\bra{w_{\alpha}} \sigma_{j,{\alpha}}]\right) \bigg) \nonumber\\
% &\overset{e}{\leq}\tr[\Pi_2\sigma_2] + \sum_{{\alpha} \in \G}\bigg(2\sqrt{2}\eps^{\frac{1}{4}}\tr[\ket{w^{\perp}_{\alpha}}\bra{w^{\perp}_{\alpha}} \sigma_{2,{\alpha}}] \nonumber\\
% &\hspace{60pt}+ 2\sqrt{2}\eps^{\frac{1}{4}}\left(\tr[\ket{w_{\alpha}}\bra{w^{\perp}_{\alpha}} \sigma_{2,{\alpha}}] + \tr[\ket{w^{\perp}_{\alpha}}\bra{w_{\alpha}} \sigma_{2,{\alpha}}]\right) \bigg) \nonumber\\
&\overset{d}{\leq}\tr[\Pi_B\sigma_j] + \sum_{{\alpha} \in \G^{(t)}}\bigg(8t\eps^{\beta}\tr[\ket{w^{\perp}_{\alpha}}\bra{w^{\perp}_{\alpha}} \sigma_{j,{\alpha}}] \nonumber\\
&\hspace{60pt}+ \cos(\theta_{\alpha}^{(t)})\sin(\theta_{\alpha}^{(t)})\left(\tr[\ket{w_{\alpha}}\bra{w^{\perp}_{\alpha}} \sigma_{j,{\alpha}}] + \tr[\ket{w^{\perp}_{\alpha}}\bra{w_{\alpha}} \sigma_{j,{\alpha}}]\right) \bigg) \nonumber\hspace{65pt}\\
&\overset{e}{\leq}\tr[\Pi_B\sigma_j] + \sum_{{\alpha} \in \G^{(t)}}\bigg(8t\eps^{\beta}\tr[\ket{w^{\perp}_{\alpha}}\bra{w^{\perp}_{\alpha}} \sigma_{j,{\alpha}}] \nonumber + 2\sqrt{2}\sqrt{t}\eps^{\frac{\beta}{2}}\left(\tr[\sigma_{j,{\alpha}}] \right) \bigg) \nonumber\\
&\overset{f}{\leq} 2^{-{k_j}} + c({t-1})\eps^{\frac{\beta}{2}} + \sum_{{\alpha} \in \G^{(t)}}\bigg(8t\eps^{\beta}\tr[\sigma_{j,{\alpha}}]  + 2\sqrt{2}\sqrt{t}\eps^{\frac{\beta}{2}}\tr[\sigma_{j,{\alpha}}] \bigg) \nonumber\\
&\leq 2^{-{k_j}}+ c({t-1})\eps^{\frac{\beta}{2}} + 2\sqrt{2}\sqrt{t}\eps^{\frac{\beta}{2}} + 8t\eps^{\beta}\hspace{225pt}\nonumber\\
&\overset{g}{\leq} 2^{-{k_j}}+ c({t-1})\eps^{\frac{\beta}{2}} + 4\sqrt{2}\sqrt{t}\eps^{\frac{\beta}{2}}\nonumber\\
&\leq 2^{-{k_j}}+ c({t})\eps^{\frac{\beta}{2}},\label{gub_case2}
\end{align}
% &\overset{h}\geq2^{-{k_2}}(1-\eps^{\frac{1}{2}})(1-16\eps^{\frac{1}{2}}) \sum_{{\alpha} \in \G}\left(\tr[\rho_{\alpha}]\right)\nonumber\\
% \label{mrf}
% %&\geq  2^{-{k_2}}(1-\eps^{\frac{1}{2}})\sum_{{\alpha} \in \G}\cos(2\theta_{\alpha}^{(t)})\tr[\rho_{\alpha}]\\
% &\overset{i}\geq 2^{-{k_2}}(1-\eps^{\frac{1}{2}})(1-16\eps^{\frac{1}{2}})(1-3\eps^{\frac{1}{2}})\nonumber\\
% &\geq 2^{-{k_2}}(1-20\eps^{\frac{1}{2}}),
% \\
where, $a$ follows from the definition of $\Pi^{\star}$ (see \eqref{tilda_gub}), $b$ follows from \eqref{valphatilda_gub}, $c$ follows because $\cos^2({\theta_{\alpha}^{(t)}}) \leq 1$, $d$ follows from the fact that from \eqref{good1_gub}, for all $\alpha \in \G^{(t)}$, $\sin^{2}(\theta_{\alpha}^{(t)}) \leq 8t\eps^{\beta}$ , $e$ follows from the following series of inequalities:
\begin{align}
   &\cos(\theta_{\alpha}^{(t)})\sin(\theta_{\alpha}^{(t)})\left(\tr[\ket{w_{\alpha}}\bra{w^{\perp}_{\alpha}} \sigma_{j,{\alpha}}] + \tr[\ket{w^{\perp}_{\alpha}}\bra{w_{\alpha}} \sigma_{j,{\alpha}}]\right)\nonumber\\
   &\leq\abs{\cos(\theta_{\alpha}^{(t)})\sin(\theta_{\alpha}^{(t)})\left(\tr[\ket{w_{\alpha}}\bra{w^{\perp}_{\alpha}} \sigma_{j,{\alpha}}] + \tr[\ket{w^{\perp}_{\alpha}}\bra{w_{\alpha}} \sigma_{j,{\alpha}}]\right)}\nonumber\\
   &= \abs{\cos(\theta_{\alpha}^{(t)})\sin(\theta_{\alpha}^{(t)})}\abs{\left(\tr[\ket{w_{\alpha}}\bra{w^{\perp}_{\alpha}} \sigma_{j,{\alpha}}] + \tr[\ket{w^{\perp}_{\alpha}}\bra{w_{\alpha}} \sigma_{j,{\alpha}}]\right)}\nonumber\\
   &= \abs{\cos(\theta_{\alpha}^{(t)})}\abs{\sin(\theta_{\alpha}^{(t)})}\abs{\left(\tr[\ket{w_{\alpha}}\bra{w^{\perp}_{\alpha}} \sigma_{j,{\alpha}}] + \tr[\ket{w^{\perp}_{\alpha}}\bra{w_{\alpha}} \sigma_{j,{\alpha}}]\right)}\nonumber\\
   &\leq 2\sqrt{2}\sqrt{t}\eps^{\frac{\beta}{2}} \tr[\sigma_{j,{\alpha}}],\label{extra_term_bound}
\end{align}
where, the last inequality follows from the facts that for all $\alpha \in \G^{(t)}$, $|\sin(\theta_{\alpha}^{(t)})| \leq 2\sqrt{2}\sqrt{t}\eps^{\frac{\beta}{2}}, \abs{\cos(\theta_{\alpha}^{(t)})} \leq 1$ and from \eqref{positivity_res_1} and \eqref{positivity_res_2} of Lemma \ref{positivity} since $\sigma_{j,{\alpha}} \succeq 0$. Inequality $f$ follows from \eqref{ind_gub_4} (see the statement of Lemma \ref{generalubound_seq}) and $g$ follows since $8t\eps^{\beta} \in (0,1]$ is arbitrarily close to zero, $8t\eps^{\beta} \leq \sqrt{8t\eps^{\beta}} = 2\sqrt{2}\sqrt{t}\eps^{\frac{\beta}{2}}$.

% We again construct intersection projectors $\Pi_{\{1,2,3,4\}},\Pi_{\{5,6,7,8\}},\cdots,\Pi_{\{2^{t}-3,2^{t}-2, 2^{t}-1, 2^{t}\}}$ corresponding to the pairs $\left\{\Pi_{\{1,2\}},\Pi_{\{3,4\}}\right\},\left\{\Pi_{\{5,6\}},\Pi_{\{7,8\}}\right\},\cdots,\left\{\Pi_{\{2^{t}-3,2^{t}-2\}}, \Pi_{\{2^{t}-1,2^{t}\}}\right\}$ respectively and thus $\forall i \in \{1,2,\cdots$\quad$,2^{t-2}\}$, the intersection projector $\Pi_{\{4i-3,\cdots,4i\}}$ has the following properties,
% \begin{align*}
%     \tr[\Pi_{\{4i-3,\cdots,4i\}}\rho] &\geq 1 - 2^2\eps^{(1 - 2.\beta)},\\
%     % \tr[\Pi_{\{2i-1,2i\}}\sigma^{\otimes n}_] &\leq \tr[\Pi_1 \sigma^{\otimes n}_1]\\
%     % &\leq 2^{-nD(\rho || \sigma_1)}\\
%     \tr[\Pi_{\{4i-3,\cdots,4i\}}\sigma_{j}] 
%     &\leq 2^{-k_{j}} + c_2.\eps^{\frac{\beta}{2}}, \forall j \in \{4i-3,\cdots,4i\}.
% \end{align*}

% where, $c_2 \in [0,8\sqrt{2}]$.
Combinining \cref{gub_lowb,gub_case1,gub_case2}, we have the following
\begin{align*}
    \tr[\Pi^{\star}\rho] &\geq 1 - (t+1)\eps^{(1 - t\beta)},\\
    % \tr[\Pi_{\{2i-1,2i\}}\sigma^{\otimes n}_] &\leq \tr[\Pi_1 \sigma^{\otimes n}_1]\\
    % &\leq 2^{-nD(\rho || \sigma_1)}\\
    \tr[\Pi^{\star}\sigma_{j}] 
    &\leq 2^{-k_{j}} + c(t)\eps^{\frac{\beta}{2}}, \forall j \in \{1,2,\cdots,2^{t}\}.
\end{align*}

 This completes the proof of Lemma \ref{generalubound} using a parallel construction of $\Pi^{\star}$.\hfill\qed 
 
 The following corollary follows straightforwardly from Lemma \ref{generalubound}. This corollary helps us in the error analysis of \ref{s1} of our decoding strategy for the authentication channel problem.
\begin{corollary}\label{decode_s0}
    Assuming the problem setup mentioned in Section \ref{authenticated_section}, consider a collection of quantum states $\{\rho_{X^n}^{(s)}\}_{s \in \cS}$ over a Hilbert space $\cH_{\cB}^{\otimes n}$ such that $\forall s \in \cS$, $\rho_{X^n}^{(s)} := \bigotimes_{i=1}^{n} \rho_{X_i}^{(s)}.$
    Further, also consider a collection of projectors $\{\Pi^{(s)}_{(n)}\}_{s \in \cS\setminus\{s_0\}}$ such that $\forall s \in \cS\setminus\{s_0\}$, for $\delta \in (0,1)$ and for $n$ large enough,
    \begin{align*}
    \Pi^{(s)}_{(n)} := \bigg\{ {\rho^{(s)}}^{\otimes n} &\preceq  2^{-n(D(\rho^{(s_0)} || \rho^{(s)}) - \delta)}{\rho^{(s_0)}}^{\otimes n} \bigg\}, \\
\tr\left[\Pi^{(s)}_{(n)} {\rho^{(s_0)}}^{\otimes n}\right] &= \mathbb{E}_{X^n}\left[\tr\left[\Pi^{(s)}_{(n)}\rho_{X^n}^{(s_0)}\right]\right] \geq 1 - \eps,\\
\tr\left[\Pi^{(s)}_{(n)} {\rho^{(s)}}^{\otimes n}\right] &= \mathbb{E}_{X^n}\left[\tr\left[\Pi^{(s)}_{(n)}\rho_{X^n}^{(s)}\right]\right] \leq 2^{-n(D(\rho^{(s_0)} || \rho^{(s)}) - \delta)},
    \end{align*}
    where, $\forall s \in \cS, \rho^{(s)} := \mathbb{E}_{X}\left[\rho_{X}^{(s)}\right] $. The existence of such $\{\Pi^{(s)}_{(n)}\}_{s \in \cS\setminus\{s_0\}}$ is guaranteed by  \eqref{F42} and \eqref{DM2} of Fact \ref{F4}. Then, we can construct a projector $\Pi_{(n)}^{\neg{\perp}}$ which satisfies the following inequalities:
    \begin{align}
        % \mathbb{E}_{X^n}\left[P_{Y^n|X^n}^{(s_0)}(D^{(s_0)})\right] &\geq 1 - (\ceil{\log|\cS|} + 1)\eps^{1 - \ceil{\log|\cS|}.\beta},\\
        \mathbb{E}_{X^n}\left[\tr\left[\Pi_{(n)}^{\neg{\perp}}\rho_{X^n}^{(s_0)}\right]\right] &\geq 1 -\left(\ceil{\log{(|\cS| -1)}}+1\right)\eps^{\left({1 - \ceil{\log(|\cS|-1)}}\beta\right)},\label{decode_s0_eq1}\\
        % \mathbb{E}_{X^n}\left[P_{Y^n|X^n}^{(s)}(D^{(s_0)})\right] &\leq 
        % 2^{-n(D(P_{Y^n}^{(s_0)}||P_{Y^n}^{(s)}) + \delta)}+O(\eps^{\frac{1}{4}}), \quad \forall s \in \cS
        \mathbb{E}_{X^n}\left[\tr\left[\Pi_{(n)}^{\neg{\perp}}\rho_{X^n}^{(s)}\right]\right] &\leq 
        2^{-n\left(\underset{\substack{s \in \cS: s \neq s_0}}{\min}D(\rho^{(s_0)} || \rho^{(s)}) - \delta\right)} + O\left(\left(\ceil{\log(|\cS|-1)}\right)^{\frac{3}{2}}\right)\eps^{\frac{\beta}{2}},  \forall s \in \cS\setminus\{s_0\},\label{decode_s0_eq2}
    \end{align}
    where, $\beta$ is such that $\left(\ceil{\log{(|\cS| -1)}}+1\right)\eps^{\left({1 - \ceil{\log(|\cS|-1)}}\beta\right)} \ll 1$
\end{corollary}
\begin{proof}
    See subsection \ref{corr_decode_s0_proof} in Appendix.
\end{proof}

 \subsection{Proof of Lemma \ref{generalubound_seq}}\label{proof_generalubound_seq}
We will prove Lemma \ref{generalubound_seq} using `proof by induction'. We start with the collection of $|\cS|= 2^{t}$ projectors ${\{\Pi_i\}}_{i\in\{1,2,\cdots,|\cS|\}}$. We then, consider the pair of projectors $\left\{\Pi_{1},\Pi_{2}\right\}$.  Thus, we have the following properties for the pair $\left\{\Pi_{1},\Pi_{2}\right\}$,
\begin{align*}
    &\tr[\Pi_{1}\rho] \geq 1- \eps,\\
     &\tr[\Pi_{1}\sigma_{1}] \leq 2^{-k_{1}},\\
      &\tr[\Pi_{2}\rho] \geq 1- \eps,\\
     &\tr[\Pi_{2}\sigma_{2}] \leq 2^{-k_{2}}.
\end{align*}
By replacing $\Pi_1$ as $\Pi_2$ and $\Pi_2$ as $\Pi_1$ in the proof of Lemma \ref{lbound} and replacing the set $\G$ defined in \eqref{good1} with the set $\G^{(1)}$ defined in \eqref{goodstep1}, we can construct an intersection projector $\Pi_{\{1,2\}}$ corresponding to $\left\{\Pi_{1},\Pi_{2}\right\}$, which satisfies the following properties,
\begin{align*}
    \tr[\Pi_{\{1,2\}}\rho] &\geq 1 - 2\eps^{(1 - \beta)}, \\
    % \tr[\Pi_{\{2i-1,2i\}}\sigma^{\otimes n}_] &\leq \tr[\Pi_1 \sigma^{\otimes n}_1]\\
    % &\leq 2^{-nD(\rho || \sigma_1)}\\
    \tr[\Pi_{\{1,2\}}\sigma_{1}] 
    &\leq 2^{-k_{1}} + 4\sqrt{2}.\eps^{\frac{\beta}{2}}\\
    \tr[\Pi_{\{1,2\}}\sigma_{2}] 
    &\leq 2^{-k_{2}}.
\end{align*}
We can combine the last two of the above equations as follows,
\begin{align*}
    \tr[\Pi_{\{1,2\}}\sigma_{j}] 
    &\leq 2^{-k_{j}} + d(j,2)\eps^{\frac{\beta}{2}}, \forall j \in \{1,2\}, 
\end{align*}
where, $d(1,2) = 4\sqrt{2}$ and $d(2,2) = 0$. This gives the proof for the base case. In the next step we again construct intersection projectors $\Pi_{\{1,2,3\}}$ corresponding to the pair $\left\{\Pi_{\{1,2\}},\Pi_{3}\right\}$. We keep doing the same for $\abs{\cS}-2 = (2^t-2)$ times and at each $k$-th step we consider $\G^{(k)}$ as defined in \eqref{goodi} for the corresponding pair of projectors.

Finally, we are left with the pair $\left\{\Pi_{\{1,\cdots,2^{t}-1\}},\Pi_{2^{t}}\right\}$. We assume that for ($2^t-2$)-th step the statement of Lemma \ref{generalubound_seq} holds i.e. $\Pi_{\{1,\cdots,2^{t}-1\}}$ satisfies the following:
\begin{align}
    \tr[\Pi_{\{1,\cdots,2^{t}-1\}}\rho] &\geq 1 - 2\eps^{1-(2^t-2)\beta},\label{ind_gub_1seq}\\
    \tr[\Pi_{\{1,\cdots,2^{t}-1\}}\sigma_u] &\leq 2^{-k_u} + d(u,2^t-1)\eps^{\frac{\beta}{2}}, \forall u \in \{1,\cdots,2^t-1\}\label{ind_gub_2seq},
\end{align}
where, $\forall u \in [2^t-2], d(u,2^t-1) = 4\sqrt{2}\sum_{j=u}^{2^t-2}\sqrt{j}$ and $d(2^t-1,2^t-1) =0$. For simplicity of notations, we denote $\Pi_{2^{t}}$ and $\Pi_{\{1,\cdots,2^{t}-1\}}$ as $\Pi_{A}$ and $\Pi_{B}$. We will apply Fact \ref{jordan} on the pair $\{\Pi_A, \Pi_B\}.$ Consider $\left\{\mathbf{P}_{\alpha}\right\}_{{\alpha}=1}^{{\bar{k}}}$ (where $\bar{k}$ is some natural number depending on $\Pi_A,\Pi_B$) as the set of orthogonal projectors obtained by Fact \ref{jordan} applied with respect to the pair $\{\Pi_A, \Pi_B\}.$ Thus, $\sum_{{\alpha} =1}^{{\bar{k}}}\mathbf{P}_{\alpha} = \mathbb{I}.$ Furthermore, for every ${\alpha} \in [1:{\bar{k}}]$ we define  $\Pi_{A,{\alpha}} := \mathbf{P}_{\alpha}\Pi_A\mathbf{P}_{\alpha}$ as the following one dimensional projector: 
%%\vspace{5pt}
\begin{align}
\Pi_{A,{\alpha}} := \ket{v_{\alpha}}\bra{v_{\alpha}}\nonumber,
\end{align}
for some $\ket{v_{\alpha}}$ in the range of $\mathbf{P}_{\alpha}.$ Similarly, for every ${\alpha} \in [1:{\bar{k}}]$, we have
\begin{align}
\Pi_{B,{\alpha}} := \ket{w_{\alpha}}\bra{w_{\alpha}}\nonumber,
\end{align}
%%\vspace{5pt}
for some $\ket{w_{\alpha}}$ in the range of $\mathbf{P}_{\alpha}.$
Thus, it now follows from the property of $\{\ket{v_{\alpha}}\}_{{\alpha}=1}^{\bar{k}}$ and $\{\ket{w_{\alpha}}\}_{{\alpha}=1}^{\bar{k}}$ that 
%%\vspace{5pt}
\begin{align*}
\Pi_A &:= \sum_{{\alpha} =1}^{{\bar{k}}} \ket{v_{\alpha}}\bra{v_{\alpha}},
\end{align*}
\begin{align*}
\Pi_B&:= \sum_{{\alpha} =1}^{{\bar{k}}} \ket{w_{\alpha}}\bra{w_{\alpha}}.
\end{align*}
%%\vspace{5pt}

Further, for every ${\alpha} \in [1:{\bar{k}}],$ let 
%%\vspace{5pt}
\begin{align}
\rho_{\alpha} &:= \mathbf{P}_{\alpha} \rho \mathbf{P}_{\alpha},\label{rhoalphaseq}\\
\sigma_{j,{\alpha}} &:= \mathbf{P}_{\alpha} \sigma_j \mathbf{P}_{\alpha}, \forall j \in {1,\cdots,\abs{\cS}}.
\end{align}
%%\vspace{5pt}
% Thus, from the definitions of $\Pi_1$ and $\Pi_2$, we have the following pair of inequalities for every ${\alpha} \in [1:{\bar{k}}],$
% %%\vspace{5pt}
% \begin{align}
% \label{property1}
% \bra{v_{\alpha}} \sigma_{1,{\alpha}}\ket{v_{\alpha}} &\leq 2^{-{k_1}} \bra{v_{\alpha}}\rho_{\alpha}\ket{v_{\alpha}}, \\
% \label{property2}
% \bra{w_{\alpha}} \sigma_{2,{\alpha}}\ket{w_{\alpha}} &\leq 2^{-{k_2}} \bra{w_{\alpha}}\rho_{\alpha}\ket{w_{\alpha}}, 
% \end{align}
% %%\vspace{5pt}

% For our future discussions it will be useful to make the following observations, for every ${\alpha} \in [1:{\bar{k}}],$ $\rho_{\alpha}\succeq 0$ and 
% $\sum_{{\alpha}=1}^{{\bar{k}}}\tr[\rho_{\alpha}]= 1.$ Thus, $\{\tr[\rho_{\alpha}]\}_{{\alpha}=1}^{{\bar{k}}}$ forms a valid probability distribution over ${\alpha} \in [1:{{\bar{k}}}]$. 
% To construct an intersection projector $\Pi^\star$ that satisfies \eqref{r1}, \eqref{r2}, \eqref{r3} we first make few observations. 
From Jordan's lemma (Fact \ref{jordan}) we have that for every ${\alpha} \in [1:\bar{k}]$ the vectors $\ket{w_{\alpha}}$ and $\ket{v_{\alpha}}$ lie in either a two-dimensional subspace or one-dimensional subspace. 
For the blocks where both $\ket{w_{\alpha}}$ and $\ket{v_{\alpha}}$ lie in a two-dimensional subspace we have the following,  
\begin{equation}
\label{valphatilda_gub_seq}
\ket{v_{\alpha}} = \cos(\theta^{(2^t-1)}_{\alpha})\ket{w_{\alpha}} + \sin(\theta^{(2^t-1)}_{\alpha})\ket{w^{\perp}_{\alpha}}.
\end{equation}

For the blocks, where both 
$\ket{w_{\alpha}}$ and $\ket{v_{\alpha}}$ lie in one dimensional subspace then in that case $\cos(\theta^{(2^t-1)}_{\alpha})=1,$ i.e., $\ket{w_{\alpha}} = \ket{v_{\alpha}}$ for these cases. Now we consider the following set
%%\vspace{5pt}
\begin{align}
\G^{(2^t-1)}&:= \left\{{\alpha}: \cos^2(\theta^{(2^t-1)}_{\alpha}) \geq 1- 8(2^t-1)\eps^{\beta}\right\},\label{good1_gub_seq}
\end{align}
 where, $\cos^2(\theta^{(2^t-1)}_{\alpha})= {|\langle v_{\alpha}| w_{\alpha}\rangle|}^2$. 
Then, from Lemma \ref{c23_gub}, fixing $\eps_1 = \eps, \eps_2 = 2\eps^{1 -(2^t-2)\beta}$ and $\delta = 8(2^t-1)\eps^{\beta}$, the probability of the set $\G^{(2^t-1)}$ (defined in \eqref{good1_gub_seq}) under the distribution $\{\tr[\rho_\alpha]\}$ has the following lowerbound,

\begin{align}
    \sum_{{\alpha}\in\mbox{\G}^{(2^t-1)}}\tr[\rho_{\alpha}] = \Pr\left\{\G^{(2^t-1)}\right\} &\geq 1 - \frac{4(\eps + 2\eps^{1 -(2^t-2)\beta})}{8(2^t-1)\eps^{\beta}}\nonumber\\
    &\geq 1 - \frac{4(3\eps^{1 -(2^t-2)\beta})}{8(2^t-1)\eps^{\beta}}\nonumber\\
    &\geq 1 - \frac{3\eps^{1 -(2^t-1)\beta}}{2(2^t-1)}\nonumber\\
    &\geq 1 - \eps^{1 -(2^t-1)\beta}.
    \label{Good_prob_gub_seq}
\end{align}

We now construct $\Pi^{\star}$ as follows, 
%%\vspace{5pt}
\begin{equation}
\label{tilda_gub_seq}
\Pi^{\star}:= \sum_{{\alpha} \in \mbox{\G}^{(t)}} \ket{v_{\alpha}}\bra{v_{\alpha}}.
\end{equation}
%%\vspace{5pt} \\

In the following discussion, we will show that $\Pi^\star$ satisfies \eqref{g1seq} and \eqref{g3seq}.
%%\vspace{5pt}
% We now show that $\Pi^{\star}$ has properties similar to that of $\cA^\star = \cA_1 \cap \cA_2$ as discussed in the introduction. In particular we show that $\Pi^{\star}$ satisfies \eqref{r1}, \eqref{r2}, \eqref{r3}. 
    \subsubsection*{Proof of \eqref{g1seq}}
%%\vspace{5pt}
Consider the following set of inequalities,

\begin{align}
\tr[\Pi^{\star}\rho] &\overset{a}=\sum_{{\alpha}\in \mbox{\G}^{(2^t-1)}}\tr[\ket{v_{\alpha}}\bra{v_{\alpha}} \rho_{\alpha}]\nonumber\\ 
&= \sum_{{\alpha}=1}^{{\bar{k}}}\tr[\ket{v_{\alpha}}\bra{v_{\alpha}} \rho_{\alpha}] - \sum_{{\alpha}\notin \mbox{\G}^{(2^t-1)}}\tr[\ket{v_{\alpha}}\bra{v_{\alpha}} \rho_{\alpha}] \nonumber\\
&\overset{b} \geq 1- \eps - \sum_{{\alpha}\notin\mbox{\G}^{(2^t-1)}}\tr[\rho_{\alpha}]\nonumber\\
&\overset{c} \geq 1-\eps -\eps^{1-(2^t-1)\beta}\geq 1 - 2\eps^{1-(2^t-1)\beta},\label{gub_lowb_seq}
\end{align}
where, $a$ follows from the definition of \eqref{tilda_gub}, $b$ follows from the fact that $\sum_{{\alpha}=1}^{{\bar{k}}}\tr[\ket{v_{\alpha}}\bra{v_{\alpha}} \rho_{\alpha}] = \tr[\Pi_A \rho] = \tr[\Pi_{2^t} \rho] \geq 1 - \eps$ (where the last inequality follows from the definition of $\Pi_{2^t}$ mentioned in the statement of Lemma \ref{generalubound_seq}) and $c$ follows from  \eqref{Good_prob_gub_seq}. 
%%\vspace{5pt}
\subsubsection*{Proof of \eqref{g3seq}}
We provide the proof of \eqref{g3seq} under two cases. We first consider the state $\sigma_{2^t}$. Then the following set of inequalities holds,
\begin{align}
\tr[\Pi^{\star}\sigma_{2^t}] &\overset{a}=
 \sum_{{\alpha}\in \mbox{\G}^{(2^t-1)}} \tr[\ket{v_{\alpha}}\bra{v_{\alpha}} \sigma_{2^t,{\alpha}} ]\nonumber\\
&\overset{b}{\leq}  \sum_{\alpha=1}^{\bar{k}} \tr[\ket{v_{\alpha}}\bra{v_{\alpha}} \sigma_{2^t,{\alpha}} ]\nonumber\\
&= \tr[\Pi_{A}\sigma_{2^t}]\nonumber\\
&= \tr[\Pi_{2^t}\sigma_{2^t}]\nonumber\\
&\overset{c}{\leq} 2^{-k_{2^t}},\label{gub_case1_seq}
%&\geq  2^{-{k_2}}(1-\eps^{\frac{1}{2}})\sum_{{\alpha} \in \G}\cos(2\theta^{(2^t-1)}_{\alpha})\tr[\rho_{\alpha}]\\
% &\overset{d}\lgeq 2^{-{k_1}}
\end{align}
% \begin{align*}
% \tr[\Pi^{\star}\sigma_1] &\overset{a}= \sum_{{\alpha}\in \mbox{\G}}\tr[\ket{v_{\alpha}}\bra{v_{\alpha}} \sigma_{1,{\alpha}}] \\
% &\overset{b} \leq 2^{-{k_1}}\sum_{{\alpha} \in \G}\tr[\ket{v_{\alpha}}\bra{v_{\alpha}} \rho_{\alpha}]\\
% &\overset{c}\leq  2^{-{k_1}}\sum_{{\alpha} \in \G}\tr[\rho_{\alpha}]\\
% %(1-\eps^{\frac{1}{2}})(1-3\eps^{\frac{1}{2}})\nonumber\\
% &\overset{d}\leq 2^{-{k_1}}
% ,
% \end{align*}
where, $a$ follows from the definition of of $\Pi^{\star}$ in $\eqref{tilda_gub_seq},$ $b$ follows from the fact that $\G^{(2^t-1)} \subseteq \{1,\cdots,\bar{k}\}$ and $c$ follows from the the definition of $\Pi_{2^t}$ mentioned in the statement of Lemma \ref{generalubound_seq}.
%\vspace{15pt}

We now consider a state $\sigma_j$, where $j \in \{1,\cdots,2^t-1\}$. Then the following set of inequalities holds,
\begin{align}
    \tr[\Pi^{\star}\sigma_j] &\overset{a}= \sum_{{\alpha}\in \mbox{\G}^{(2^t-1)}} \tr[\ket{v_{\alpha}}\bra{v_{\alpha}}\sigma_{j,{\alpha}}] \nonumber\\
&\overset{b}=\sum_{{\alpha} \in \G^{(2^t-1)}}\bigg(\cos^2(\theta^{(2^t-1)}_{\alpha})\tr[\ket{w_{\alpha}}\bra{w_{\alpha}}\sigma_{j,{\alpha}}] + \sin^2(\theta^{(2^t-1)}_{\alpha})\tr[\ket{w^{\perp}_{\alpha}}\bra{w^{\perp}_{\alpha}} \sigma_{j,{\alpha}}] \nonumber\\
&\hspace{60pt}+ \cos(\theta^{(2^t-1)}_{\alpha})\sin(\theta^{(2^t-1)}_{\alpha})\left(\tr[\ket{w_{\alpha}}\bra{w^{\perp}_{\alpha}} \sigma_{j,{\alpha}}] + \tr[\ket{w^{\perp}_{\alpha}}\bra{w_{\alpha}} \sigma_{j,{\alpha}}]\right) \bigg)\hspace{120pt} \nonumber\hspace{5pt}\\
&\overset{c} \leq \sum_{{\alpha} \in \G^{(2^t-1)}}\bigg(\tr[\ket{w_{\alpha}}\bra{w_{\alpha}} \sigma_{j,{\alpha}}] +  \sin^2(\theta^{(2^t-1)}_{\alpha})\tr[\ket{w^{\perp}_{\alpha}}\bra{w^{\perp}_{\alpha}} \sigma_{j,{\alpha}}] \nonumber\\
&\hspace{60pt}+ \cos(\theta^{(2^t-1)}_{\alpha})\sin(\theta^{(2^t-1)}_{\alpha})\left(\tr[\ket{w_{\alpha}}\bra{w^{\perp}_{\alpha}} \sigma_{j,{\alpha}}] + \tr[\ket{w^{\perp}_{\alpha}}\bra{w_{\alpha}} \sigma_{j,{\alpha}}]\right) \bigg)\nonumber
\end{align}
\begin{align}
% &\overset{e}{\leq}\tr[\Pi_2\sigma_2] + \sum_{{\alpha} \in \G}\bigg(2\sqrt{2}\eps^{\frac{1}{4}}\tr[\ket{w^{\perp}_{\alpha}}\bra{w^{\perp}_{\alpha}} \sigma_{2,{\alpha}}] \nonumber\\
% &\hspace{60pt}+ 2\sqrt{2}\eps^{\frac{1}{4}}\left(\tr[\ket{w_{\alpha}}\bra{w^{\perp}_{\alpha}} \sigma_{2,{\alpha}}] + \tr[\ket{w^{\perp}_{\alpha}}\bra{w_{\alpha}} \sigma_{2,{\alpha}}]\right) \bigg) \nonumber\\
&\overset{d}{\leq}\tr[\Pi_B\sigma_j] + \sum_{{\alpha} \in \G^{(2^t-1)}}\bigg(8(2^t-1)\eps^{\beta}\tr[\ket{w^{\perp}_{\alpha}}\bra{w^{\perp}_{\alpha}} \sigma_{j,{\alpha}}] \nonumber\\
&\hspace{60pt}+ \cos(\theta^{(2^t-1)}_{\alpha})\sin(\theta^{(2^t-1)}_{\alpha})\left(\tr[\ket{w_{\alpha}}\bra{w^{\perp}_{\alpha}} \sigma_{j,{\alpha}}] + \tr[\ket{w^{\perp}_{\alpha}}\bra{w_{\alpha}} \sigma_{j,{\alpha}}]\right) \bigg) \nonumber\\
&\overset{e}{\leq}\tr[\Pi_B\sigma_j] + \sum_{{\alpha} \in \G^{(2^t-1)}}\bigg(8(2^t-1)\eps^{\beta}\tr[\ket{w^{\perp}_{\alpha}}\bra{w^{\perp}_{\alpha}} \sigma_{j,{\alpha}}] \nonumber + 2\sqrt{2}\sqrt{2^t-1}\eps^{\frac{\beta}{2}}\left(\tr[\sigma_{j,{\alpha}}] \right) \bigg) \nonumber\\
&\overset{f}{\leq} 2^{-{k_j}} + d(j,2^t-1)\eps^{\frac{\beta}{2}} + \sum_{{\alpha} \in \G^{(2^t-1)}}\bigg(8(2^t-1)\eps^{\beta}\tr[\sigma_{j,{\alpha}}]  + 2\sqrt{2}\sqrt{2^t-1}\eps^{\frac{\beta}{2}}\tr[\sigma_{j,{\alpha}}] \bigg) \nonumber\\
&\leq 2^{-{k_j}}+ d(j,2^t-1)\eps^{\frac{\beta}{2}} + 2\sqrt{2}\sqrt{2^t-1}\eps^{\frac{\beta}{2}} + 8(2^t-1)\eps^{\beta}\nonumber\\
&\overset{g}{\leq} 2^{-{k_j}}+ d(j,2^t-1)\eps^{\frac{\beta}{2}} + 4\sqrt{2}\sqrt{2^t-1}\eps^{\frac{\beta}{2}}\nonumber\\
&\leq 2^{-{k_j}}+ d(j,2^t)\eps^{\frac{\beta}{2}},\label{gub_case2_seq}
\end{align}
% &\overset{h}\geq2^{-{k_2}}(1-\eps^{\frac{1}{2}})(1-16\eps^{\frac{1}{2}}) \sum_{{\alpha} \in \G}\left(\tr[\rho_{\alpha}]\right)\nonumber\\
% \label{mrf}
% %&\geq  2^{-{k_2}}(1-\eps^{\frac{1}{2}})\sum_{{\alpha} \in \G}\cos(2\theta^{(2^t-1)}_{\alpha})\tr[\rho_{\alpha}]\\
% &\overset{i}\geq 2^{-{k_2}}(1-\eps^{\frac{1}{2}})(1-16\eps^{\frac{1}{2}})(1-3\eps^{\frac{1}{2}})\nonumber\\
% &\geq 2^{-{k_2}}(1-20\eps^{\frac{1}{2}}),
where, $a$ follows from the definition of $\Pi^{\star}$ (see \eqref{tilda_gub_seq}), $b$ follows from \eqref{valphatilda_gub_seq}, $c$ follows because $\cos^2({\theta^{(2^t-1)}_{\alpha}}) \leq 1$, $d$ follows from the fact that from \eqref{good1_gub_seq}, for all $\alpha \in \G^{(2^t-1)}$, $\sin^{2}(\theta^{(2^t-1)}_{\alpha}) \leq 8(2^t-1)\eps^{\beta}$ , $e$ follows from the following series of inequalities:
\begin{align}
   &\cos(\theta^{(2^t-1)}_{\alpha})\sin(\theta^{(2^t-1)}_{\alpha})\left(\tr[\ket{w_{\alpha}}\bra{w^{\perp}_{\alpha}} \sigma_{j,{\alpha}}] + \tr[\ket{w^{\perp}_{\alpha}}\bra{w_{\alpha}} \sigma_{j,{\alpha}}]\right)\nonumber\\
   &\leq\abs{\cos(\theta^{(2^t-1)}_{\alpha})\sin(\theta^{(2^t-1)}_{\alpha})\left(\tr[\ket{w_{\alpha}}\bra{w^{\perp}_{\alpha}} \sigma_{j,{\alpha}}] + \tr[\ket{w^{\perp}_{\alpha}}\bra{w_{\alpha}} \sigma_{j,{\alpha}}]\right)}\nonumber\\
   &= \abs{\cos(\theta^{(2^t-1)}_{\alpha})\sin(\theta^{(2^t-1)}_{\alpha})}\abs{\left(\tr[\ket{w_{\alpha}}\bra{w^{\perp}_{\alpha}} \sigma_{j,{\alpha}}] + \tr[\ket{w^{\perp}_{\alpha}}\bra{w_{\alpha}} \sigma_{j,{\alpha}}]\right)}\nonumber\\
   &= \abs{\cos(\theta^{(2^t-1)}_{\alpha})}\abs{\sin(\theta^{(2^t-1)}_{\alpha})}\abs{\left(\tr[\ket{w_{\alpha}}\bra{w^{\perp}_{\alpha}} \sigma_{j,{\alpha}}] + \tr[\ket{w^{\perp}_{\alpha}}\bra{w_{\alpha}} \sigma_{j,{\alpha}}]\right)}\nonumber\\
   &\leq 2\sqrt{2}\sqrt{2^t-1}\eps^{\frac{\beta}{2}} \tr[\sigma_{j,{\alpha}}],\label{extra_term_bound}
\end{align}
where, the last inequality follows from the facts that for all $\alpha \in \G^{(2^t-1)}$, $|\sin(\theta^{(2^t-1)}_{\alpha})| \leq 2\sqrt{2}\sqrt{2^t-1}\eps^{\frac{\beta}{2}},$ $ \abs{\cos(\theta^{(2^t-1)}_{\alpha})} \leq 1$ and from \eqref{positivity_res_1} and \eqref{positivity_res_2} of Lemma \ref{positivity} since $\sigma_{j,{\alpha}} \succeq 0$. Inequality $f$ follows from \eqref{ind_gub_2seq} (see the statement of Lemma \ref{generalubound_seq}) and $g$ follows since $8(2^t-1)\eps^{\beta} \in (0,1]$ is arbitrarily close to zero, $8(2^t-1)\eps^{\beta} \leq \sqrt{8(2^t-1)\eps^{\beta}} = 2\sqrt{2}\sqrt{2^t-1}\eps^{\frac{\beta}{2}}$.

% We again construct intersection projectors $\Pi_{\{1,2,3,4\}},\Pi_{\{5,6,7,8\}},\cdots,\Pi_{\{2^{t}-3,2^{t}-2, 2^{t}-1, 2^{t}\}}$ corresponding to the pairs $\left\{\Pi_{\{1,2\}},\Pi_{\{3,4\}}\right\},\left\{\Pi_{\{5,6\}},\Pi_{\{7,8\}}\right\},\cdots,\left\{\Pi_{\{2^{t}-3,2^{t}-2\}}, \Pi_{\{2^{t}-1,2^{t}\}}\right\}$ respectively and thus $\forall i \in \{1,2,\cdots$\quad$,2^{t-2}\}$, the intersection projector $\Pi_{\{4i-3,\cdots,4i\}}$ has the following properties,
% \begin{align*}
%     \tr[\Pi_{\{4i-3,\cdots,4i\}}\rho] &\geq 1 - 2^2\eps^{(1 - 2.\beta)},\\
%     % \tr[\Pi_{\{2i-1,2i\}}\sigma^{\otimes n}_] &\leq \tr[\Pi_1 \sigma^{\otimes n}_1]\\
%     % &\leq 2^{-nD(\rho || \sigma_1)}\\
%     \tr[\Pi_{\{4i-3,\cdots,4i\}}\sigma_{j}] 
%     &\leq 2^{-k_{j}} + c_2.\eps^{\frac{\beta}{2}}, \forall j \in \{4i-3,\cdots,4i\}.
% \end{align*}

% where, $c_2 \in [0,8\sqrt{2}]$.

Combinining \cref{gub_lowb_seq,gub_case1_seq,gub_case2_seq}, we have the following
\begin{align*}
    \tr[\Pi^{\star}\rho] &\geq 1 - 2\eps^{(1 - (2^t-1)\beta)},\\
    % \tr[\Pi_{\{2i-1,2i\}}\sigma^{\otimes n}_] &\leq \tr[\Pi_1 \sigma^{\otimes n}_1]\\
    % &\leq 2^{-nD(\rho || \sigma_1)}\\
    \tr[\Pi^{\star}\sigma_{j}] 
    &\leq 2^{-{k_j}}+ d(j,\abs{\cS})\eps^{\frac{\beta}{2}}, \forall j \in \{1,2,\cdots,2^{t}\}.
\end{align*}

 This completes the proof of Lemma \ref{generalubound_seq} using a sequential construction of $\Pi^{\star}$.\hfill\qed
 
 \subsection{Proof of Lemma \ref{ssrho}}\label{ssrhop}
\begin{proof}
    The proof of Lemma \ref{ssrho} follows by replacing $k_i$ by $n(D(\rho||\sigma_i))$, and $\rho$ by $\rho^{\otimes n}$ and $\sigma_i$ by $\sigma_i^{\otimes n},$ $\forall i \in \cS$ in \eqref{gubeq} and $t$ by $\ceil{\log\abs{\cS}}$ in \eqref{g1} and \eqref{g3} of Lemma \ref{generalubound}.
\end{proof}

\subsection{Proof of Corollary \ref{main}}
The proof follows techniques similar to the one used in the proof of Lemma \ref{mr}. However, we give its proof for completeness. Let $\Pi^{(1)}_n:=\{\sigma^{\otimes n} \succeq 2^{-n(D(\rho\| \sigma)+\delta)}\rho^{\otimes n}\}$ and consider $n$ large enough such that $\tr[\Pi^{(1)}_n \rho^{\otimes n}] \geq 1-\eps$ (this is guaranteed by Fact \ref{F4}). Further, consider  $\left\{\mathbf{P}_{n,\alpha}\right\}_{{\alpha}=1}^{{k}}$ as the orthogonal projectors obtained by Fact \ref{jordan} with respect to the pair $\left\{\Pi^{(1)}_n,\Pi^{(2)}_n\right\}$. For every $\alpha \in [1:k]$ we define  $ \Pi^{(1)}_{n,\alpha}:=  \mathbf{P}_{n,\alpha}\Pi^{(1)}_n \mathbf{P}_{n,\alpha}$ and $\Pi^{(2)}_{n,\alpha}:= \mathbf{P}_{n,\alpha}\Pi^{(2)}_n \mathbf{P}_{n,\alpha}$ as the following one dimensional projectors, 
\begin{align*}
\Pi^{(1)}_{n,\alpha} &= \ket{v_{n,\alpha}}\bra{v_{n,\alpha}},\\
\Pi^{(2)}_{n,\alpha} &= \ket{w_{n,\alpha}}\bra{w_{n,\alpha}},
\end{align*}
for some $\ket{v_{n,\alpha}}, \ket{w_{n,\alpha}}$ in the subspace spanned by the eigen vectors of $\mathbf{P}_{n,\alpha}$. Thus, it is now follows from the property of $\{\ket{v_{n,\alpha}}\}_{{\alpha}=1}^k$ and $\{\ket{w_{n,\alpha}}\}_{{\alpha}=1}^k$ that 
%%\vspace{5pt}
\begin{align}
% \label{pi_1_corollary1}
\Pi^{(1)}_{n} &:= \sum_{{\alpha} =1}^k \ket{v_{n,\alpha}}\bra{v_{n,\alpha}},\nonumber
\end{align}
\begin{align}
\Pi^{(2)}_{n}&:= \sum_{{\alpha} =1}^k \ket{w_{n,\alpha}}\bra{w_{n,\alpha}}.\label{pi_2_corollary1}
\end{align}
%%%\vspace{5pt}

Further, for every ${\alpha} \in [1:k],$ let 
%%\vspace{5pt}
\begin{align*}
\rho^{n}_{\alpha} &:= \mathbf{P}_{n,\alpha}\rho^{\otimes n} \mathbf{P}_{n,\alpha},\\
% \sigma_{1,{\alpha}} &:= \Tilde{\Pi}_{\alpha} \sigma_1 \Tilde{\Pi}_{\alpha},\\
% \sigma_{2,{\alpha}} &:= \Tilde{\Pi}_{\alpha} \sigma_2 \Tilde{\Pi}_{\alpha}\\
\sigma^{n}_{\alpha} &:= \mathbf{P}_{n,\alpha} \sigma^{\otimes n} \mathbf{P}_{n,\alpha}.
\end{align*}

Now for ${\alpha} \in [1:k],$ $\ket{w_{n,\alpha}}$ can be defined in the following way,
\begin{equation}
    \ket{w_{n,\alpha}} = \cos(\theta_{n,\alpha})\ket{v_{n,\alpha}} + \sin(\theta_{n,\alpha})\ket{v^{\perp}_{n,\alpha}}.\label{v_n_alpha}
\end{equation}

Similar to \eqref{good_lb}, in the proof of Lemma \ref{mr}, we define $\G$ as follows,
\begin{align}
{\G}&:=\left\{{\alpha}: \cos^2(\theta_{\alpha}) \geq 1- 8\eps^{\frac{1}{2}}\right\},\label{good_lb_corollary}
\end{align}
where, $\cos^2(\theta_{n,\alpha}) = {|\langle v_{n,\alpha}| w_{n,\alpha}\rangle|}^2.$ 
% Using arguments similar to that of the one used in the proof of Lemma \ref{mr}, we can show that $\Pr\{\G\} \geq 1- 2\eps^{\frac{1}{2}}.$
%\begin{claim}
%\label{c5}
%For every $\alpha \in [1:k]$ the vectors $\ket{w_{\alpha}}$ and $\ket{v_{\alpha}}$ lie in a two dimensional subspace spanned by the eigen vectors of $\mathbf{P}_{n,\alpha}.$ Thus,  
%\begin{equation}
%\label{walpha}
%\ket{w_{n,\alpha}} = \cos(\theta_{n,\alpha})\ket{v_{n,\alpha}} + \sin(\theta_{n,\alpha})\ket{v^{\perp}_{n,\alpha}}
%\end{equation}

We now prove the corollary using the following set of inequalities, 
\begin{align*}
&\tr[\Pi^{(2)}_n\sigma^{\otimes n}] \overset{a}= \sum_{\alpha=1}^k\tr[\ket{w_{n,\alpha}}\bra{w_{n,\alpha}} \sigma^{n}_{\alpha}] \\
& \overset{b}\geq \sum_{\alpha \in \G}\tr[\ket{w_{n,\alpha}}\bra{w_{n,\alpha}} \sigma^{n}_{\alpha}]\\
&\overset{c}{=} \sum_{\alpha \in \G}\bigg(\cos^2(\theta_{n,\alpha})\tr[\ket{v_{n,\alpha}}\bra{v_{n,\alpha}} \sigma^{n}_{\alpha}] + \sin^2(\theta_{n,\alpha}) \tr[\ket{v^{\perp}_{n,\alpha}}\bra{v^{\perp}_{n,\alpha}} \sigma^{n}_{\alpha}] \hspace{95pt}\\
&\hspace{60pt}+ \cos(\theta_{n,\alpha})\sin(\theta_{n,\alpha}) ( \tr[\ket{v_{n,\alpha}}\bra{v^\perp_{n,\alpha}} \sigma^{n}_{\alpha}] 
 +\tr[\ket{v^\perp_{n,\alpha}}\bra{v_{n,\alpha}} \sigma^{n}_{\alpha}]) \bigg)\\
&\overset{d}{\geq} \sum_{\alpha \in \G}\bigg((1-8\eps^{\frac{1}{2}})\tr[\ket{v_{n,\alpha}}\bra{v_{n,\alpha}} \sigma^{n}_{\alpha}] +  \cos(\theta_{n,\alpha})\sin(\theta_{n,\alpha}) ( \tr[\ket{v_{n,\alpha}}\bra{v^\perp_{n,\alpha}} \sigma^{n}_{\alpha}] 
 +\tr[\ket{v^\perp_{n,\alpha}}\bra{v_{n,\alpha}} \sigma^{n}_{\alpha}]) \bigg)\\
 &\overset{e}{\geq} \sum_{\alpha \in \G}\bigg((1-8\eps^{\frac{1}{2}})\tr[\ket{v_{n,\alpha}}\bra{v_{n,\alpha}} \sigma^{n}_{\alpha}] - \abs{ \cos(\theta_{n,\alpha})\sin(\theta_{n,\alpha}) ( \tr[\ket{v_{n,\alpha}}\bra{v^\perp_{n,\alpha}} \sigma^{n}_{\alpha}] 
 +\tr[\ket{v^\perp_{n,\alpha}}\bra{v_{n,\alpha}} \sigma^{n}_{\alpha}])} \bigg)\\
 &\overset{f}{\geq} \sum_{\alpha \in \G}\bigg((1-8\eps^{\frac{1}{2}})\tr[\ket{v_{n,\alpha}}\bra{v_{n,\alpha}} \sigma^{n}_{\alpha}] -  2\sqrt{2}\eps^{\frac{1}{4}}\tr[\sigma^{n}_{\alpha}]\bigg)
\end{align*}
% &= 2^{-n(D(\rho\| \sigma)+\delta)}\cos(2\theta_{n,\alpha})\tr[\Pi^{(2)}_n\rho^{\otimes n}] \\
% &\overset{f}\geq \sum_{\alpha \in \G}\left(2^{-n(D(\rho\| \sigma)+\delta)}\cos(2\theta_{n,\alpha})(1-\eps^{\frac{1}{2}})\tr[\rho^{n}_{\alpha}]\right)\\
% &\overset{g}\geq2^{-n(D(\rho\| \sigma)+\delta)}(1-\eps^{\frac{1}{2}})(1-16\eps^{\frac{1}{2}}) \sum_{\alpha \in \G}\left(\tr[\rho^{n}_{\alpha}]\right)\\
% %&\geq  2^{-n(D(\rho\| \sigma)+\delta)}(1-\eps^{\frac{1}{2}})\sum_{\alpha \in \G}\cos(2\theta_{n,\alpha})\tr[\rho^{n}_{\alpha}]\\
\begin{align*}
&\overset{g}\geq \sum_{\alpha \in \G}\bigg(2^{-n(D(\rho\| \sigma)+\delta)}(1-8\eps^{\frac{1}{2}})\tr[\ket{v_{n,\alpha}}\bra{v_{n,\alpha}} \rho^{n}_{\alpha}]\bigg) -2\sqrt{2}\eps^{\frac{1}{4}} \hspace{240pt}\\ 
&= 2^{-n(D(\rho\| \sigma)+\delta)}(1-8\eps^{\frac{1}{2}})\sum_{\alpha \in \G}\left(\tr[\ket{v_{n,\alpha}}\bra{v_{n,\alpha}} \rho^{n}_{\alpha}]\right) - 2\sqrt{2}\eps^{\frac{1}{4}}\\
&\overset{h}\geq 2^{-n(D(\rho\| \sigma)+\delta)}(1-8\eps^{\frac{1}{2}})(1-2\eps^{\frac{1}{2}}) - 2\sqrt{2}\eps^{\frac{1}{4}}\\
&= 2^{-n(D(\rho\| \sigma)+\delta)}(1-10\eps^{\frac{1}{2}}) - 2\sqrt{2}\eps^{\frac{1}{4}},
\end{align*}
where, $a$ follows from the definition of \eqref{pi_2_corollary1}, $b$ follows because each term in the summation is non-negative, $c$ follows from \eqref{v_n_alpha}, $d$ follows from $\cos(\theta_{n,\alpha}) \geq 1-8\eps^{\frac{1}{2}}$ for every $\alpha \in \G$ due to \eqref{good_lb_corollary} and the trivial bound $\sin^2(\theta_{n,\alpha}) \tr[\ket{v^{\perp}_{n,\alpha}}\bra{v^{\perp}_{n,\alpha}} \sigma^{n}_{\alpha}] \geq 0$, $e$ follows from facts that for any  real number $x$, $x \geq -\abs{x}$ and $\cos(\theta_{n,\alpha})\sin(\theta_{n,\alpha}) ( \tr[\ket{v_{n,\alpha}}\bra{v^\perp_{n,\alpha}} \sigma^{n}_{\alpha}] 
 +\tr[\ket{v^\perp_{n,\alpha}}\bra{v_{n,\alpha}} \sigma^{n}_{\alpha}])$ is a real number, $f$ follows from \eqref{extra_term_bound}, $g$ follows from \eqref{property2} and $\sum_{{\alpha} \in  {\G}}\tr[\sigma^{n}_{\alpha}] \leq \sum_{\alpha = 1}^{k}\tr[\sigma^{n}_{\alpha}] = 1$ and $h$ follows from the fact that $\sum_{\alpha \in \G}\tr[\ket{v_{n,\alpha}}\bra{v_{n,\alpha}} \rho^{n}_{\alpha}] \geq 1-2\eps^{\frac{1}{2}}$ due to \eqref{prop_rlb1}. This completes the proof of Corollary \ref{main}.\hfill\qed

% where, $a$ follows from the definition of \eqref{pi_2_corollary1}, $b$ follows because each term in the summation is non-negative, $c$ follows from \eqref{v_n_alpha}, $d$ follows from the inequality \eqref{property}, $e$ follows from \eqref{positivity_res_2} of lemma \eqref{positivity} and because $\sigma^{n}_{\alpha}\succeq 0$, $f$ follows from the fact that $\cos(2\theta_{n,\alpha})= \cos^2(\theta_{n,\alpha}) - \sin^2(\theta_{n,\alpha})$ and for $\alpha \in \G,$ $\cos^2(\theta_{n,\alpha}) \geq 1-8 \eps^{\frac{1}{2}}$ and $g$ trivially follows from the fact that $tr[\ket{v_{n,\alpha}}\bra{v_{n,\alpha}} \rho^{n}_{\alpha}] \geq 1-2\eps^{\frac{1}{2}}$ from proof of \ref{rlb1} in subsection \ref{subsec_rlb1}.
\begin{remark}
    It appears that the proof of Corollary \ref{main} is exactly similar to that of Lemma \ref{mr}. In Lemma \ref{mr} we need operator inequalities with respect to both $\Pi_1$ and $\Pi_2$. However, this is not the case in Corollary \ref{main}.
\end{remark}
\subsection{Proof of Corollary \ref{decode_s0}}\label{corr_decode_s0_proof}

We consider a collection of states $\left\{{\rho^{(s)}}^{\otimes n}\right\}_{s \in \cS\setminus\{s_0\}}$ and a state ${\rho^{(s_0)}}^{\otimes n}$, where $\forall s \in \cS$ ${\rho^{(s)}}^{\otimes n} = \mathbb{E}_{X^n}\left[\rho^{(s)}_{X^n}\right]$. From the statement of Corollary \ref{decode_s0}, for each $s \in \cS$, $\Pi^{(s)}_{(n)}$, satisfies the following properties,
\begin{align*}
\tr\left[\Pi^{(s)}_{(n)} {\rho^{(s_0)}}^{\otimes n}\right] &= \mathbb{E}_{X^n}\left[\tr\left[\Pi^{(s)}_{(n)}\rho_{X^n}^{(s_0)}\right]\right] \geq 1 - \eps,\\
\tr\left[\Pi^{(s)}_{(n)} {\rho^{(s)}}^{\otimes n}\right] &= \mathbb{E}_{X^n}\left[\tr\left[\Pi^{(s)}_{(n)}\rho_{X^n}^{(s)}\right]\right] \leq 2^{-n(D(\rho^{(s_0)} || \rho^{(s)}) - \delta)}.
    \end{align*}
    
Now we can directly apply Lemma \ref{generalubound} on  $\left\{{\rho^{(s)}}^{\otimes n}\right\}_{s \in \cS\setminus\{s_0\}}$ and ${\rho^{(s_0)}}^{\otimes n}$ and we get the following result
\begin{align}
        % \mathbb{E}_{X^n}\left[P_{Y^n|X^n}^{(s_0)}(D^{(s_0)})\right] &\geq 1 - (\ceil{\log|\cS|} + 1)\eps^{1 - \ceil{\log|\cS|}.\beta},\\
        \mathbb{E}_{X^n}\left[\tr\left[\Pi_{(n)}^{\neg{\perp}}\rho_{X^n}^{(s_0)}\right]\right] &\geq 1 -\left(\ceil{\log{(|\cS| -1)}}+1\right)\eps^{\left({1 - \ceil{\log(|\cS|-1)}}\beta\right)},\nonumber\\
        % \mathbb{E}_{X^n}\left[P_{Y^n|X^n}^{(s)}(D^{(s_0)})\right] &\leq 
        % 2^{-n(D(P_{Y^n}^{(s_0)}||P_{Y^n}^{(s)}) + \delta)}+O(\eps^{\frac{1}{4}}), \quad \forall s \in \cS
        \mathbb{E}_{X^n}\left[\tr\left[\Pi_{(n)}^{\neg{\perp}}\rho_{X^n}^{(s)}\right]\right] &\leq 
        2^{-n\left(D(\rho^{(s_0)} || \rho^{(s)}) - \delta\right)} + c\left(\ceil{\log(|\cS|-1)}\right)\eps^{\frac{\beta}{2}},  \forall s \in \cS\setminus\{s_0\},\nonumber\\
        &\overset{a}{\leq} 
        2^{-n\left(D(\rho^{(s_0)} || \rho^{(s)}) - \delta\right)} + O\left(\left(\ceil{\log(|\cS|-1)}\right)^\frac{3}{2}\right)\eps^{\frac{\beta}{2}},  \forall s \in \cS\setminus\{s_0\},\nonumber\\\label{decode_so_proof_1}
    \end{align}
     where, $a$ follows from the fact that $c\left(\ceil{\log(|\cS|-1)}\right) := \sum_{i=1}^{\ceil{\log(|\cS|-1)}}\sqrt{i} = O(\left(\ceil{\log(|\cS|-1)}\right)^\frac{3}{2})$ and since $|\cS|-1$ may not be in the form of $2^{t}$ (for $t > 0$), we have used the term $\ceil{\log(|\cS|-1)}$ in the above two inequalities, as discussed in the paragraph below Lemma \ref{gen_cht} (mentioned in bold font). Finally, we can rewrite \eqref{decode_so_proof_1} as follows,
    \begin{align*}
        \mathbb{E}_{X^n}\left[\tr\left[\Pi_{(n)}^{\neg{\perp}}\rho_{X^n}^{(s)}\right]\right] &\leq 
        2^{-n\left(\underset{\substack{s \in \cS: s \neq s_0}}{\min}D(\rho^{(s_0)} || \rho^{(s)}) - \delta\right)} + O\left(\left(\ceil{\log(|\cS|-1)}\right)^\frac{3}{2}\right)\eps^{\frac{\beta}{2}},  \forall s \in \cS\setminus\{s_0\}.
    \end{align*}

This completes the proof of Corollary \ref{decode_s0}.\hfill\qed
% \subsection*{Proof of claim \ref{c2}}
% The proof of claim \ref{c2} follows from the following set of inequalities, 

% \begin{align}
% 1-\eps & \overset {a} \leq \tr[\Pi_1 \rho] \nonumber \\
% & \overset{b}= \sum_{{\beta} =1}^k \tr[\ket{v_{\beta}}\bra{v_{\beta}} \rho_{\beta}] \nonumber\\
% & = \sum_{{\beta} \in \cE_2} \tr[\ket{v_{\beta}}\bra{v_{\beta}} \rho_{\beta}] + \sum_{{\beta} \in \cE_2^c} \tr[\ket{v_{\beta}}\bra{v_{\beta}} \rho_{\beta}] \nonumber\\
% & \overset{c}\leq \sum_{{\beta} \in \cE_2} \tr[\rho_{\beta}] + (1-\eps^{\frac{1}{2}}) \sum_{{\beta} \in \cE^c_2} \tr[\rho_{\beta}] \nonumber\\
% & = \Pr\{\cE_2\} + (1-\eps^{\frac{1}{2}})(1- \Pr\{\cE_2\})\nonumber \\
% \label{ep2}
% & = 1-\eps^{\frac{1}{2}} + \eps^{\frac{1}{2}}\Pr\{\cE_2\},
% \end{align}
% where, $a$ follows from the property of the projector $\Pi_1$ given in Lemma \ref{main}, $b$ follows from the Jordan's lemma and $c$ from the definition of the set $\cE_2$ and from montonicity property of the trace and the fact that $\ket{v_{\beta}}\bra{v_{\beta}} \leq \overline{\Pi}_{\beta}.$ claim \ref{c2} now easily follows from \eqref{ep2}.
\end{document}